\documentclass[11pt, a4paper, oneside]{Thesis} 

\graphicspath{{Pictures/}} 

\usepackage{cite}
\usepackage{mathrsfs}
\usepackage{fancyhdr}
\usepackage{amsmath,latexsym}
\usepackage{bm}
\usepackage{float}
\usepackage{caption}
\usepackage{relsize}
\usepackage{subcaption}
\usepackage{tikz}
\newtheorem*{theorem*}{\bf{Theorem}}
\usepackage{hyperref}
\hypersetup{colorlinks,citecolor=red}
\hypersetup{colorlinks=true,urlcolor=black}

\newcommand{\CLC}[3]{ \left\lbrace {}^{#1}{}_{#2 #3} \right\rbrace}

\newcommand{\RLC}{ \overset{ \scalebox{.7}{$\mathclap[\scriptscriptstyle]{\mathrm{LC}}$} }{R}{}}
\newcommand{\RWC}{ \overset{ \scalebox{.7}{$\mathclap[\scriptscriptstyle]{\mathrm{WC}}$} }{R}{}}
\newcommand{\RSTG}{ \overset{ \scalebox{.7}{\mathclap[\scriptscriptstyle]{\mathrm{STG}}} }{R}{}}

\newcommand{\TLC}{ \overset{ \scalebox{.7}{$\mathclap[\scriptscriptstyle]{\mathrm{LC}}$} }{T}{}}

\newcommand{\TSTG}{ \overset{ \scalebox{.7}{$\mathclap[\scriptscriptstyle]{\mathrm{STG}}$} }{T}{}}

\newcommand{\QLC}{ \overset{ \scalebox{.7}{$\mathclap[\scriptscriptstyle]{\mathrm{LC}}$} }{Q}{}}
\newcommand{\QWC}{ \overset{ \scalebox{.7}{$\mathclap[\scriptscriptstyle]{\mathrm{WC}}$} }{Q}{}}

\newcommand{\GSTG}{ \overset{ \scalebox{.7}{$\mathclap[\scriptscriptstyle]{\mathrm{STG}}$} }{\Gamma}{}}

\newcommand{\DWC}{ \overset{ \scalebox{.7}{$\mathclap[\scriptscriptstyle]{\mathrm{WC}}$} }{\nabla}{}}

\allowdisplaybreaks[1]

\title{\ttitle} 
\DeclareUnicodeCharacter{2212}{-}
\DeclareUnicodeCharacter{2212}{+}

\begin{document}
\setstretch{1.3} 

\fancyhead{} 
\rhead{\thepage} 
\lhead{} 

%

\thesistitle{Cosmological Models with Symmetric Teleparallel Gravity and its Extension}
\documenttype{\textbf{THESIS}}
\supervisor{Prof. BIVUDUTTA MISHRA}
\supervisorposition{Professor}
\supervisorinstitute{BITS-Pilani, Hyderabad Campus}
\examiner{}
\degree{Ph.D. Research Scholar}
\coursecode{\textbf{DOCTOR OF PHILOSOPHY}}
\coursename{Thesis}
\authors{\textbf{Shubham Atmaram Narawade}}
\IDNumber{\textbf{2021PHXF0039H}}
\addresses{}
\subject{}
\university{\texorpdfstring{\href{http://www.bits-pilani.ac.in/} 
                {\textbf{BIRLA INSTITUTE OF TECHNOLOGY AND SCIENCE, PILANI}}} 
                {\textbf{BIRLA INSTITUTE OF TECHNOLOGY AND SCIENCE, PILANI}}}
\UNIVERSITY{\texorpdfstring{\href{http://www.bits-pilani.ac.in/} 
                {\textbf{BIRLA INSTITUTE OF TECHNOLOGY AND SCIENCE, PILANI}}} 
                {\textbf{BIRLA INSTITUTE OF TECHNOLOGY AND SCIENCE, PILANI}}}


\department{\texorpdfstring{\href{http://www.bits-pilani.ac.in/pilani/Mathematics/Mathematics} 
                {Mathematics}} 
                {Mathematics}}
\DEPARTMENT{\texorpdfstring{\href{http://www.bits-pilani.ac.in/pilani/Mathematics/Mathematics} 
                {Mathematics}} 
                {Mathematics}}
\group{\texorpdfstring{\href{Research Group Web Site URL Here (include http://)}
                {Research Group Name}} 
                {Research Group Name}}
\GROUP{\texorpdfstring{\href{Research Group Web Site URL Here (include http://)}
                {RESEARCH GROUP NAME (IN BLOCK CAPITALS)}}
                {RESEARCH GROUP NAME (IN BLOCK CAPITALS)}}
\faculty{\texorpdfstring{\href{Faculty Web Site URL Here (include http://)}
                {Faculty Name}}
                {Faculty Name}}
\FACULTY{\texorpdfstring{\href{Faculty Web Site URL Here (include http://)}
                {FACULTY NAME (IN BLOCK CAPITALS)}}
                {FACULTY NAME (IN BLOCK CAPITALS)}}
\maketitle

\clearpage

\pagestyle{empty} 
\pagenumbering{gobble}

\frontmatter
\Certificate

\Declaration

%
\pagestyle{empty}

\Dedicatory{\bf \begin{LARGE}
\textit{\textbf{To}} 
\end{LARGE} 
\\
\vspace{1.5cm}
\begin{Large} \textit{\textbf{My Parents, Wife \& Friends}} \end{Large}\\
 \vspace{1cm}
 }
\addtocontents{toc}{\vspace{2em}}


\begin{acknowledgements}
It gives me great joy to have the chance to offer my sincere gratitude to everyone whose inspirational influences have supported me as I pursue my doctorate.

My supervisor, \textbf{Prof. Bivudutta Mishra}, Professor, Department of Mathematics, BITS-Pilani, Hyderabad Campus, is to be sincerely thanked. His tenacity and unwavering excitement motivate me. The chance to collaborate with him has been a wonderful privilege. Along with his superior mathematical knowledge and teaching abilities, his humanism has served as a wonderful model for me. This dissertation would not exist without his committed help and direction. 

I would also like to thank my DAC members, \textbf{Prof. Pradyumn Kumar Sahoo} and \textbf{Prof. Prasant Samantray}, for their insightful questions, insightful comments and constant feedback throughout this research.
 
I would like to extend my sincere gratitude to the Head of the Department, the DRC Convener and the entire staff of the Department of Mathematics at BITS-Pilani Hyderabad Campus for their assistance, support and encouragement in carrying out my research.

I am also grateful to the Associate Dean of AGRSD BITS-Pilani, Hyderabad Campus. 

I gratefully acknowledge BITS-Pilani, Hyderabad Campus, for providing the necessary facilities and financial support during my research endeavors.

I would like to take this opportunity to thank, \textbf{Prof. Sunil Kumar Tripathy, Prof. Tomohiro Inagaki} and \textbf{Mr. Muhammad Azzam Alwan} for providing me with the opportunity to collaborate with them. I have gained a great deal of knowledge from them, for which I am quite grateful.

I would like to express my gratitude to my research team {\bf Dr. Amar, Dr. Siddheshwar, Dr. Santosh, Dr. Lokesh, Rahul, Kalpana, Shivam, Priyobarta}. I would also like to acknowledge all my friends at BITS-Pilani, Hyderabad Campus for their unwavering support and encouragement throughout this journey. While not all are mentioned by name, each one has played a vital role in my success.

Most importantly, I would like to thank my parents, wife, brothers, sisters, family members and friends for their love, care and support for my personal life.
\end{acknowledgements}


\clearpage

\begin{abstract}       
This thesis investigates late-time cosmic acceleration using modified gravity theories with a focus on $f(Q)$ gravity, as an alternative to the $\Lambda$CDM model. The standard cosmological model attributes the acceleration to a cosmological constant, but it faces issues like the unexplained nature of dark matter and dark energy and discrepancies with certain observations. Modified gravity including $f(Q)$ gravity, offers a potential solution by incorporating dynamic dark energy or changes to gravitational interactions, avoiding the need for a constant cosmological term. Also, thesis evaluates the viability of $f(Q)$ gravity by analyzing observational data from Type Ia Supernovae, Hubble parameter measurements and other cosmological datasets. Using statistical tools like Markov Chain Monte Carlo (MCMC) analysis, this work constrains the parameters of $f(Q)$ gravity and compares it to the $\Lambda$CDM model.

The chapter \ref{Chapter1} provides an overview of background formulation, fundamental gravity theories and cosmological observations. Chapters \ref{Chapter2}-\ref{Chapter5} explore the dark energy sector of the Universe within the context of modified gravity, using MCMC methods and extensive datasets obtained from measurements of the background expansion. Our analysis includes the reconstruction as well as the stability of the cosmological model.

In chapter \ref{Chapter2}, we present a comprehensive analysis of two distinct forms of $f(Q)$, specifically the log-square-root model and the exponential model, within the framework of cosmological evolution driven by dark energy. The equation of state (EoS) parameter for dark energy in both models is derived as a dynamical quantity, with log-square-root model exhibiting quintessence behavior at present and converges to the $\Lambda$CDM model at late-time, while exponential model demonstrates phantom behavior. A dynamical system analysis is employed to examine the stability of critical points and phase portraits, revealing the evolutionary behavior of the matter, radiation and dark energy phases.

The chapter \ref{Chapter3} delves into the $f(Q)$ gravity framework, investigating the cosmological model by constraining its parameters using cosmological datasets, including recent Hubble data and the Pantheon+SH0ES datasets. An MCMC analysis is performed to constrain the free parameter of the Hubble function, with the results being validated against the BAO dataset. The parametrization of cosmographic parameters results the early deceleration and late-time acceleration transition, with the equation of state parameter indicating a phantom behavior, further supported by the $Om(z)$ diagnostics. Additionally, the current age of the Universe is determined.

In chapter \ref{Chapter4}, we explore the reconstruction of the $q(z)$ and its impact on the $f(Q)$ gravity, utilizing OHD, SNe Ia and BAO datasets. Two parametrization of $q(z)$ are fitted to the cosmological models using MCMC analysis. The model I, parameterized by a power-law function of redshift, shows strong alignment with recent higher $H_0$ measurements, while model II, incorporating logarithmic terms, provides tighter constraints on matter density. Both models successfully recover observationally consistent estimates for the Hubble constant $H_0$ and matter density $\Omega_{m_0}$, with varying precision in alignment with the $\Lambda$CDM model. Statistical comparisons with the Akaike Information Criteria (AIC) and Bayesian Information Criterion (BIC) indicate that these $f(Q)$ models are competitive alternatives to the $\Lambda$CDM model in explaining dark energy and cosmic acceleration.

The chapter \ref{Chapter5}, focuses on the reconstruction of the cosmological model within the framework of $f(Q)$ gravity through non-trivial connection. The dynamic behavior of two models, particularly model within coincident gauge and the reconstructed model are thoroughly studied using the Hubble parameter $H(z)$ and various observational datasets for the comparative analysis. The findings confirm that both the model exhibits quintessence behavior at the present epoch and asymptotically approaches the $\Lambda$CDM model at late-time. Additionally, we explore the energy conditions and the violation of strong energy condition supports the accelerating behavior of the models. The scalar perturbation analysis confirms the stability of the reconstructed model with respect to the Hubble parameter. These results provide strong support for $f(Q)$ gravity as a viable alternative to the $\Lambda$CDM model, offering compelling insights into the nature of cosmic acceleration and the evolution of the Universe.

The concluding chapter \ref{Chapter6} of the thesis summarizes the key findings and discusses the research presented throughout the thesis, highlighting potential future directions for research and applications in the field.
\end{abstract}


\pagebreak

\pagestyle{empty} 



\addtocontents{toc}{\vspace{1em}}
\tableofcontents 
\addtocontents{toc}{\vspace{1em}}
\lhead{\emph{List of Tables}}
\listoftables 
\addtocontents{toc}{\vspace{1em}}
\lhead{\emph{List of Figures}}
\listoffigures 
\addtocontents{toc}{\vspace{1em}}


\setstretch{1.5}
\lhead{\emph{List of Symbols}}
\listofsymbols{ll}{
\begin{tabular}{cp{0.5\textwidth}}
$H$ & : Hubble parameter\\
$q$ & : Deceleration parameter\\
$s$ & : Snap parameter\\
$\mathrm{j}$ & : Jerk parameter\\
$z$ &: Redshift\\
$\mathcal{L}_{\text{m}}$ & : Matter Lagrangian\\
$g_{ij}$ &: Lorentzian metric\\
$g$ &: Determinant of $g_{ij}$\\ 
$G_{ij}$ & : Einstein tensor\\
$\Lambda$ & : Cosmological constant\\
$\Gamma^{k}_{ij}$  &: General affine connection \\ 
$\{^{k}_{~ij}\}$ &: Levi-Civita connection\\  
$\nabla_{i}$ &: Covariant derivative \\ 
$R^{k}_{\sigma ij}$ &: Riemann tensor \\
$R_{ij}$ &: Ricci tensor \\
$R$ &: Ricci scalar \\
$T$ & : Torsion \\
$Q$ & : Non-metricity scalar\\
$S_{M}$ & : Matter action\\
$T_{ij}$ & : Energy-momentum tensor\\
$\mathcal{T}$ & : Trace of the energy-momentum tensor \\
$L^{k}_{ij}$  &: Disformation tensor \\ 
$P^{k}_{ij}$  &: Super potential \\
$H^{\lambda}_{\mu\nu}$ &: Hyper momentum tensor\\
$\chi^2$ &: Chi-square\\
\end{tabular}
}
\addtocontents{toc}{\vspace{1em}}
\clearpage

\setstretch{1.5}
\lhead{\emph{Acronyms}}
\listofacronyms{ll}{
\begin{tabular}{cp{0.6\textwidth}}
\textbf{GR} &: \textbf{G}eneral \textbf{R}elativity\\
\textbf{TEGR} &: \textbf{T}eleparallel \textbf{E}quivalent to \textbf{G}eneral \textbf{R}elativity\\
\textbf{STEGR} &: \textbf{S}ymmetric \textbf{T}eleparallel \textbf{E}quivalent to \textbf{G}eneral \textbf{R}elativity\\
\textbf{$\Lambda$CDM} &: $\Lambda$ \textbf{C}old \textbf{D}ark \textbf{M}atter\\
\textbf{EoS} &: \textbf{E}quation \textbf{o}f \textbf{S}tate\\
\textbf{CMB} &: \textbf{C}osmic \textbf{M}icrowave \textbf{B}ackground\\
\textbf{BAO} &: \textbf{B}aryon \textbf{A}coustic \textbf{O}scillations\\
\textbf{MCMC} &: \textbf{M}arkov \textbf{C}hain \textbf{M}onte \textbf{C}arlo\\
\textbf{DE} &: \textbf{D}ark \textbf{E}nergy\\
\textbf{DM} &: \textbf{D}ark \textbf{M}atter\\
\textbf{MG} &: \textbf{M}odified \textbf{G}ravity\\
\textbf{FLRW} &: \textbf{F}riedmann \textbf{L}ema\^{i}tre \textbf{R}obertson \textbf{W}alker\\
\textbf{MCMC} &: \textbf{M}arkov \textbf{C}hain \textbf{M}onte \textbf{C}arlo\\
\textbf{SNe Ia} &: Type \textbf{I}a \textbf{S}uper\textbf{n}ova\\
\textbf{CC} &: \textbf{C}osmic \textbf{C}hronometers\\
\textbf{OHD} &: \textbf{O}bservational \textbf{H}ubble \textbf{D}ata\\
\textbf{EC} &: \textbf{E}nergy \textbf{C}onditions\\
\textbf{WEC} &: \textbf{W}eak \textbf{E}nergy \textbf{C}ondition\\
\textbf{NEC} &: \textbf{N}ull \textbf{E}nergy \textbf{C}ondition\\
\textbf{SEC} &: \textbf{S}trong \textbf{E}nergy \textbf{C}ondition\\
\textbf{DEC} &: \textbf{D}ominant \textbf{E}nergy \textbf{C}ondition\\
\textbf{LC} &: \textbf{L}evi \textbf{C}ivita\\
\textbf{WC} &: \textbf{W}eitzenb\"ock \textbf{C}onnection\\
\textbf{STG} &: \textbf{S}ymmetric \textbf{T}eleparallel \textbf{G}eometry
\end{tabular}
}
\addtocontents{toc}{\vspace{1em}}
\clearpage 
\lhead{\emph{Glossary}} 

 


\mainmatter 
\setstretch{1.2}
\pagestyle{fancy} 



 \chapter{Introduction} 
\label{Chapter1}

\lhead{Chapter 1. \emph{Introduction}} 


The study of cosmology aims to explain the history and future of the Universe, from its birth to ultimate fate. The standard cosmological model is based on the cosmological principle, which states that the Universe is homogeneous and isotropic on large scales. Einstein’s General Relativity (GR) Theory expanded upon this by providing a more comprehensive framework for understanding gravity, especially in regions of intense gravitational forces \cite{Moffat_2021_2021_017, Vilhena_2023_2023_044}. GR not only predicted black holes but also anticipated the existence of gravitational waves, both of which have since been observed, confirming Einstein’s theories. In $1917$, he proposed a static model of the Universe, introducing a cosmological constant to counteract gravitational forces \cite{Einstein_1917_142}. However, the model was abandoned when it became clear that the Universe was not static but expanding. Russian mathematician, Alexander Friedmann found solutions to Einstein’s equations that showed an expanding Universe, which set the stage for the Big Bang theory. According to this model, the Universe is believed to have begun from an incredibly hot and dense state approximately 13 billion years ago, which is radically transforming our understanding of its origin \cite{Friedman_1922_10_377}. The expansion of the Universe was first observed by Edwin Hubble in 1929 \cite{Hubble_1929_15_168}, who found that galaxies were receding from us at velocities proportional to their distance, a relationship now known as Hubble's law, $v = H_0 d$ \cite{Hubble_1931_74_43}. Following this, Friedmann \cite{Friedman_1922_10_377, Friedman_1924_21_326}, Lemaître \cite{Lemaitre_1927_49_1635}, Robertson \cite{Robertson_1935_82_284} and Walker \cite{Walker_1937_42_90} presented new solutions to Einstein’s field equation for an expanding Universe. These contributions led to the development of the FLRW metric, which became the fundamental aspect of modern cosmology, describing the Universe that is both homogeneous and isotropic at large scales and expanding over time. 

This discovery, along with the later detection of the CMB radiation in 1965 by Penzias and Wilson, provided crucial evidence supporting the Big Bang theory and the expansion of the Universe. However, when examining the smaller scales, the Universe is far from the uniform can be observed. It is filled with galaxies, galaxy clusters and vast voids, revealing an inhomogeneous structure at scales smaller than about 350 Mpc \cite{Yadav_2005_364_601, Clowes_2013_429_2910}. Moreover, the CMB radiation, a remnant of the Big Bang shows minute fluctuations that suggest slight variations in temperature, but these are incredibly small on the order of one part in $10^{5}$ \cite{Hu_2002_40_171}. Following that, significant advances were made in both theoretical and observational fields. Starobinsky \cite{Starobinsky_1979_30_682, Starobinsky_1982_117_175}, Guth \cite{Guth_1981_23_347} and Linde \cite{Linde_1982_108_389} proposed cosmic inflation in the late 1970s and early 1980s to explain the origin of large-scale structures of the Universe. An exponential expansion is described in this theory, which also addresses the problems of horizons, flatness and monopoles. The first inflationary models were applied by Guth in 1981. He proposed that large-scale structures were formed from the gravitational collapse of perturbations after quantum fluctuations during inflationary periods \cite{Guth_1981_23_347}. The standard cosmological model describes how matter and energy influence the curvature of spacetime, dictating the behavior of the Universe. 

In the early 20$^{\text{th}}$ century, advancements in technology played a crucial role in confirming the existence of galaxies beyond our own, with observations revealing that these galaxies were all moving away from us \cite{Slipher_1917_56_403}. The late $1990$s brought a new understanding of the evolution of the Universe. According to observations of SNe Ia in $1998$, the Universe is expanding more rapidly than previously believed. According to two groups, Riess and Schmidt \cite{Riess_1998_116_1009, Schmidt_1998_507_46, Riess_1999_117_707} (High-Z Supernova Search Team), measured the luminosity distances of SNe Ia, which are candles with a fixed intrinsic brightness and Perlmutter (Supernova Cosmology Project) \cite{Perlmutter_1999_517_565}, galaxies and clusters are moving apart at an accelerated rate. This discovery which contradicted previous expectations that gravity should slow the expansion, suggested the presence of an unknown force later considered as Dark Energy. More recent observations from the Planck satellite \cite{Ade_2016_594_A13} confirmed that ordinary matter constitutes only about $4–5\%$ of the total content of the Universe, with DM and DE making up the rest approximately $25\%$ and $70\%$, respectively. These findings led to the development of the $\Lambda$CDM model, which assumes that the expansion of the Universe is driven by DE and that DM plays a crucial role in structuring the cosmic landscape.

A brief overview of the accelerated expansion of the Universe is presented in this chapter, followed by an introduction to statistical analysis methods and the stability analysis used for reconstruction in cosmology. For more information on modern cosmology, see standard literature \cite{Weinberg_1972, Raychaudhuri_1979, Padmanabhan_2000, Rich_2009, Liddle_2015, Ryden_2017}.

\section{Geometric Setup}\label{Geometric Setup}
The development of the theory of gravitation in the initial phase focuses on establishing a mathematical framework that allows us to express the laws of physics in a covariant manner. This is achieved through the use of vectors and tensors, which are mathematical entities that transform in line with the principles of covariance. This section examines the key differential operations derived from these vectors and tensors. One can find more information on these concepts in \cite{Carroll_2003_68_023509}.

\begin{definition}[\textbf{Metric tensor}] The metric tensor is a mathematical object that describes the geometry of a coordinate system or manifold. The components of the metric describe lengths, angles and other geometric quantities. The components of the metric tensor are defined as the dot product between basis vectors and is given by
\begin{equation}
    g_{\mu\nu}= \vec{e_{\mu}}~\cdot~\vec{e_{\nu}}~,
\end{equation}
where, $g_{\mu\nu}$ are the components of the metric tensor and these $e$'s are the basis vectors in whatever coordinate system we are using.    
\end{definition}

\begin{definition}[\textbf{Christoffel symbol}] The Christoffel symbol, derived from the metric tensor, is a key component in differential geometry. It is defined as,
\begin{equation}\label{CS}
\big\{^{\alpha}_{~\mu\nu}\big\} = \frac{1}{2} g^{\alpha \lambda}\left(\partial_{\mu} g_{\nu \lambda}+ \partial_{\nu} g_{\lambda \mu}- \partial_{\lambda} g_{\mu \nu} \right)~.
\end{equation} 
\end{definition}
Christoffel symbol is not a tensor itself, despite having three indices and resembling a tensor. By definition, it is symmetric in the lower pair of indices $\mu$ and $\nu$. Therefore, in $n$ dimensions, it has $\frac{n(n+1)}{2}$ symmetric components for the lower indices. The third index $\alpha$ can take any value independently, leading to a total of $\frac{n^2(n+1)}{2}$ independent components in $n$ dimensions. The Christoffel symbol plays a crucial role in the computation of covariant derivatives, which extend the concept of partial derivatives to curved spaces.

\begin{definition}[\textbf{Covariant derivative}]
The covariant derivative mathematically defined as,
\begin{equation}\label{CD}
\nabla_{\mu} A^{\nu}= \partial_{\mu} A^{\nu} + \big\{^{\nu}_{~\mu\lambda}\big\} A^{\lambda}~.
\end{equation}
An operator $\nabla$ known as covariant derivative, acts on tensors to yield a tensor in a manner that is entirely coordinate independent. Furthermore, the general expression for the covariant derivative is 
\begin{multline}
\nabla_{\lambda} A^{\mu_{1} \mu_{2}...\mu_{n}}_{~~~~~~~~~\nu_{1} \nu_{2}...\nu_{m}} = \partial_{\lambda}A^{\mu_{1} \mu_{2}...\mu_{n}}_{~~~~~~~~~\nu_{1} \nu_{2}...\nu_{m}} + \big\{^{\mu_{1}}_{~~\lambda \sigma}\big\}  A^{\sigma \mu_{2}...\mu_{n}}_{~~~~~~~~~\nu_{1} \nu_{2}...\nu_{m}} +\ldots+ \big\{ ^{\mu_{n}}_{~~\lambda \sigma}\big\}  A^{\mu_{1} \mu_{2}...\sigma}_{~~~~~~~~~\nu_{1} \nu_{2}...\nu_{m}}\\[5pt]
 - \big\{ ^{\sigma}_{~~\lambda \nu_{1}}\big\}  A^{\mu_{1} \mu_{2}...\mu_{n}}_{~~~~~~~~~\sigma \nu_{2}...\nu_{m}} -\ldots- \big\{ ^{\sigma}_{~~\lambda \nu_{m}}\big\}  A^{\mu_{1} \mu_{2}...\mu_{n}}_{~~~~~~~~~\nu_{1}\nu_{2}...\sigma}
\end{multline}
\end{definition}

\begin{definition}[\textbf{Parallel transport}]
The process of moving a vector along a curve in such a way that its orientation remains unchanged relative to the underlying manifold is known as parallel transport. Consider a path parameterized by $x^{\mu}(\lambda)$ with a tangent vector given by $\frac{dx^{\mu}}{d\lambda}$. If a vector $A^{\mu}$ satisfies the condition
\begin{equation}\label{PT}
\frac{dx^{\mu}}{d\lambda}. \nabla_{\mu} A^{\nu} = 0~.
\end{equation}
then it is said to be parallel transported along the path $x^{\mu}(\lambda)$. 
\end{definition}
This condition implies that there is no ``change" in the components of the vector due to the curvature of the space.

\begin{definition}[\textbf{Geodesic}]
A geodesic is a mathematical concept that extends the notion of a straight line in Euclidean geometry to curved spaces. Given two points connected by a curve, the arc length allows us to quantify the distance traversed along that path. The focus is on those curves, termed geodesics, that minimize this distance, as they are critical in understanding the intrinsic geometry of the space in question. If the curve or path $x^{\mu}(\lambda)$ satisfies the equation
\begin{eqnarray}\label{geodesic}
\frac{d^{2} x^{\mu}}{d \lambda^{2}} + \big\{^{\mu}_{~\delta \sigma}\big\} \frac{dx^{\delta}}{d \lambda} \frac{dx^{\sigma}}{d \lambda}&=&0~,
\end{eqnarray}
then the curve or path $x^{\mu}(\lambda)$ is said to be a geodesic. Where $\big\{^{\mu}_{~~ \delta \sigma}\big\}$ is the Christoffel symbol.
\end{definition}

\begin{definition}[\textbf{Non-metricity Tensor}] The non-metricity tensor is defined as,
\begin{equation}\label{NMT1}
    Q_{\lambda\mu\nu} = \nabla_{\lambda}g_{\mu\nu}~.
\end{equation}
The non-metricity tensor is symmetric in the last two indices by definition as $Q_{\lambda\mu\nu} = Q_{\lambda(\mu\nu)}$. Expanding the covariant derivative, the dependence of non-metricity in both the metric tensor $g_{\mu\nu}$ and the connection coefficients $\Gamma^{\alpha}_{\phantom{\alpha}\lambda\mu}$ becomes apparent. This relationship is expressed as,
\begin{equation}\label{NMT}
    Q_{\lambda\mu\nu} = \partial_{\lambda}g_{\mu\nu} - \Gamma^{\alpha}_{~~\lambda\mu}g_{\alpha\nu} - \Gamma^{\alpha}_{~~\lambda\nu}g_{\alpha\mu}~.
\end{equation}
It is crucial to exercise caution while raising and lowering indices in the context of non-metricity, as the operations of covariant differentiation and index manipulation are no longer commutative. Consequently, one must apply the Leibniz rule whenever covariant derivatives are involved to ensure accuracy in calculations.
\end{definition}

\begin{definition}[\textbf{Superpotential tensor}] The conjugate of the Non-metricity tensor is a tensor of the type (1, 2), called as superpotential tensor and is given by,
\begin{equation}\label{ST}
    P^{\lambda}_{~~\mu\nu} = -\frac{1}{4}Q^{\lambda}_{~\mu \nu} + \frac{1}{4}\left(Q^{~\lambda}_{\mu~\nu} + Q^{~\lambda}_{\nu~~\mu}\right) + \frac{1}{4}Q^{\lambda}g_{\mu \nu}- \frac{1}{8}\left(2 \tilde{Q}^{\lambda}g_{\mu \nu} + {\delta^{\lambda}_{\mu}Q_{\nu} + \delta^{\lambda}_{\nu}Q_{\mu}} \right),
\end{equation}
where, $Q_{\alpha} = Q_{\alpha~~\mu}^{~~\mu}$ and $\tilde{Q}_{\alpha}=Q^{\mu}_{~\alpha\mu}$ are two traces of non-metricity.
\end{definition}
\begin{definition}[\textbf{Non-metricity scalar}] The non-metricity scalar that will play a central role is given by $Q = Q_{\lambda\mu\nu}P^{\lambda\mu\nu}$. A different definition of $Q = -Q_{\lambda\mu\nu}P^{\lambda\mu\nu}$ has been proposed in the literature, which changes the sign of the non-metricity scalar $Q$. This is important to consider while comparing different $f(Q)$ results.
\end{definition}
So far, key mathematical concepts and techniques necessary for developing the mathematical framework have been explored. The focus will now shift to some cosmological aspects.

\section{Foundation of Cosmological Dynamics} \label{Foundation of Cosmological Dynamics}
This section provides an overview of the standard cosmological model. For a thorough introduction to cosmology one may refer to Weinberg \cite{Weinberg_1972}, which delves into critical concepts such as FLRW metric, Hubble parameter, redshift, fluid equations governing cosmic dynamics and an analysis of the successes and limitations inherent in the $\Lambda$CDM framework.

\subsection{The FLRW metric}\label{The FLRW Metric}
The cosmological principle states that, on sufficiently large scales, the Universe is both homogeneous and isotropic. Homogeneity means that the Universe appears consistent to observers regardless of their location, while isotropy indicates that there are no favored directions from any given viewpoint. Assuming a homogeneous Universe, any two particles can be transformed to a different coordinate system known as comoving coordinates. These are coordinates which are carried along with the expansion. Because the expansion is uniform, the relationship between real distance $\vec{r}$ and the comoving distance, which can called as $\vec{x}$, can be written $\vec{r} = a(t)~\vec{x}$. The scale factor, $a(t)$ describes how physical distances evolve over time. The work of Friedmann, Lemaître, Robertson and Walker leads to the following expression for the line element in spherical coordinates, reflecting the geometric characteristics of homogeneity and isotropy while accounting for the expansion of the Universe,
\begin{equation}\label{FLRW}
ds^{2}= -c^{2}dt^{2} + a^{2}(t) \left[\frac{dr^{2}}{1-\kappa r^{2}} + r^{2} d\theta^{2} + r^{2} \sin^{2}\theta d\phi^{2}\right].
\end{equation}
In spherical geometry, $(r,\theta,\phi)$ represent the comoving spatial coordinates and the spatial curvature is denoted by $\kappa$. For convenience, $c=1$ is set.\\
Additionally, depending on the value of $\kappa$ (zero, positive or negative) there are three possible geometries for the Universe.
\begin{itemize}
    \item \textbf{Flat geometry:} These laws are based on the Euclidean geometry axioms (e.g. a straight line is the shortest distance between two points) plus one more axiom, which says parallel straight lines must remain a fixed distance apart. A Universe with the geometry ($\kappa=0$) is called a flat Universe.
    \item \textbf{Spherical geometry:} A Universe with spherical geometry similar to the surface of the Earth, has a finite size but no boundaries. In spherical geometry, traveling in a straight line would not result in an endless journey. Instead, after covering a certain distance, the traveler would eventually return to the starting point, coming from the opposite direction. This is similar to how a journey starting at the North pole leads back to the same spot after traveling outward, but from the opposite direction. A Universe with $\kappa>0$ is referred to as a closed Universe because of its finite size.
    \item\textbf{Hyperbolic geometry:} The choice $\kappa<0$ is referred as the hyperbolic geometry. The hyperbolic geometry is much less familiar than spherical geometry. It is typically represented by a saddle-like surface. While it may be difficult to see how this is consistent with isotropy. In this type of geometry, parallel lines never meet; they violate Euclid's axiom by diverging from each other. The situation $\kappa<0$ is known as an open Universe.
\end{itemize}
\begin{table}[ht]
{\small
\addtolength{\tabcolsep}{-2pt}
\renewcommand\arraystretch{1.5}
\centering
\caption{A summary of possible geometries.}\label{table1.1}
\begin{tabular}{|c|c|c|c|c|}
\hline
~Curvature~ &~~Geometry~~&~~ Angles of triangle~~&~~Circumference of circle~~&~~Type of Universe~~ \\
\hline\hline
~$\kappa>0$~ &~~Spherical~~&~~$>180$~~&~~$<2\pi r$~~&~~Closed~~ \\
\hline
~$\kappa=0$~ &~~Flat~~&~~$=180$~~&~~$=2\pi r$~~&~~Flat~~ \\
\hline
~$\kappa>0$~ &~~Hyperbolic~~&~~$<180$~~&~~$>2\pi r$~~&~~Open~~ \\
\hline
\end{tabular}
}
\end{table}
Additional parameters essential for understanding the dynamics of the expansion of the Universe will be introduced before deriving the Friedmann equations and their solutions.

\subsection{Redshift in cosmology} \label{Redshift in Cosmology}
The observation that nearly everything in the Universe is moving away from us and that this movement accelerates with distance is a key finding in cosmology. As photons travel through space, they are affected by the expansion of the Universe, which stretches their wavelengths. This phenomenon causes photons to lose energy. When these photons start in the visible spectrum, they initially appear blue and energetic but shift to red as their energy decreases. This change in wavelength is known as redshift.\\
In cosmological terms, by utilizing the observed wavelength ($\lambda_{\text{ob}}$) and the wavelength at emission ($\lambda_{\text{em}}$), the redshift is denoted as $z$ and defined by the equation
\begin{equation}\label{redshift}
z = \frac{\lambda_{\text{ob}} - \lambda_{\text{em}}}{\lambda_{\text{em}}}~.
\end{equation}
According to the Doppler effect, the difference in wavelength relates to the velocity of the object
\begin{equation}\label{redshiftvelocity}
\frac{\lambda_{\text{ob}} - \lambda_{\text{em}}}{\lambda_{\text{em}}} = \frac{dv}{c}~,
\end{equation}
where $\lambda_{\text{ob}} > \lambda_{\text{em}}$. The time it takes for light to travel from emission to observation is given by $dt = \frac{dr}{c}$. This leads to
\begin{equation}\label{redshiftscalefactor1}
\frac{\lambda_{\text{ob}} - \lambda_{\text{em}}}{\lambda_{\text{em}}} = \frac{\dot{a}}{a} \frac{dr}{c} = \frac{\dot{a}}{a} dt = \frac{da}{a}~.
\end{equation}
From this, it follows that $\lambda \propto a$, where $\lambda$ is the instantaneous wavelength. Therefore, redshift can be expressed in terms of the scale factor as:
\begin{equation}\label{redshiftscalefactor}
1 + z = \frac{\lambda_{\text{ob}}}{\lambda_{\text{em}}} = \frac{a(t_{\text{ob}})}{a(t_{\text{em}})}~.
\end{equation}
The relationship between redshift and the scale factor is crucial for understanding the expansion of the Universe. In the next section, we will delve into the concept of energy conservation in cosmology.

\subsection{Energy conservation in cosmology}\label{Energy Conservation in Cosmology}
The Friedmann equation provides a framework for understanding cosmic evolution, but it alone cannot fully describe how the scale factor $a(t)$ and energy density $\rho$ change over time. To resolve this, a concept from thermodynamics specifically the first law can be use as,
\begin{equation}\label{firstlaw}
d\mathcal{Q} = dE + p~dV~,
\end{equation}
where, $p$ is pressure, $d\mathcal{Q}$ is the heat exchange, $dE$ is the change in internal energy and $dV$ is the change in volume. In a homogeneous Universe, $d\mathcal{Q} = 0$ indicating no bulk heat flow is assumed. This allows us to express energy conservation as,
\begin{equation}\label{conservation}
\dot{E} + p \dot{V} = 0~.
\end{equation}
Consider a sphere with a comoving radius $r_s$ that expands with the Universe, giving it a proper radius of $R_s(t) = a(t) r_s$. The volume of this sphere is given by,
\begin{equation}\label{volume}
V(t) = \frac{4}{3} \pi r_s^3 a^3(t)~.
\end{equation}
Thus, the rate of change of the volume is
\begin{equation}\label{vdot}
\dot{V} = \frac{4}{3} r_s^3 \big(3a^2 \dot{a}\big) = V \left(3 \frac{\dot{a}}{a}\right).
\end{equation}
The internal energy of the sphere can be expressed as,
\begin{equation}\label{IE}
E(t) = V(t)\rho(t)~.
\end{equation}
Taking the time derivative, the result is found to be
\begin{equation}\label{IED}
\dot{E} = V \dot{\rho} + \dot{V} \rho = V \left(\dot{\rho} + 3 \frac{\dot{a}}{a} \rho\right).
\end{equation}
Substituting equations for $\dot{E}$ and $\dot{V}$ into the energy conservation equation, the result is
\begin{equation}\label{conservation1}
V \left(\dot{\rho} + 3 \frac{\dot{a}}{a} \Big(\rho + p\Big)\right) = 0~.
\end{equation}
This leads to the fluid equation, which governs the dynamics of the expansion of the Universe
\begin{equation}\label{continuityeqn1}
\dot{\rho} + 3 \frac{\dot{a}}{a} \Big(\rho + p\Big) = 0~.
\end{equation}
This equation must apply to each cosmological component independently, provided there is no interaction between them. Hence for each component $i$, the following holds
\begin{equation}\label{continuityeqn}
\dot{\rho}_i + 3H\big(\rho_i + p_i\big) = 0~.
\end{equation}
If the EoS parameter $\omega_i$ is introduced for each component, the continuity equation can be rewritten as
\begin{equation}\label{continuityeqn2}
\dot{\rho}_i + 3H\big(1 + \omega_i\big) \rho_i = 0~.
\end{equation}
Rearranging this leads to
\begin{equation}\label{continuityeqn3}
\frac{d\rho_i}{\rho_i} = -3\big(1 + \omega_i\big) \frac{da}{a}~.
\end{equation}
The continuity equation can be integrated to find the dependence of energy density on the scale factor, assuming $\omega_i$ is constant.
\begin{equation}\label{rhoarelation1}
\rho_i(a) = \rho_{i_0} a^{-3(1 + \omega_i)}~.
\end{equation}
For specific cases, the energy densities for nonrelativistic matter $(\omega_m = 0)$ and radiation $\left(\omega_r = \frac{1}{3}\right)$ are respectively given as
\begin{equation}\label{roarelation}
\rho_m(a) = \frac{\rho_{m_0}}{a^3} =\rho_{m_0}(1+z)^{3} ~, \quad\quad\quad \rho_r(a) = \frac{\rho_{r_0}}{a^4} = \rho_{r_0}(1+z)^{4}~.
\end{equation}
This framework enables the modeling of energy density evolution for different components in an expanding Universe.\\
The next section will focus on exploring how the specific energy-momentum tensor characterizes different categories of matter sources.

\subsection{The stress-energy-momentum tensor} \label{The Stress Energy-Momentum Tensor}
This section analyzes the properties of the stress-energy-momentum tensor for a Universe characterized by homogeneity and isotropy, filled with a perfect fluid\footnote{In the context of a continuous matter component, the absence of shear stress implies that viscosity can be disregarded.}. The equation for the stress-energy-momentum tensor is 
\begin{equation}\label{PF}
\mathcal{T}_{\mu \nu} = \left(\rho + p \right) u_{\mu} u_{\nu} + p g_{\mu \nu}~.
\end{equation}
In the context of isotropic fluid dynamics, the symbols $\rho$ represent the energy density and $p$ represent the pressure of the fluid, while $u_{\mu}$ denotes the covariant 4-velocity of an observer moving with the fluid. It is important to note that the functions $\rho$ and $p$ are purely time dependent $(t)$ and do not have any spatial dependence. Friedmann Universe is characterized by isotropy and homogeneity, which restrict the possibility of vector-like dependence. Since isotropy forbids vector-like dependence and homogeneity eliminates spatial dependence, therefore only time dependence remains. Again since mathematical quantities are time dependent scalars in an isotropic and homogeneous Universe, they are represented by $f(t)$.

Furthermore, it is important to note that the stress-energy-momentum tensor adheres to the conservation condition represented by the equation $\nabla^{\mu} \mathcal{T}_{\mu \nu} = 0$, which leads to
\begin{equation}\label{contieqn}
\dot{\rho} + 3\frac{\dot{a}}{a}\left(\rho + p\right)  = 0~.
\end{equation}
After outlining the properties of the stress-energy-momentum tensor, attention is now directed to the energy conditions, which impose further constraints on the behavior of matter and energy in the Universe.

\subsection{Energy conditions} \label{Energy Conditions}
An energy conditions play a pivotal role in cosmology as they govern how matter and energy behave in spacetime by providing constraints on the stress-energy tensor $T_{\mu\nu}$. These conditions, which are formulated to be coordinate invariant, allow us to understand gravitational systems without detailed knowledge of the matter content. They are essential for the study of singularities, black holes and the dynamics of the Universe, as demonstrated in Penrose and Hawking's singularity theorems \cite{Penrose_1965_14_57, Hawking_1966_294_511}.

The following energy conditions are derived from the well-known Raychaudhuri equations \cite{Ehlers_2006_15_1573, Nojiri_2007_4_115, Hawking_1973},
\begin{eqnarray}
\frac{d\theta}{d\tau} &=& -\frac{1}{3}\theta^2 - \sigma_{\mu\nu}\sigma^{\mu\nu} + \omega_{\mu\nu}\omega^{\mu\nu} - R_{\mu\nu}u^{\mu}u^{\nu}~, \label{Ray1}\\
\frac{d\theta}{d\tau} &=& -\frac{1}{2}\theta^2 - \sigma_{\mu\nu}\sigma^{\mu\nu} + \omega_{\mu\nu}\omega^{\mu\nu} - R_{\mu\nu}n^{\mu}n^{\nu}~,\label{Ray2}
\end{eqnarray}
where $\theta$ is the expansion scalar, $\sigma^{\mu\nu}$ represents shear and $\omega_{\mu\nu}$ denotes rotation associated with the geodesic congruences defined by vectors $u^{\mu}$ and $n^{\mu}$. These conditions are typically expressed by constructing scalars from $\mathcal{T}_{\mu \nu}$ using arbitrary timelike vectors $t^{\mu}$ or null vectors $n^{\mu}$ \cite{Kontou_2020_37_193001, Capozziello_2014_730_280}.
\begin{table}[ht]
    \centering
    \renewcommand\arraystretch{1.2}
    \begin{tabular}{|c|c|c|c|} \hline
    Energy conditions & Physical form & Geometrical form & Perfect fluid \\ \hline\hline
     \begin{tabular}{@{}c@{}} WEC\end{tabular} & $ \mathcal{T}_{\mu\nu}t^{\mu}t^{\nu}\geq 0 $ & $ G_{\mu\nu}t^{\mu}t^{\nu}\geq 0 $ & \begin{tabular}{@{}c@{}}$\rho\geq 0$,\\ $\rho +p\geq 0$ \end{tabular}\\ \hline
     \begin{tabular}{@{}c@{}}NEC\end{tabular}` & $ \mathcal{T}_{\mu\nu}\xi^{\mu}\xi^{\nu}\geq 0 $ & $ R_{\mu\nu}\xi^{\mu}\xi^{\nu}\geq 0 $ & $ \rho +p\geq 0 $ \\ \hline
     \begin{tabular}{@{}c@{}}SEC \end{tabular}& $ \big(\mathcal{T}_{\mu\nu}-\frac{\mathcal{T}}{2}g_{\mu\nu}\big)t^{\mu}t^{\nu}\geq 0 $ & $ R_{\mu\nu}t^{\mu}t^{\nu}\geq 0 $ & \begin{tabular}{@{}c@{}} $\rho +p\geq 0$,\\ $\rho +3p\geq 0$ \end{tabular} \\ \hline
     \begin{tabular}{@{}c@{}}DEC \end{tabular} & \begin{tabular}{@{}c@{}}$\mathcal{T}_{\mu\nu}t^{\mu}t^{\nu}\geq 0$ and \\ $\mathcal{T}_{\mu\nu}t^{\mu}$ is non-space-like \end{tabular} & \begin{tabular}{@{}c@{}}$G_{\mu\nu}t^{\mu}t^{\nu}\geq 0$ and \\ $G_{\mu\nu}t^{\mu}$ is non-space-like \end{tabular} & $ \rho \geq |p| $ \\ \hline
    \end{tabular}
    \caption{Energy conditions with their physical and geometrical significance.}
    \label{table1.2}
\end{table}
In cosmology, EC are used to define normal matter and help establish general properties of the Universe. The four main EC are the Weak Energy Condition, Null Energy Condition, Strong Energy Condition and Dominan Energy Condition. In terms of validity, significance and interpretation, each of these EC has its own advantages and limitations. Additionally, each condition imposes different constraints on the energy density, $\rho$ and pressure, $p$ of matter.
\begin{itemize}
    \item \textbf{WEC} states that the energy density must be non-negative $\big(\rho \geq 0 \big)$ and the pressure must satisfy $\rho + p \geq 0$, ensuring that energy is physically meaningful and not excessively negative.
    \item \textbf{NEC} requires that $\rho + p \geq 0$. It allows for negative energy density if compensated by sufficiently large pressure, playing a role in the study of lightlike geodesics and black hole horizons.
    \item \textbf{SEC} is key for understanding the large-scale structure and evolution of the Universe. It implies that $\rho + p \geq 0$ and $\rho + 3p \geq 0$, which suggests that gravity is attractive and leads to a decelerating expansion of the Universe under normal matter conditions. It is also important in analyzing cosmic phenomena such as the formation of singularities in black holes and the Big Bang.
    \item \textbf{DEC} combines WEC with a requirement that the energy flux be non-spacelike, meaning that $\rho \geq |p|$. This ensures that the energy density is non-negative and at least as large as the absolute value of pressure, implying that matter behaves in a physically reasonable way.
\end{itemize}
The SEC is particularly important for understanding the acceleration phenomena in the Universe. Therefore, the upcoming discussion will concentrate on the SEC.\\
The following section discusses the cosmological parameters that are commonly considered.

\subsection{Cosmographic parameters}\label{Cosmographic Parameters}
Cosmologists aim to determine the scale factor $a(t)$, which describes the expansion of the Universe. In models where the contents of the Universe are well understood, this scale factor can be derived from the Friedmann equations. However, establishing $a(t)$ for the actual Universe is complex. The scale factor itself is not directly measurable; it must be inferred from our limited and sometimes imperfect observations \cite{Ryden_2017}. According to the cosmological principle, the scale factor represents a fundamental degree of freedom that governs the dynamics of the Universe. The scale factor $a(T)$ is determined as a Taylor expansion \cite{Weinberg_1972} around the present time $t_0$. This gives
\begin{eqnarray}\label{ATS}
a(t)  \approx~ & a(t_0)+ \dot{a}(t_0) \Delta t + \frac{\ddot{a}(t_0)}{2!} \Delta t^2+
\frac{\dddot{a}(t_0)}{3!} \Delta t^3 + \frac{a^{(iv)}(t_0)}{4!} \Delta t^4 +\frac{a^{(v)}(t_0)}{5!} \Delta t^5+\ldots~,
\end{eqnarray}
which recovers signal causality if one assumes $\Delta t\equiv t-t_0>0$. From the above expansion of $a(t)$, one define
\begin{equation}\label{CP}
H \equiv \frac{1}{a} \frac{d a}{d t}~,\quad
q \equiv -\frac{1}{a H^2} \frac{d^2a}{d t^2}~,\quad
j  \equiv \frac{1}{a H^3} \frac{d^3a}{d t^3}~,\quad
s \equiv \frac{1}{a H^4} \frac{d^4a}{d t^4}~,\quad
l  \equiv \frac{1}{a H^5} \frac{d^5a}{d t^5}~.
\end{equation}
These functions can be directly constrained by observations, as they are model independent quantities by construction, i.e. they do not rely on the specific form of the DE fluid. They are known in the literature as the Hubble parameter ($H$), the deceleration parameter ($q$), the jerk parameter ($j$), the snap parameter ($s$) and the lerk parameter ($l$)~\cite{Alam_2003_344_1057, Sahni_2003_77_201}. Once these functions are determined at the present time, they are referred to as the cosmographic series. This is the set of coefficients usually derived in cosmography from observations. Rewriting $a(t)$ in terms of the cosmographic series gives
\begin{equation}\label{ATS1}
a(t)  \approx 1+  H_0 \Delta t - \frac{q_0}{2!}  H_0^2\Delta t^2+\frac{j_0}{3!} H_0^3 \Delta t^3 +   \frac{s_0}{4!}  H_0^4\Delta t^4 +\frac{l_0}{5!}  H_0^5\Delta t^5+\ldots\ ,
\end{equation}
The scale factor is normalize by setting $a(t_0) = 1$. Reformulating Eq. \eqref{ATS} into the form shown in Eq. \eqref{ATS1} describes the significance of each parameter within the cosmographic series. Each term reveals distinct dynamical characteristics, with the parameter $s$ and the parameter $l$ specifically influencing the profile of Hubble's flow in the high redshift regime. It is essential for the Hubble parameter to be positive to ensure the continued expansion of the Universe. Furthermore, the parameters $q$ and $j$ characterize the kinematics at lower redshift domains. Notably, the value of $q$ at any instant provides insight into whether the Universe is undergoing acceleration or deceleration and hints at the nature of the cosmological fluid driving these dynamics.\\
Attention is now directed to the simplest model of the Universe, which will serve as the foundation for the discussion.

\subsection{Overview of the $\Lambda$CDM model}\label{Overview of the LCDM Model}
The $\Lambda$CDM model is designed to explain the dynamics of the Universe, specifically the phenomenon of accelerated expansion. This framework includes GR while incorporating dark energy characterized by a negative EoS parameter, conventionally represented as a cosmological constant. Within GR, the introduction of a cosmological constant permits accelerated expansion; however, the fundamental origins of this DE remains unexplained. This topic will be examined in further detail in the subsequent sections. For this discussion, natural units are employed where $c$ and $\hbar$ (the reduced Planck constant) are set to unity, unless stated otherwise.

In 1915, Einstein's theory of GR was derived from the Einstein-Hilbert action, which is given by
\begin{equation}\label{EHaction}
S = \int d^{4}x ~ \sqrt{-g} \left[\frac{1}{2\kappa}(R+2\Lambda) + L_{m}\right],
\end{equation}
where $\kappa = 8\pi G$ and $\Lambda$ is the cosmological constant. The Einstein field equation can be obtained by the variation of the Einstein-Hilbert action with respect to the metric tensor, leads to
\begin{equation}\label{LCDM}
G_{\mu \nu} + \Lambda g_{\mu \nu} = \kappa~ \mathcal{T}_{\mu \nu}~.
\end{equation}
Additionally, the cosmological equations of the $\Lambda$CDM model, commonly referred to as the Friedmann equations, are explored. The equations are derived using the FLRW metric \eqref{FLRW}, the stress-energy-momentum \eqref{PF} and Eq. \eqref{LCDM}, resulting in
\begin{eqnarray}
H^{2} &=& \frac{8 \pi G}{3} \rho - \frac{k}{a^{2}} + \frac{\Lambda}{3}~,\label{EHF1}\\ 
2 \dot{H} + 3 H^{2} &=& -8 \pi G p - \frac{k}{a^{2}} + \Lambda~.\label{EHF2}
\end{eqnarray}
Combining equations \eqref{EHF1} and \eqref{EHF2} gives
\begin{equation}\label{accelerationeqn}
\frac{\ddot{a}}{a} = -\frac{4 \pi G}{3} \left(\rho+3p\right) + \frac{\Lambda}{3}~,
\end{equation}
which is known as the acceleration equation. If $\ddot{a}>0$, the Universe experiences acceleration, while $\ddot{a}<0$ indicates a decelerating Universe. For the geometry of the Universe to be flat corresponding to $k=0$, there is a specific density that must be present for a given value of the Hubble parameter, $H$. This density is referred to as the critical density $\rho_{c}$ and it is calculated as,
 \begin{equation}\label{criticaldensity}
 \rho_{c}= \frac{3H^{2}}{8 \pi G}~.
 \end{equation}
The critical density is a key concept, but it does not always match the actual density of the Universe because the Universe does not have to be perfectly flat. As a result, it is often more useful to express the density of the Universe in relation to the critical density, rather than quoting its absolute value directly. This quantity is known as the density parameter $\Omega$ and defined as
\begin{equation}\label{densityparameter}
\Omega_{i}(t) = \frac{\rho_{i}(t)}{\rho_{c}(t)} = \frac{8 \pi G}{3} \frac{\rho_{i}(t)}{H^{2}(t)}~,
\end{equation}
where $i$ denotes the sum over matter, radiation, spatial curvature and vacuum. One can rewrite the Friedmann equation \eqref{EHF1} as
\begin{align}\label{equality}
    1 &= \Omega_{m}(t) + \Omega_{r}(t) + \Omega_{k}(t) + \Omega_{\Lambda}(t)~.\nonumber\\[10pt]
\Omega_{m}(t) = \frac{8 \pi G}{3H^{2}}\rho_{m}(t)~, ~~ \Omega_{r}(t) &= \frac{8 \pi G}{3H^{2}}\rho_{r}(t)~, ~~ \Omega_{k}(t) = -\frac{k}{H^{2}a^{2}}~,~~  \Omega_{\Lambda}(t) = \frac{8 \pi G}{3 H^{2}} \rho_{\Lambda}~, ~~ \rho_{\Lambda} = \frac{\Lambda}{8 \pi G}
\end{align}
For flat geometry ($k=0$), $\Omega_{m}+\Omega_{r}+\Omega_{\Lambda} =1$, i.e. $\Omega_{total}=1$ at all times. The values of $\Omega_{i}$ change over time, so at the present moment, they are referred to as $\Omega_{i_0}$. This allows us to understand how different components of the Universe such as matter, dark energy and radiation contribute to the overall density relative to the critical density at the specific time.
\begin{equation}\label{generalH}
H(z) = H_{0}\sqrt{\Big(\Omega_{r_0}~(1+z)^{4} + \Omega_{m_0}~(1+z)^{3} +  \Omega_{k_0}~(1+z)^{2} + \Omega_{\Lambda_0}\Big)}~.
\end{equation}
The Planck collaboration \cite{Aghanim_2020_641_A6} findings provided precise values for the parameters in Eq. \eqref{generalH}, refining our understanding of the composition and evolution of the Universe. The next section explores both the successes and limitations of the standard cosmological model.

\subsection{Successes and shortcomings of standard cosmological model}\label{Successes and Shortcomings}
The standard cosmological model $\Lambda$CDM has been remarkably successful in describing the evolution of the Universe and structure. It incorporates both the theory of GR and the concept of a cosmological constant ($\Lambda$) to account for various observed phenomena. Here are some of the major successes of the $\Lambda$CDM model:
\begin{itemize}
    \item \textbf{Cosmic evolution and structure formation:} The $\Lambda$CDM model successfully explains the large-scale structure of the Universe, including the formation and distribution of galaxies and galaxy clusters. It incorporates the effects of DM and DE to describe how structures form and evolve over time through gravitational interactions.
    \item \textbf{Primordial nucleosynthesis:} The model accurately predicts the abundances of light elements such as Helium, Deuterium and Lithium, as observed in the Universe today. These predictions stem from the processes of primordial nucleosynthesis that occurred during the first few minutes after the Big Bang.
    \item \textbf{CMB:} The $\Lambda$CDM model aligns well with observations of the CMB, which provides a snapshot of the Universe approximately 380,000 years after the Big Bang. The model reproduces the observed temperature fluctuations and the angular power spectrum of the CMB with high precision.
    \item \textbf{Accelerated expansion of the Universe:} The model accounts for the observed accelerated expansion of the Universe, attributed to DE. Observations of distant supernovae and the large-scale structure of the Universe support this explanation, fitting well with the predictions made by $\Lambda$CDM.
\end{itemize}
Despite these successes, the $\Lambda$CDM model faces several significant challenges:
\begin{itemize}
    \item \textbf{Cosmological constant problem:} There is a profound discrepancy between the theoretical and observed values of the cosmological constant, which differs by about 60 orders of magnitude. Theoretical estimates based on quantum field theory suggest a value of $10^{-60}~ M_{Pl}^{4}$, whereas observations indicate a value of $10^{-120}~ M_{Pl}^{4}$ \cite{Joyce_2015_568_1}. This discrepancy is known as the cosmological constant problem.
    \item \textbf{Horizon problem:} The $\Lambda$CDM model encounters the horizon problem, which refers to the difficulty in explaining the uniform temperature of the Universe across regions that are not in causal contact. The widely accepted resolution is the theory of cosmic inflation, which proposes that the Universe underwent exponential expansion during its early stages \cite{Tsujikawa_2003}.
    \item \textbf{Cosmological coincidence problem:} The model suggests that the Universe is in a transitional phase between the matter-dominated era and a period of accelerated expansion driven by DE. Observations show that the densities of DE and DM are of comparable magnitude, which raises questions about why these quantities are so similar \cite{Velten_2014_74_3160}.
    \item \textbf{DM:} While $\Lambda$CDM assumes the presence of DM, which is non-baryonic and interacts only through gravity, no direct detection of DM particles has been achieved. The model relies on DM to explain phenomena such as the rotational curves of galaxies and the formation of large-scale structures \cite{Astesiano_2022_106_044061, Katsuragawa_2017_95_044040, Zaregonbadi_2016_94_084052}.
    \item \textbf{$H_{0}$ tension:} There is a notable discrepancy between the local measurements of the Hubble constant ($H_{0}$) and those derived from CMB observations. This discrepancy, known as $H_{0}$ tension, amounts to approximately $4.4\sigma$ tension \cite{Valentino_2021_131_102605}.
    \item \textbf{$\sigma_{8}$ tension:} There is also tension regarding the amplitude of cosmic structure growth. The Planck collaboration estimates $S_{8} = \sigma_{8} \sqrt{\Omega_{m}/0.3} = 0.834 \pm 0.016$ \cite{Valentino_2021_131_102604, Benisty_2021_31_100766}, while the KIDS-450 collaboration finds $S_{8} = 0.745 \pm 0.039$ \cite{Joudaki_2017_471_1259}, resulting in about $2\sigma$ tension.
\end{itemize}
These challenges highlight the need for alternative theories or modifications to the $\Lambda$CDM model. The following sections will explore various models that attempt to address these limitations and extend beyond the standard $\Lambda$CDM framework.

\section{Exploring Alternative Cosmological Theories}\label{Alternatives to LCDM}
Given the limitations of the $\Lambda$CDM model discussed previously, it is essential to explore alternative approaches that extend beyond $\Lambda$CDM and GR. This section will discuss potential alternatives to GR, driven by the shortcomings associated with the cosmological constant. These alternatives generally fall into two categories: DE models and Modified Gravity models.

\subsection{Dark energy models}\label{Dark Energy Models}
These models introduce a DE component characterized by either a static or dynamic equation of state. Rather than relying on the cosmological constant $\Lambda$, DE models modify the matter side of Einstein's equation, typically by incorporating additional fluid components.
\begin{itemize}
\item There have been several dynamical dark energy models proposed with time dependent EoS parameter during the evolution of the Universe. Some common dynamical dark energy models, as functions of redshift $z$, are Chevallier-Polarski-Linder (CPL) model \cite{Linder_2003_90_091301}, Jassal-Bagla-Padmanbhan (JBP) model \cite{Jassal_2010_405_2639}, Barboza-Alcaniz parameterization \cite{Barboza_2008_666_415} and Wetterich parameterization \cite{Wetterich_2004_594_17} and many more.
\item The quintessence scalar field model is a prominent framework for describing dynamical DE. This model involves a scalar field that is minimally coupled to matter, characterized by a time dependent scalar field $\phi$ and an associated potential $V(\phi)$. The negative pressure generated by the scalar field drives a decrease in the potential over time. Quintessence models are classified according to the form of the potential. In scenarios where $V(\phi) \ll \dot{\phi}^2$, the EoS parameter approximates to $\omega \approx 1$, resembling stiff matter and yielding negligible contributions to DE. Conversely, when $V(\phi) \gg \dot{\phi}^2$, the EoS parameter approaches $\omega \approx -1$, akin to a cosmological constant. For intermediate cases where $-1 < \omega < 1$, the energy density behaves as $\rho \propto a^{-m}$, facilitating accelerated expansion for $0 \leq m < 2$. For a deeper exploration of the diverse quintessence potentials and their implications for late-time cosmic acceleration \cite{Gupta_2015_92_123003, Chiba_2013_87_083505, Roy_2014_129_162, Pantazis_2016_93_103503}.
\item Caldwell \cite{Caldwell_2002_545_23} introduced the phantom field model of DE was proposed to explain the late-time cosmic acceleration. In this model, the kinetic term carries a negative sign, leading to negative kinetic energy, which leads to an accelerated expansion of the Universe to an infinite size within a finite amount of time, a scenario known as the big rip. Scalar field models whose EoS evolution mimics that of the phantom field are called quintom models. These models combine features of both quintessence and phantom fields, allowing for a more flexible description of DE behavior, particularly in the transition between different epochs of cosmic acceleration \cite{Cai_2008_25_165014, Feng_2005_607_35}.
\item The K-essence scalar field model describes an inflationary model of the early Universe, also known as K-inflation \cite{Picon_1999_458_209, Picon_2001_63_103510, Chiba_2000_62_023511}. Unlike quintessence models, which primarily rely on the potential energy term to drive accelerated expansion, K-essence models are characterized by a dominant kinetic contribution to the energy density, which plays a key role in explaining late-time cosmic acceleration.
\end{itemize}

\subsection{Modified gravity models}\label{Modified Gravity Models}
There is an alternative approach to explaining the late-time acceleration of the Universe that involves modifying Einstein's GR. Various modified gravity theories propose mechanisms for accelerated expansion without relying on a dark energy component \cite{Tsujikawa_2010, Clifton_2012_513_1}. These theories include $f(R)$ gravity \cite{Carroll_2005_71_063513, Nojiri_2007_652_343, Motohashi_2019_2019_025, Olmo_2019_100_044020, Odintsov_2019_99_064049, Odintsov_2020_101_044009, Oikonomou_2021_103_044036}, scalar-tensor models \cite{Sanders_1997_480_492, Das_2008_78_043512, Riazuelo_2002_66_023525, Boisseau_2000_85_2236, Crisostomi_2018_97_084004, Elizalde_2004_70_043539}, $f(T)$ gravity \cite{Capozziello_2015_91_124037, Cai_2015_79_106901, Nunes_2018_2018_052, Anagnostopoulos_2019_100_083517, Bahamonde_2020_2020_024}, $f(Q)$ gravity \cite{Jimenez_2018_98_044048, Lazkoz_2019_100_104027, Barros_2020_30_100616, Anagnostopoulos_2021_822_136634, Frusciante_2021_103_044021, Hu_2022_106_044025} and other extended gravitational theories \cite{Harko_2011_84_024020, Myrzakulov_2012_72_2203, Harko_2010_70_373, Xu_2019_79_708, Xu_2020_80_449}. As a result of these alternative frameworks present different mechanisms to explain cosmic acceleration, each highlighting distinct aspects of fundamental properties and dynamics of the gravity.
\begin{itemize}
\item\textbf{$f(R)$ gravity} is a generalization of the Einstein-Hilbert action, where the Ricci scalar $R$ is replaced by a more general function $f(R)$. This allows for different gravitational behaviors and dynamics. $f(R)$ gravity can provide alternative explanations for the accelerated expansion of the Universe, without relying on dark energy. Different choices of $f(R)$ can lead to varied cosmological models and predictions.
\item\textbf{$f(T)$ gravity} is a modified theory of gravity based on the TEGR. It uses the torsion scalar $T$ instead of the Ricci scalar to describe gravitational interactions. $f(T)$ gravity offers different predictions for cosmic expansion and structure formation compared to $f(R)$ gravity, potentially providing new insights into the accelerated expansion of the Universe.
\item \textbf{$f(Q)$ gravity} is a modified theory that extends GR by considering a function of the non-metricity scalar $Q$. This theory utilizes the non-metricity of spacetime, which refers to the variation of the metric tensor. $f(Q)$ gravity provides an alternative framework for understanding cosmic acceleration by modifying the role of non-metricity, leading to different cosmological models and predictions.
\end{itemize}
We discuss one of the popular theory of modified gravity in the next section.

\section{Symmetric Teleparallelism} \label{Symmetric Teleparallelisms}
To better understand DM and DE, various modified theories of gravity have been proposed that go beyond the standard framework of GR. These theories address a number of observed and theoretical inconsistencies. By combining primordial inflation and cosmic acceleration, modified gravity offers a promising solution. It could also help reconcile gravity with quantum mechanics and the electromagnetic, weak nuclear and strong nuclear interactions as well.\\
GR can be represented through two equivalent geometric frameworks: the teleparallel representation and the curvature representation. In the teleparallel representation, non-metricity and curvature vanish while torsion remains non-zero; conversely, in the curvature representation, torsion and non-metricity are zero, but curvature is non-zero. Another equivalent approach is the symmetric teleparallel gravity, where the basic geometry of the gravitational action is represented by the non-metricity $Q$ of the metric, which defines the variation of the length of a vector during the parallel transport around a close loop.

\subsection{Geometrical representation}\label{Geometrical Representation}
The geometry associated with non-metricity will be explored in this section. Consider the vectors defined on a differential manifold equipped with a connection and a metric $g_{\mu \nu}$ are $u^{\mu}$ and $v^{\mu}$. The $u.v = u^{\mu} v^{\nu} g_{\mu \nu}$ is the inner product of these vectors. When parallel transporting two vectors along a curve C with the parametrization $x^{\mu} = x^{\mu}(\lambda)$, the implications of non-metricity during the transport process can be analyzed by examining the covariant derivative of the metric tensor.
\begin{equation}\label{innerproduct}
\tilde{\nabla}_{\lambda}(u.v) =  \frac{dx^{\alpha}}{d\lambda} \left(\tilde{\nabla}_{\alpha} u^{\mu} \right) v_{\mu} + \frac{dx^{\alpha}}{d\lambda} \left(\tilde{\nabla}_{\alpha} v^{\nu} \right) u_{\nu} + \frac{dx^{\alpha}}{d\lambda} \left(\tilde{\nabla}_{\alpha} g_{\mu \nu} \right) u^{\mu} v^{\nu}~. 
\end{equation}
The condition of parallel transport of $u^{\mu}$ and $v^{\mu}$ leads to 
\begin{equation}\label{CPT}
\frac{dx^{\alpha}}{d\lambda} \left(\tilde{\nabla}_{\alpha} u^{\mu} \right) = 0~, \quad\quad \mathrm{and} \quad\quad \frac{dx^{\alpha}}{d\lambda} \left(\tilde{\nabla}_{\alpha} v^{\mu} \right) = 0~,
\end{equation}
and hence
\begin{equation}\label{paralleltransport}
\tilde{\nabla}_{\lambda}(u.v) = Q_{\alpha \mu \nu} \frac{dx^{\alpha}}{d\lambda}  u^{\mu} v^{\nu}~.
\end{equation}
This shows that parallel transporting two vectors around a curve changes their inner product. Equalizing the two vectors gives 
\begin{equation}\label{magnitude}
\tilde{\nabla}_{\lambda}(|u|^{2}) = Q_{\alpha \mu \nu} \frac{dx^{\alpha}}{d\lambda} u^{\mu} u^{\nu}~.
\end{equation}
The discussion highlights the variation in the magnitude of a vector during parallel transport along a specified curve. Consequently, in a space characterized by non-metricity, the length of a vector is not invariant under parallel transport.

\subsection{Decomposition of Affine connection}\label{Decomposition of Affine Connection}
In metric-affine spacetime, the connection $\Gamma^{\lambda}{}_{\sigma\rho}$ and the metric $g_{\mu\nu}$ are treated independently. This allows for more general formulations of covariant derivatives and parallel transport compared to the standard metric-compatible cases found in GR. The covariant derivative of a $\mathcal{T}^\lambda{}_\nu$ in a metric-affine spacetime is given by the expression
\begin{equation}\label{Covariantderivative1}
\nabla_\mu \mathcal{T}^\lambda{}_\nu =  \partial_\mu \mathcal{T}^\lambda{}_\nu  + \Gamma^\lambda{}_{\mu \alpha} \mathcal{T}^\alpha{}_\nu  -  \Gamma^\alpha{}_{\mu \nu} \mathcal{T}^\lambda{}_\alpha~.
\end{equation}
Here, $\nabla_\mu$ denotes the covariant derivative, $\partial_\mu$ is the partial derivative. The term $\Gamma^\lambda{}_{\mu \alpha} \mathcal{T}^\alpha{}_\nu$ adjusts for the change in the tensor component $\mathcal{T}^\lambda{}_\nu$ due to the connection, while $\Gamma^\alpha{}_{\mu \nu} \mathcal{T}^\lambda{}_\alpha$ accounts for the variation in the covariant indices. As known from differential geometry (see, e.g., \cite{Hehl_1995_258_1,Ortin_2004}), generic affine connection can be decomposed into three parts \cite{Jimenez_2018_2018_039},
\begin{equation}\label{Connectiondecomposition}
\Gamma^{\lambda}_{\phantom{\alpha}\mu\nu} =
\left\lbrace {}^{\lambda}_{\phantom{\alpha}\mu\nu} \right\rbrace +
K^{\lambda}_{\phantom{\alpha}\mu\nu}+
 L^{\lambda}_{\phantom{\alpha}\mu\nu}~,
\end{equation}
the Levi-Civita connection of the metric $g_{\mu\nu}$~,
\begin{equation}\label{LeviCivita}
 \CLC{\lambda}{\mu}{\nu} \equiv \frac{1}{2} g^{\lambda \beta} \left( \partial_{\mu} g_{\beta\nu} + \partial_{\nu} g_{\beta\mu} - \partial_{\beta} g_{\mu\nu} \right)~,
\end{equation}
contortion
\begin{equation}\label{Contortion}
K^{\lambda}{}_{\mu\nu} \equiv \frac{1}{2} g^{\lambda \beta} \left( T_{\mu\beta\nu}+T_{\nu\beta\mu} +T_{\beta\mu\nu} \right) = - K_{\nu\mu}{}^{\lambda}~,
\end{equation}
and disformation
\begin{equation}\label{Disformation}
L^{\lambda}{}_{\mu\nu} \equiv \frac{1}{2} g^{\lambda \beta} \left( -Q_{\mu \beta\nu}-Q_{\nu \beta\mu}+Q_{\beta \mu \nu} \right) = L^{\lambda}{}_{\nu\mu}~.
\end{equation}
In metric-affine spacetimes, the connection $\Gamma^{\lambda}{}_{\mu\nu}$ governs various geometric properties, including torsion, non-metricity and curvature. These quantities are defined as follows:

Torsion is given by,
\begin{equation}\label{torsion}
  T^{\lambda}{}_{\mu\nu} \equiv \Gamma^{\lambda}{}_{\mu\nu} - \Gamma^{\lambda}{}_{\nu\mu}~.  
\end{equation}
Torsion measures the failure of the connection to be symmetric in its lower indices.

Curvature is given by the Riemann curvature tensor
\begin{equation}\label{riccitensor}
    R^{\sigma}{}_{\rho\mu\nu} \equiv \partial_{\mu} \Gamma^{\sigma}{}_{\nu\rho} - \partial_{\nu} \Gamma^{\sigma}{}_{\mu\rho} + \Gamma^{\alpha}{}_{\nu\rho} \Gamma^{\sigma}{}_{\mu\alpha} - \Gamma^{\alpha}{}_{\mu\rho} \Gamma^{\sigma}{}_{\nu\alpha}~.
\end{equation}
Curvature describes how the connection fails to be locally flat.

\begin{figure}[H]
\centering
	\begin{tikzpicture}
	\tikzset{venn circle/.style={draw,circle,minimum width=5cm,fill=#1,opacity=0.1}}
	
	\node [venn circle = blue] (A) at (0,0) {};
	\node [venn circle = magenta] (B) at (0:2.5cm) {};
	\node [venn circle = red] (C) at (-60:2.5cm) {};
	\node[above] at (barycentric cs:A=-3.0,B=1/2,C=1/2) {Riemann-Cartan}; 
	\node[below] at (barycentric cs:A=-3.0,B=1/2,C=1/2) {$\scriptstyle{Q_{\rho\mu\nu}=0}$}; 
	\node[above] at (barycentric cs:B=-3.0,A=1/2,C=1/2) {torsion free}; 
	\node[below] at (barycentric cs:B=-3.0,A=1/2,C=1/2) {$\scriptstyle{T^\lambda{}_{\mu\nu}=0}$}; 
	\node[above] at (barycentric cs:C=-3.0,A=1/2,B=1/2) {teleparallel}; 
	\node[below] at (barycentric cs:C=-3.0,A=1/2,B=1/2) {$\scriptstyle{R^\sigma{}_{\rho\mu\nu}=0}$};
	\node at (barycentric cs:A=1,B=1,C=-0.6) {\begin{tabular}{c} Riemann \\$\scriptstyle{\QLC_{\rho\mu\nu}=0}$,\\ $\scriptstyle{\TLC^\lambda{}_{\mu\nu}=0}$ \end{tabular}};
	\node at (barycentric cs:A=1,C=1,B=-2/3) {\begin{tabular}{c}Weitzenb\"ock \\ $\scriptstyle{\QWC_{\rho\mu\nu}=0,}$ \\ $\scriptstyle{\RWC^\sigma{}_{\rho\mu\nu}=0}$ \end{tabular}}; 
	\node at (barycentric cs:B=1,C=1,A=-2/3) {\begin{tabular}{c} symmetric \\ teleparallel \\ $\scriptstyle{\RSTG^\sigma{}_{\rho\mu\nu}=0,}$ \\ $\scriptstyle{\TSTG^\lambda{}_{\mu\nu}=0}$\end{tabular}}; 
	\node[above] at (barycentric cs:A=1/3,B=1/3,C=1/3 ){Minkowski};
	
	\end{tikzpicture}  
	\caption{Depending on the properties of connection, the subclasses of metric-affine geometry. [Credit: Phys. Rev. D. 97, 124025 (2018)]}
	\label{fig:geometries}
\end{figure}
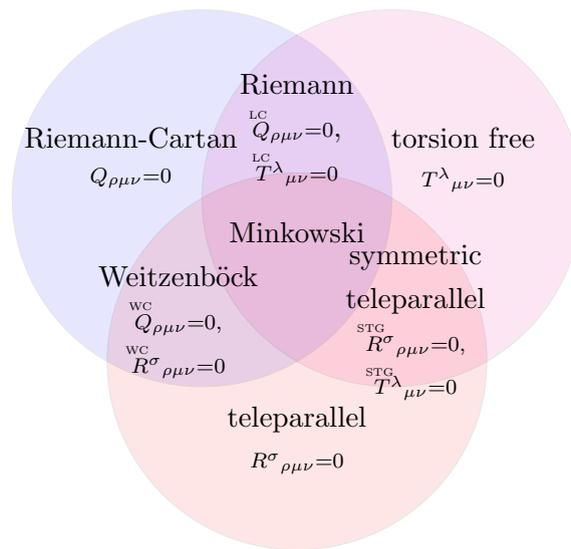

By imposing constraints on the connection, different geometric frameworks emerge shown in Fig. \ref{fig:geometries}. These frameworks are categorized as \cite{Jarv_2018_97_124025},
\begin{itemize}
    \item \textbf{Riemann-Cartan geometry}: This geometry is obtained by assuming that the non-metricity $Q_{\rho \mu \nu}$ vanishes, but torsion $T^{\lambda}{}_{\mu\nu}$ and curvature $R^{\sigma}{}_{\rho\mu\nu}$ are generally non-zero.
    \item \textbf{Teleparallel geometry}: In this case, curvature $R^{\sigma}{}_{\rho\mu\nu}$ is zero, leading to parallel transport of vectors being path-independent. Torsion and non-metricity are generally non-zero.
    \item \textbf{Torsion-free geometry}: This geometry is characterized by a vanishing torsion $T^{\lambda}{}_{\mu\nu}$, but non-metricity and curvature are allowed to be non-zero.
    \item \textbf{LC connection and Riemann geometry}: By setting both torsion $T^{\lambda}{}_{\mu\nu}$ and non-metricity $Q_{\rho \mu \nu}$ to zero, the connection is simplified to the Levi-Civita connection. This connection is metric-compatible and torsion-free, which is a defining characteristic of Riemannian geometry.
    \item \textbf{WC}: This is characterized by vanishing non-metricity $Q_{\rho \mu \nu}$ and curvature $R^{\sigma}{}_{\rho\mu\nu}$, leading to a specific class of teleparallel geometries.
    \item \textbf{STG}: In this case, torsion $T^{\lambda}{}_{\mu\nu}$ and curvature $R^{\sigma}{}_{\rho\mu\nu}$ are both zero, leading to symmetric teleparallel geometry.
    \item \textbf{Minkowski Space}: This is the case where torsion $T^{\lambda}{}_{\mu\nu}$, non-metricity $Q_{\rho \mu \nu}$ and curvature $R^{\sigma}{}_{\rho\mu\nu}$ are all zero, representing flat spacetime.
\end{itemize}
To denote specific situations based on these properties, the labeled notation are used, such as $\GSTG{}^\lambda{}_{\mu\nu}$ for a geometry with generalized symmetric teleparallel properties, $\DWC_\mu$ for a geometry with the Weitzenböck connection and $\RLC^{\sigma}{}_{\rho\mu\nu}$ for the Riemann curvature tensor in Levi-Civita connection. The best approach to relax Riemannian constraints is now clear from the discussion above. Specifically, to explore non-Riemannian geometry, a generic affine connection that accommodates both torsion and non-metricity is needed. This type of connection naturally incorporates the degrees of freedom associated with torsion and non-metricity. Let us discuss more about non-metricity teleparallel theory.

\subsection{Symmetric teleparallel equivalent to GR} \label{Symmetric Teleparallel Equivalent to GR}
In the presence of a metric $g_{\mu\nu}$, not only the curvature and torsion can be define, but also the non-metricity. The most general even-parity second order quadratic form of the non-metricity is \cite{Einstein_1915_1915_844}
\begin{equation}\label{quadraticform}
Q =   \frac{1}{4}Q_{\alpha\beta\gamma}Q^{\alpha\beta\gamma} -  \frac{1}{2}Q_{\alpha\beta\gamma}Q^{\beta\alpha\gamma} 
  -   \frac{1}{4}Q_\alpha Q^\alpha + \frac{1}{2}Q_\alpha\tilde{Q}^\alpha~.
\end{equation}
The general quadratic action incorporating appropriate Lagrange multipliers is then
\begin{equation}\label{quadraticaction}
S_{Q}=\int d^4x\left[\frac{1}{16\pi G}\sqrt{-g}~Q+\lambda_{\alpha}{}^{\beta\mu\nu} R^\alpha{}_{\beta\mu\nu}+\lambda_\alpha{}^{\mu\nu} T^\alpha{}_{\mu\nu}\right]~.
\end{equation}
Here, Eq. \eqref{quadraticaction} have a 5-parameter family of quadratic theories. The next step is to explore this space of theories and identify if any specific case can reproduce an equivalent of GR. However, the existence of an equivalent to GR for a torsion-free connection can be demonstrated again, which leads to
\begin{equation}\label{RtoQ}
R(\Gamma)   =  \mathcal{R}(g)  + Q +   \nabla_\alpha \big( Q^\alpha - \tilde{Q}^\alpha \big)~.
\end{equation}
The second postulate of STEGR asserts that the curvature of the affine connection vanishes (i.e. $R(\Gamma)=0$) implies that the relation $\mathcal{R}(g)=-Q-\nabla_\alpha ( Q^\alpha - \tilde{Q}^\alpha )$ and consequently, the action 
\begin{equation}\label{STGRaction}
\mathcal{S}_{\rm STEGR}=\frac{1}{16\pi G}\int d^4x\sqrt{-g}~Q~.
\end{equation}
According to the generalized Gauss theorem, the term $\nabla_\alpha ( Q^\alpha - \tilde{Q}^\alpha )$ is discarded, as a term amounts to a mere boundary term which can thus have no influence on the field equations \cite{Heisenberg_2024_1066_1}. The STEGR \cite{Nester_1999_37_113}, is characterized by its distinction from the Hilbert action, which differs solely by a total derivative. This property ensures that the dynamics of STEGR align with those of GR. Notably, in the TEGR framework, the quadratic form $Q$ exhibits a unique symmetry realized up to a total derivative.

The difference between the invariant $Q$ and the Ricci scalar is simply a boundary term. The framework defined by $Q$, which excludes this boundary contribution, represents a specific class of special STG models, analogous to an enhanced version of GR.

\subsection{$f(Q)$ extension}\label{f(Q) Extension}
The general $f(Q)$ gravity can be achieved by extending STEGR through the incorporation of an action defined by an arbitrary $f(Q)$ function as
\begin{equation}\label{fqaction}
S = \int d^{4}x ~ \sqrt{-g} \left[\frac{1}{2\kappa}f(Q) + L_{m}\right].
\end{equation}
A particular choice of non-metricity scalar and the associated action results in the recovery of General Relativity when $f(Q) = Q$.
Varying action \eqref{fqaction} with respect to $g_{\mu \nu}$ yields the field equation \cite{Jimenez_2018_98_044048}
\begin{equation}\label{fqfe}
\frac{2}{\sqrt{-g}} \nabla_{\alpha}\left(\sqrt{-g}f_{Q} P^{\alpha}_{~~\mu \nu}\right) + \frac{1}{2}g_{\mu \nu} f + f_{Q}\left(P_{\mu \alpha \beta} Q_{\nu}^{~~ \alpha \beta} - 2 Q_{\alpha \beta \mu} P^{\alpha \beta}_{~~~\nu}\right) = \kappa\mathcal{T}_{\mu \nu}~.
\end{equation}
Recently, the covariant formulation to this field equation has been developed and successfully applied to study geodesic deviations and various phenomena within the cosmological context \cite{Zhao_2022_82_303}. So Eq. \eqref{fqfe} can be rewritten as
\begin{equation}\label{fqgre}
    f_{Q}\stackrel{\circ}{G}_{\mu\nu}+\frac{1}{2}g_{\mu\nu}(Qf_{Q}-f)+2f_{QQ}P^{\lambda}_{~~\mu\nu}\stackrel{\circ}{\nabla}_{\lambda}Q = \kappa \mathcal{T}_{\mu \nu}~,
\end{equation}
where, $\stackrel{\circ}{G}_{\mu\nu} = R_{\mu\nu}-\frac{1}{2}g_{\mu\nu}R$, with $R$ and $R_{\mu\nu}$ are the Riemannian Ricci scalar and tensor respectively and $f_{Q}$ is derivative of $f$ with respect to $Q$. Varying the action with respect to the connection leads to
\begin{equation}\label{hme}
\nabla _{\mu}\nabla _{\nu}(\sqrt{-g}f_{Q}P_{~~\mu\nu}^{\lambda} + H_{~~\mu\nu}^{\lambda})=0~,
\end{equation}
where $H_{~~\mu\nu}^{\lambda}=-\frac{1}{2}\frac{\delta(\sqrt{-g}\mathcal{L}_{m})}{\delta\Gamma _{~~\mu\nu}^{\lambda}}$ denotes the hyper momentum tensor density. Furthermore, the additional restriction over the connection may deducted, $\nabla _{\mu}\nabla _{\nu}( H_{~~\mu\nu}^{\lambda})=0$. As follows from Eq. \eqref{hme}
\begin{equation}\label{connectioneqn}
\nabla _{\mu}\nabla _{\nu}(\sqrt{-g}f_{Q}P_{~~\mu\nu}^{\lambda} )=0~.
\end{equation}
While considering the variation of action with respect to connection, there are two ways to incorporate symmetric teleparallelism. The first approach involves the use of inertial variation \cite{Golovnev_2017_34_145013}, where the connection is set in its pure-gauge form within the action. The second approach involves the consideration of a general connection in the action, but with the introduction of Lagrange multipliers to compensate for curvature and torsion \cite{Jimenez_2018_2018_039}.

\subsubsection{Coincident gauge}\label{Coincident Gauge}
The flatness condition imposes that the connection remains purely inertial, which allows it to be expressed in terms of a general element $\Lambda^\alpha{}_\beta$ from $GL(4,~\mathbb{R})$. This formulation, combined with the absence of torsion, results in the additional constraint $\partial_{[\mu}\Lambda^\alpha{}_{\nu]}=0$. Consequently, the general element from $GL(4,\mathbb{R})$ that governs the connection can be parameterized by a set of functions $\xi^\lambda$ as,
\begin{equation}\label{generalaffineeqn}
\Gamma^\alpha{}_{\mu\nu}=\frac{\partial x^\alpha}{\partial\xi^\lambda}\partial_\mu\partial_\nu\xi^\lambda~.
\end{equation}
The term $\frac{\partial x^{\alpha}}{\partial \xi ^{\lambda}}$ is interpreted as the inverse of the Jacobian matrix $\frac{\partial \xi ^{\lambda}}{\partial  x^{\alpha}}$ . This assertion indicates that within any coordinate system $\{x^{0}, x^{1}, x^{2}, x^{3}\}$, A four independent functions $\{\xi^{0}, \xi^{1}, \xi^{2}, \xi^{3}\}$ for which the Jacobian matrix $\frac{\partial \xi^{\mu}}{\partial x^{\nu}}$ is invertible can be selected, characterized by a non-zero determinant. This configurational freedom permits the construction of a flat and torsionless connection as expressed in Eq. \eqref{generalaffineeqn}. It is important to note that the construction of such connections is not unique; a whole family of connections is permissible under this framework. While this freedom to choose a connection holds no significant consequence in STEGR, it does bear substantial implications in modified theories like $f(Q)$ gravity. 

Moreover, Eq. \eqref{generalaffineeqn} illustrates an intriguing attribute of flat and torsionless connections: they can be globally set to zero through an appropriate coordinate choice. Specifically, for any given flat and torsionless connection, it invariably conforms to the form outlined in Eq. \eqref{generalaffineeqn} with certain functions $\xi^{\mu}$. Thus, by selecting coordinates such that $x^{\mu} = \xi^{\mu}$, a scenario where the connection is identically zero can be achieved, as $\partial_\mu\partial_\nu\xi^\lambda=0$. This scenario is referred to as the coincident gauge. In this context, the covariant derivative $\nabla _{\alpha }$ simplifies to the partial derivative $\partial_{\alpha}$, meaning that $Q_{\alpha \mu \nu }=\partial _{\alpha}g_{\mu\nu}$. This indicates that the Levi-Civita connection can be expressed in terms of the disformation tensor with the relationship $\left\{_{~\mu \nu}^\lambda\right\}=-L_{\ \mu \nu }^{\alpha }$.

Notably, excluding the boundary terms, this formulation specifically leads to the Einstein-Hilbert action for GR. A key advantage of this approach is its reliance solely on first derivatives of the metric, which ensures a well-posed variational principle that circumvents the need for Gibbons-Hawking-York boundary terms \cite{Gibbons_1977_15_2752}. Despite the implementation of diffeomorphism invariance, this principle is realised only up to a total derivative, leading to an action that is contingent on the choice of coordinates. It might seem surprising that the diffeomorphisms while using them to define the coincident gauge, but there are no inconsistencies in doing so. This is because, just like the TEGR is special due to its unique symmetry, the theory being discussed is also special among quadratic formulations because it has a stronger four-parameter gauge symmetry. Consequently, the complete framework actually possesses an eight-parameter gauge symmetry. In the coincident gauge, an additional symmetry appears as a diffeomorphism symmetry. Unlike in TEGR, where the metric and connection are interrelated, the non-metricity formulation of GR treats the connection as fundamentally a pure gauge. As a result, all dynamics are described by the metric, which operates in a trivially connected spacetime. It is important to note that the fields $\xi^\alpha$ that parameterize the connection play a role akin to St\"uckelberg fields associated with invariance under coordinate transformations. In this context, the coincident gauge can be viewed as the corresponding unitary gauge.

To derive the generalized Friedmann equations, start with a flat FLRW metric (i.e. $\kappa=0$). The most commonly used matter component considered in this framework is a perfect cosmic fluid, characterized by its stress-energy-momentum tensor \eqref{PF}. Incorporating the equation presented in \eqref{fqfe} yields the corresponding set of equations as,
\begin{eqnarray}
Qf_{Q}-\frac{f}{2} &=& \rho~,\label{fqr}\\
f_{Q}\dot{H}+\dot{f_{Q}}H &=& -\frac{1}{2}(\rho+p)~.\label{fqp} 
\end{eqnarray}
It is worth noting that the standard Friedmann equations of GR can be obtained if $f(Q)=Q$ is substituted. Some of the relevant studies can be seen in Refs. \cite{Harko_2018_98_084043, Hohmann_2019_99_024009, Soudi_2019_100_044008, Lazkoz_2019_100_104027, Bajardi_2020_135_912, Lin_2021_103_124001, Khyllep_2021_103_103521, Frusciante_2021_103_044021, Anagnostopoulos_2021_822_136634} within the framework of coincident gauge.

\subsubsection{Non-coincident gauge}\label{Non-coincident Gauge}
In the context of gravitational theories, adopting a non-coincident gauge, where $\Gamma_{\ \mu \nu}^{\alpha}\neq 0$, indicates the selection of a coordinate system that diverges from the standard Riemannian structure found in GR. This choice can significantly influence the formulation of the gravitational field equations and the definition of various physical quantities. The non-metricity tensor is crucial for capturing the effects of non-metricity on gravitational dynamics. By using a non-coincident gauge, researchers can explore the implications of these modifications in detail. This approach can lead to different physical interpretations and solutions compared to conventional GR, particularly in terms of how the gravitational field interacts with matter.

These connections can be classified into three types, based on $K_{1}$, $K_{2}$ and $K_{3}$ given by \cite{Dimakis_2022_106_123516},
\begin{eqnarray}\label{connectioncmnt}
 \Gamma^t_{~tt} &=& K_{1}~, \quad \Gamma^t_{~rr} = K_{2}~, \quad \Gamma^t_{~\theta\theta} = K_{2}r^2~, \nonumber\\
 \Gamma^r_{~tr} &=& \Gamma^r_{~rt} = \Gamma^{\theta}_{~t\theta} = \Gamma^{\theta}_{~\theta t} = \Gamma^\phi_{~t\phi} = \Gamma^\phi_{~\phi t} = K_{3}~, \nonumber\\
 \Gamma^\theta_{~r\theta} &=& \Gamma^\theta_{~\theta r} = \Gamma^\phi_{~r\phi} = \Gamma^\phi_{~\phi r} = \frac{1}{r}~, \quad \Gamma^r_{~\theta\theta} = -r~, \nonumber\\
 \Gamma^r_{~\phi\phi} &=&  -r\sin^2\theta~, \quad \quad \Gamma^t_{~\phi\phi} = K_{2}r^2\sin^2\theta~, \nonumber\\
 \Gamma^\phi_{~\theta\phi} &=& \Gamma^\phi_{~\phi\theta} = \cot\theta, \quad \Gamma^{\theta}_{~\phi\phi} = -\cos\theta\sin\theta~.
\end{eqnarray}
Using the definition of non-metricity scalar, Eqs. \eqref{NMT} and \eqref{connectioncmnt}, the non-metricity scalar can be obtained as,
\begin{eqnarray}\label{gnrlQ}
    Q = -6H^{2}+9HK_{3}+3K_{3}\left(K_{1}-K_{3}\right)+\frac{3K_{2}H}{a^{2}}-\frac{3K_{2}\left(K_{1}+K_{3}\right)}{a^{2}}~.
\end{eqnarray}
Meanwhile this connection should satisfy the curvature-free condition and this gives the constraints about three free parameters $K_{1}$, $K_{2}$ and $K_{3}$
\begin{eqnarray*}
    K_{3}(K_{1}-K_{3})-\dot{K_{3}} &=& 0~, \\
    K_{2}(K_{1}-K_{3})+\dot{K_{2}} &=& 0~, \\
    \kappa+K_{2}K_{3} &=& 0~.
\end{eqnarray*}
Consider three different connections for $\kappa=0$, each defined as,
\begin{itemize}
    \item Connection I: $K_{1} = \gamma(t)~, \quad K_{2} = K_{3} = 0$~;    
    \item Connection II: $K_{1} = \gamma(t) + \frac{\dot{\gamma}(t)}{\gamma(t)}, \quad K_{2} = 0~, \quad K_{3} = \gamma(t)$~;  
    \item Connection III: $K_{1} = -\frac{\dot{\gamma}(t)}{\gamma(t)}, \quad K_{2} = \gamma(t)~, \quad K_{3} = 0$~,
\end{itemize}
where $\gamma$ is the non-vanishing function of $t$. For each connection, the resulting field equations are different. The non-metricity scalar $Q$ and Friedmann equations for each connection are\\
\textbf{For Connection I:}
\begin{eqnarray}
    && Q = -6H^{2}~, \label{connection1Q}\\[5pt]
    && 3H^{2}f_{Q}+\frac{1}{2}(f-Qf_{Q}) = \rho~,\label{connection1r}\\[5pt]
    && -2\frac{d(f_{Q}H)}{dt}-3H^{2}f_{Q}-\frac{1}{2}(f-Qf_{Q}) = p~.\label{connection1p}
\end{eqnarray}
\textbf{For Connection II:}
\begin{eqnarray}
    && Q = -6H^{2}+9\gamma H+3\dot{\gamma}~, \label{connection2Q}\\[5pt]
    && 3H^{2}f_{Q}+\frac{1}{2}(f-Qf_{Q})+\frac{3\gamma}{2}\dot{Q}f_{QQ} = \rho~,\label{connection2r}\\[5pt]
    && -2\frac{d(f_{Q}H)}{dt}-3H^{2}f_{Q}-\frac{1}{2}(f-Qf_{Q})+\frac{3\gamma}{2}\dot{Q}f_{QQ} = p~.\nonumber\\
    \label{connection2p}
\end{eqnarray}
\textbf{For Connection III:}
\begin{eqnarray}
    && Q = -6H^{2}+\frac{3\gamma H}{a^{2}}+\frac{3\dot{\gamma}}{a^{2}}~, \label{connection3Q}\\[5pt]
    && 3H^{2}f_{Q}+\frac{1}{2}(f-Qf_{Q})-\frac{3\gamma}{2a^{2}}\dot{Q}f_{QQ} = \rho~,\label{connection3r}\\[5pt]
    && -2\frac{d(f_{Q}H)}{dt}-3H^{2}f_{Q}-\frac{1}{2}(f-Qf_{Q})+\frac{\gamma}{2a^{2}}\dot{Q}f_{QQ} = p~,\nonumber\\
    \label{connection3p}
\end{eqnarray}
where $\rho$ denotes the energy density, including baryonic matter, cold dark matter and radiation and $p$ represents the pressure. In the case of connection I, the non-metricity scalar $Q$ and the Friedmann equations \eqref{connection1Q} - \eqref{connection1p} are independent of the function $\gamma$, aligning with the results obtained under the coincident gauge where $\Gamma^{\lambda}_{~\mu\nu}=0$ in the flat FLRW metric.

\section{Basics of Statistics}\label{Basics of Statistics}
This section will focus on fundamental statistical techniques crucial for cosmological reconstruction. The best-fit values for the parameters can be derived through two main approaches: by minimizing a specific objective function utilizing optimization methods or by maximizing the likelihood function through marginalization. Another method involves using MCMC analysis, which efficiently marginalizes the posterior probability distribution across the parameter space to obtain constraints on the parameters of interest.

\subsection{$\chi^{2}$ model fitting}\label{chi2 Model Fitting}
The $\chi^2$ statistic is a widely used tool in statistical analysis, particularly for assessing the goodness of fit of a model to observational data. It quantifies the deviation between observed data and the corresponding theoretical predictions, considering the uncertainties in the measurements. This approach is commonly applied in various fields such as physics, astronomy and economics, where data is often subject to measurement errors and uncertainties.

Let $\{f_i\}$ represent the set of measured values at data points indexed by $i$ and $\{f_{i, \text{obs}}\}$ denote the corresponding observed data points. The corresponding theoretical predictions, $f_{i, \text{th}}\big(\{\theta\}\big)$, depend on a set of model parameters $\{\theta\}$. The standard deviation or uncertainty of each measurement is represented by $\sigma_i$. 

The $\chi^{2}$ is defined as,
\begin{equation}\label{chi2}
    \chi^2 = \sum_i \frac{\Big[f_{i, \text{obs}} - f_{i, \text{th}}\big(\{\theta\}\big)\Big]^2}{\sigma_i^2}~.
\end{equation}
This expression compares the difference between the observed and theoretical values, weighting the discrepancies by the inverse of the variance (i.e., the square of the standard deviation). This allows more reliable measurements to contribute more significantly to the overall statistic.

Moreover, in many experimental setups, the data points are not independent but may exhibit correlation due to various factors such as measurement conditions or systematic errors. In such cases, the uncertainty in the measurements is not simply described by individual $\sigma_i$'s but rather by a covariance matrix $Cov_{ij}$, which represents the correlation between different data points. Then the generalized form of the $\chi^{2}$ is give by
\begin{equation}\label{chi2gen}
\chi ^{2} = \sum_{i,j} \Big[{f_{i}}_{obs} - {f_{i}}_{th}\big(\{\theta\}\big) \Big]^{T}~ \big(Cov_{ij}\big)^{-1}~\Big[{f_{j}}_{obs} - {f_{j}}_{th}\big(\{\theta\}\big)\Big]~,
\end{equation}
where symbol $T$ as a superscript represents the transpose of a matrix. The combined $\chi^{2}_{tot}$ for statistical analysis with a combination of $n$ datasets is defined by adding up the $\chi^{2}$ associated
 with individual datasets.
\begin{equation}\label{chi2tot}
    \chi^{2}_{tot}=  \sum_{m}\chi^{2}_{n}~,
\end{equation}
where $n$ represents the datasets considered for that specific combination. This form of the $\chi^2$ statistic is necessary when dealing with correlated errors, as it properly adjusts for the dependencies between the measurements, ensuring that the covariance structure of the data is incorporated into the analysis.

To determine the optimal set of model parameters $\{\theta\}$, the goal is to minimize the $\chi^2$ function. The set of parameters $\{\theta_{best}\}$ that minimizes $\chi^2$ provides the best fit to the observed data. To assess the quality of the fit, it is useful to consider the $\chi^2_{\text{red}}$, which normalizes $\chi^2$ by the number of degrees of freedom $d$. The reduced $\chi^2$ is given by
\begin{align}\label{chi2red}
    &\chi^{2}_{red} = \frac{\chi^{2}}{d}~,\quad\quad d = N-k~.\nonumber\\[5pt]
    &N:\text{The number of data points}, \quad k: \text{The number of parameters in the model}
\end{align}
The value of $\chi^2_{\text{red}}$ provides an indication of the goodness of fit.
\begin{itemize}
    \item If $\chi^2_{\text{red}} = 1$, the model is considered to be a good fit to the data, with the observed deviations being consistent with the experimental uncertainties.
    \item If $\chi^2_{\text{red}} > 1$, the model does not fit the data well and it suggests that the uncertainties might have been overestimated or the model is inappropriate.
    \item If $\chi^2_{\text{red}} < 1$, it suggests over-fitting, where the model fits the data too perfectly, possibly due to the inclusion of too many parameters or underestimation of the uncertainties.
\end{itemize}
Additionally, the difference between the minimized $\chi^2$ and $\chi^2_{min}$ at a given set of parameter values, $\Delta \chi^2 = \chi^2 - \chi^2_{{min}}$, can be used to determine the confidence intervals for a specific mode at the $1\sigma~(68\%)$ and $2\sigma$ $(95\%)$ levels.

In the context of MCMC, $\chi^2$ minimization plays a key role in efficiently exploring the parameter space. The MCMC algorithm generates a sequence of parameter sets, forming a Markov chain, which ultimately converges to the distribution of parameters that minimize the $\chi^2$ function. This process enables a thorough exploration of the parameter space, accounting for uncertainties and correlations between parameters. By minimizing $\chi^2$, MCMC analysis yields robust parameter estimates and provides valuable insights into how well the model fits the data. This method is widely used in diverse fields such as cosmology, particle physics and statistical modeling, as it can handle complex models and large datasets effectively.

\subsection{Maximum likelihood analysis}\label{Maximum Likelihood Analysis}
Bayesian inference is a statistical approach that provides a framework for updating the probability distribution of model parameters in light of new data. It is based on Bayes' theorem, which relates the posterior probability of parameters $\{\theta\}$ to the prior probability and the likelihood of the observed data $\mathcal{D}$. With prior information $I$, the posterior probability distribution is given by,
\begin{equation}\label{PPD}
    P\big( \{\theta\} \big| \mathcal{D}, I \big) = \frac{P \big( \{\theta\}\big|I \big) P \big( \mathcal{D}\big|\{\theta\}, I \big)}{P\big( \mathcal{D}\big|I \big)}~,
\end{equation}
where, $P \big( \mathcal{D}/\{\theta\},I \big)$ denotes the probability of obtaining the data $\mathcal{D}$ given the parameters $\{\theta\}$ with respect to $I$, also referred to as the likelihood $\mathcal{L}$ and $P \big( \{\theta\}/I \big)$ represents the prior probabilities. Additionally, the global likelihood serving as a normalization factor is denoted and given by,
\begin{equation}\label{PDI}
P\big(\mathcal{D}/I \big) = \int_{\theta_{1}} .....\int_{\theta_{n}}  P \big( \{\theta\}\big|I \big)~P\big( \mathcal{D}\big|\{\theta\},I \big)~ d\theta_{1}.... d\theta_{n}~,
\end{equation}
 such that
\begin{equation}\label{PDIsum}
\int_{\theta_{1}} .....\int_{\theta_{n}}  P \big( \{\theta\}\big|\mathcal{D},I \big)~ d\theta_{1}.... d\theta_{n}=1~.
\end{equation}
To evaluate the goodness of fit between the model and the data, the likelihood function $P \big( \mathcal{D}/\{\theta\}, I \big)$ is often related to the $\chi^2$ function, which is commonly used in many practical applications. The likelihood can be expressed in terms of the $\chi^2$ statistic as
\begin{equation}\label{likelihoodgen}
\mathcal{L}\big(\{\theta\} \big) = exp\left(-\frac{\chi^{2}}{2} \right).
\end{equation}
Therefore, it can be observed that the maximized value of the likelihood function $\mathcal{L}$ corresponds to minimized value of $\chi^{2}$.

To estimate the posterior distribution of the parameters $\{\theta\}$, MCMC analysis is used, which allow for efficient sampling of the parameter space. The goal is to approximate the posterior distribution $P\big(\{\theta\}/\mathcal{D}, I\big)$ by utilizing the sampled sets and to generate a sequence of parameter sets $\{\theta\}$. One of the most widely used MCMC algorithms is the Metropolis-Hastings algorithm. The Metropolis-Hastings algorithm works by proposing a new set of parameters $\{\theta^{new}\}$ based on the current set $\{\theta\}$ and then evaluates the likelihood ratio between the new and current sets of parameters and decides whether to accept or reject the proposed set according to this ratio. Specifically, the acceptance probability is given by,
\begin{equation}\label{accepted prob}
  A = \min \left( 1, \frac{P\big(\mathcal{D}\big|\{\theta^{new}\}, I \big)}{P\big(\mathcal{D}\big| \{\theta\}, I \big)} \right),  
\end{equation}
where $P\big(\mathcal{D}\big|\{\theta\}, I \big)$ and $P\big(\mathcal{D}\big|\{\theta^{new}\}, I \big)$ are the likelihoods of the current and proposed parameter sets, respectively. If the proposed set is accepted, it becomes the next state in the chain; otherwise, the chain stays at the current set. Over time, the MCMC chain will explore the parameter space, with the probability of being in any given region of the space proportional to the likelihood of that region. This enables the chain to converge to the posterior distribution. The algorithm will tend to spend more time in regions of high likelihood, allowing for an efficient search for the best-fitting parameter values. Once the MCMC chain has run for a sufficient number of iterations and reached equilibrium, the samples from the chain to estimate the posterior distribution of the parameters can be utilize. From the MCMC samples, the marginalized posterior distributions for each parameter by integrating over the other parameters can be computed. These marginal distributions provide the parameter values along with their uncertainties. Therefore, the MCMC analysis resembles $\chi^{2}$ minimization, but it marginalizes the parameters instead of optimizing them.

Foreman-Mackey et al. \cite{Foreman-Mackey_2013_125_306} developed the ensemble sampler for MCMC in Python, emcee \footnote{\href{https://github.com/dfm/emcee}{https://github.com/dfm/emcee}}. As part of our analysis, a two-dimensional confidence contours using the Python module GetDist \footnote{\href{https://github.com/cmbant/getdist}{https://github.com/cmbant/getdist}} are visualized, developed by Lewis \cite{Lewis_2019_190.13970}. The Python gallery ChainConsumer \footnote{\href{https://github.com/Samreay/ChainConsumer}{https://github.com/Samreay/ChainConsumer}} to visualize these contours are used.

\section{Observational Datasets} \label{Observational data}
Cosmological observations have played a crucial role in understanding the expansion history and dynamics of the Universe. This discussion provides a brief overview of significant cosmological discoveries that enhance our understanding of cosmic evolution.

\subsection{Hubble measurements}\label{Hubble Measurements}
The Hubble dataset provides valuable insights into the detailed history of the expansion of the Universe. Direct measurements of $H(z)$ at various redshifts serve as a key cosmological data, which are obtained from the ages of the most massive and passively evolving galaxies. These Hubble measurements are derived using two primary techniques: the galaxy differential age method (often referred to as the CC method) and radial BAO size method \cite{Yu_2018_856_3}.

Reference \cite{Moresco_2012_2012_006, Moresco_2016_2016_014} have presented $13~H(z)$ values derived from the BC03 and MaStro stellar population synthesis models \cite{Bruzual_2003_344_1000}, which are included in the CCB and CCM compilations, respectively. Furthermore, \cite{Zhang_2014_14_1221, Maraston_2011_418_2785} have provided $5~H(z)$ measurements using the BC03 model, which have been integrated into the CCB compilation. The combined MaStro/BC03 $H(z)$ values, totaling 2 measurements, are presented in \cite{Moresco_2015_450_L16}. In addition, \cite{Stern_2010_2010_008} introduced an alternative SPS model, distinct from both MaStro and BC03, providing $11~H(z)$ measurements known as the CCH compilation, along with $26$ points derived from the BAO method \cite{Sharov_2018_6_1}. Our analysis incorporates a total of $32~H(z)$ measurements spanning a redshift range of $0.07 \leq z \leq 1.965$ \cite{Moresco_2022_25_6}. 

\subsection{Type Ia Supernovae}\label{Type Ia Supernovae}
When a white dwarf star explodes after reaching the Chandrasekhar mass limit, typically by accreting mass from a companion star then a SNe Ia occurs. These events are crucial as standard candles for measuring cosmic distances due to their predictable luminosity. In 1998, Riess et al. \cite{Riess_1998_116_1009} used observations of 16 distant and 34 nearby SNe Ia from the Hubble Space Telescope to discover the accelerated expansion of the Universe. This finding was confirmed in 1999 by Perlmutter et al. \cite{Perlmutter_1999_517_565}, who analyzed 18 nearby supernovae from the Calan-Tololo survey and 42 high-redshift SNe Ia. Since then, numerous surveys have contributed to the study of SNe Ia, including the Sloan Digital Sky Survey (SDSS) Supernova Survey \cite{Holtzman_2008_136_2306, Kessler_2009_185_32}, the Lick Observatory Supernova Search (LOSS) \cite{Leaman_2011_412_1419, Li_2011_412_1441}, the Nearby Supernova Factory (NSF) \cite{Copin_2007, Scalzo_2010_713_1073}, the Supernova Legacy Survey (SNLS) \cite{Astier_2006_447_31, Baumont_2008_491_567} and the Higher-Z Team \cite{Riess_2004_607_665, Riess_2007_659_98}. More recently, the Union 2.1 dataset, containing 580 SNe Ia, was released \cite{Visser_2004_21_2603}.

The Pantheon$^+$ \big( $\mathrm{PN}^+~\&~$SH0ES \big) analysis \cite{Scolnic_2022_938_113} is an updated version of the original Pantheon study. The primary distinction between the two lies in the inclusion of additional data in the Pantheon$^+$ compilation. While the original Pantheon analysis utilized a sample of $1048$ SNe Ia to investigate the expansion history of the Universe, the Pantheon$^+$ analysis incorporates an even larger sample of $1701$ SNe Ia, providing a more extensive dataset for cosmological studies. The term ``$\mathrm{PN}^+~\&~$SH0ES'' as referred to in the Pantheon$^+$ analysis in Ref.~\cite{Brout_2022_938_110}, incorporates the SH0ES Cepheid host distance anchors \big(\cite{Riess_2022_934_L7}\big) in the likelihood which helps to break the degeneracy between the parameter when analyzing SNe Ia alone. This comprehensive dataset includes distance moduli derived from $1701$ light curves of $1550$ spectroscopically confirmed SNe Ia, gathered from $18$ distinct surveys. The Pantheon$^+$ analysis encompasses a broader redshift range of $0.00122 < z < 2.2613$, in contrast to the original Pantheon dataset, which is limited to redshifts above $z < 0.01$. This expanded range allows for a more thorough evaluation of systematic uncertainties, resulting in better-constrained parameters.

\subsection{Baryon acoustic oscillations}\label{Baryon Acoustic Oscillations}
BAO refer to the periodic density fluctuations in baryonic matter created by sound waves propagating through the early Universe \cite{Peebles_1970_162_815, Peebles_2003_75_559}. Like SNe Ia, BAO serves as a standard ruler in cosmology, providing a means to study the expansion history of the Universe. The imprint of BAO on the matter power spectrum can be detected through galaxy cluster surveys at low redshifts ($z < 1$) \cite{Alam_2004_2004_008}. Additionally, BAO signatures can be observed in reionization-era emissions, offering valuable information about the early Universe at higher redshifts \big($1.5 \leq z \leq 2.0$\big) \cite{Jonsson_2004_2004_007}.

By analyzing the apparent magnitude of BAO in astronomical data, one can calculate the Hubble parameter and angular diameter distance, providing key cosmological insights. Numerous surveys, including the Two-degree-Field Galaxy Redshift Survey (2dFGRS) \cite{Colless_2003_astro-ph/0306581} and the SDSS \cite{York_2000_120_1579, Tegmark_2006_74_123507}, have focused on BAO measurements. Among these, SDSS has been especially successful, with regular data releases, including the eighth one (SDSS DR8) in 2011.\footnote{\href{www.sdss3.org/dr8/}{www.sdss3.org/dr8/}}

Recent observational efforts, such as the Dark Energy Spectroscopic Instrument (DESI) surveys \cite{Adame_2024_2404.03002}, SNe Ia \cite{Riess_1998_116_1009, Perlmutter_1999_517_565}, the Wilkinson Microwave Anisotropy Probe \cite{Spergel_2003_148_175}, the CMB \cite{Hinshaw_2013_208_19} and the Baryon Oscillation Spectroscopic Survey (BOSS) \cite{Alam_2017_470_2617}, have prompted further exploration into potential modifications and extensions to GR. These datasets are central to ongoing efforts to refine our understanding of cosmic acceleration and to explore alternatives to the standard cosmological model. 

\section{Stability Analysis}\label{Stability Analysis}
A dynamical system can be understood as any system that evolves over time or through iterations, described by a space (phase space) and a mathematical rule that governs its evolution. In this context, the phase space is a collection of variables that describe the state of the system and the mathematical rule (often in the form of differential equations or iterated maps) determines how the state changes over time. This abstract formulation applies to systems ranging from simple mechanical systems like a pendulum to complex phenomena such as the behavior of the human brain or the entire Universe.

There are two primary categories of dynamical system based on how time is treated: continuous and discrete. Continuous dynamical systems are governed by Ordinary Differential Equation, which describe how the state of the system changes continuously with respect to an independent variable. On the other hand, discrete dynamical systems are governed by iterated maps or difference equations, where the state evolves in discrete steps rather than continuously. For the purpose of this discussion is to focus on continuous systems, particularly in fields like cosmology, where the evolution of the Universe is described by the Einstein field equations - a set of ODE.

Consider an $n$-dimensional phase space, $X \subset \mathbb{R}^n$, where each point in this space represents the state of the system. The state is described by a vector of variables $\big(x_1, x_2, \dots, x_n \big)$, which represent positions or other physical quantities depending on the system being modeled. The time evolution of these variables is governed by a set of ODE \cite{Strogatz_2018},
\begin{equation}\label{ODES}
    \dot{x}_i = f_i(x_1, \dots, x_n), \quad i = 1, \dots, n
\end{equation}
where $f_i : X \to X$ are smooth functions defining how each state variable $x_i$ changes over time. The system is said to be autonomous if the equations do not explicitly depend on the independent variable $t$. In a more compact form, this system can be expressed as a vector field $f(x)$ on $\mathbb{R}^n$ as,
\begin{equation}\label{system}
    \dot{x} = f(x)~.
\end{equation}
Here, $\dot{x} = \big(\dot{x}_1, \dot{x}_2, \ldots, \dot{x}_n \big)$ is the derivative of the state vector $x = \big(x_1, x_2, \ldots, x_n \big)$ with respect to time and $f(x) = \big(f_{1}(x), f_{2}(x), \ldots, f_{n}(x)\big)$.

The evolution of the system can be visualized in the phase space, where each point represents a possible state of the system and its trajectory is determined by the initial conditions. A solution to the system of ODE with an initial condition $x(0) = x_0$ traces a curve (trajectory or orbit) in phase space. This curve represents the evolution of the system over time, starting from the initial state $x_0$. The behavior of the system can be studied by observing the flow of trajectories in phase space.

An important aspect of dynamical systems is the concept of critical points (or fixed points), which are points in phase space where the derivatives of all state variables are zero \big(i.e. $f(x_0) = 0$\big) \cite{Wiggins_2003}. These critical points can represent equilibrium states of the system. The stability of these points is crucial for understanding the long-term behavior of the system. A critical point is stable (or Lyapunov stable) if solutions that start close to it remain close over time. It is asymptotically stable if solutions not only remain close but also converge to the critical point. On the other hand, an unstable critical point causes nearby solutions to move away from it. The nature of stability can be analyzed through linearization, where the system is approximated by a linear model near the critical point, providing insights into the behavior of solutions in its neighborhood.

Through the study of these concepts, dynamical systems offer powerful tools for understanding a wide range of physical, biological and cosmological phenomena from simple mechanical systems to the complex evolution of the Universe itself. Analyzing the trajectories in phase space and the stability of critical points provide insights into the behavior of the system, allows prediction of future states and helps understand the underlying principles governing its evolution.

\subsection{Linear stability theory}\label{Linear Stability Theory}
To linearize the system near a critical point, a Taylor expansion of $f(x)$ around $x_{0}$ is used. Since $f(x_{0}) = 0$ by definition of a critical point, the system to the first order expanded as
\begin{equation}\label{systemapprox}
   f(x) \approx J(x_{0})(x - x_{0})~,
\end{equation}
where $J(x_{0})$ is the Jacobian matrix of the system evaluated at $x_{0}$ and $x - x_{0}$ represents small perturbations from the critical point.
Thus, the linearized system becomes
\begin{equation}\label{linearizedsys}
 \dot{x} = J(x_{0}) (x - x_{0})~.
 \end{equation}
This approximation holds when the deviations from $x_{0}$ are small and the system behaves in a linear manner around the critical point. The evolution of the state near $x_{0}$ is now governed by the Jacobian matrix, which encodes the local behavior of the system.

The Jacobian matrix $J(x_{0})$ is a matrix of partial derivatives representing how the components of the vector field $f(x)$ change with respect to each state variable at the critical point $x_{0}$ as
\begin{equation}\label{Jacobimatrix}
    J(x_{0}) = \begin{pmatrix}
\frac{\partial f_1}{\partial x_1} & \cdots & \frac{\partial f_1}{\partial x_n} \\
\vdots & \ddots & \vdots \\
\frac{\partial f_n}{\partial x_1} & \cdots & \frac{\partial f_n}{\partial x_n}
\end{pmatrix}_{x = x_{0}}.
\end{equation}
The matrix $J(x_{0})$ is crucial for determining the nature of the critical point. To understand the stability of $x_{0}$, the eigenvalues of the Jacobian matrix are examined. The eigenvalues give insight into how small perturbations evolve over time. The stability of a critical point $x_{0}$ depends on the eigenvalues of $J(x_{0})$. The classification of fixed points as,
\begin{itemize}
    \item \textbf{Hyperbolic point :} A point is considered hyperbolic if all eigenvalues of the Jacobian matrix have non-zero real parts. In this case, the linearized system adequately describes the stability of the critical point and the analysis using the Jacobian matrix suffices to determine whether the point is an attractor, repeller or saddle \cite{Strogatz_2018, Glendinning_1994, Wiggins_2003}. 
\begin{itemize}
    \item \textit{Stable point (Attractor) -} If all the eigenvalues of $J(x_{0})$ have negative real parts, the trajectories near $x_{0}$ will converge to $x_{0}$ indicating that $x_{0}$ is a stable attractor. This is the case for stable nodes.
    \item \textit{Unstable point (Repeller) -} If all the eigenvalues have positive real parts, trajectories will diverge from $x_{0}$ making it an unstable point or repeller. The system tends to move away from $x_{0}$, as seen in unstable nodes.
    \item \textit{Saddle point -} If the eigenvalues of $J(x_{0})$ have mixed signs, the point is a saddle. Trajectories will be attracted to $x_{0}$ along the directions associated with negative real parts but repelled along directions with positive real parts.
    \item \textit{Spiral -} In systems with complex eigenvalues (when the real part is non-zero and the imaginary part is also present), the fixed point may behave as a spiral. The direction of the spiral depends on the sign of the real part of the eigenvalues. A stable spiral occurs when the real part is negative, causing the trajectories to spiral inward, while an unstable spiral occurs when the real part is positive, causing the trajectories to spiral outward.
\end{itemize}
\item \textbf{Non-Hyperbolic point :} If at least one eigenvalue of the Jacobian matrix has a real part equal to zero, the system is non-hyperbolic and the standard linear stability theory is inadequate for determining the stability of the critical point. In such cases, the behavior of trajectories near the critical point can be more complicated and non-linear methods such as central manifold theory, Lyapunov methods or bifurcation analysis are often used to classify these points. Non-hyperbolic points can exhibit neutrally stable behavior or can be a center (in the case of purely imaginary eigenvalues).
\end{itemize}
This section covers the essential elements of linear stability analysis, offering a comprehensive understanding of how critical points in dynamical systems can be classified and the tools required for more advanced stability analysis when necessary.

\subsection{Cosmological scalar perturbation}\label{Cosmological Scalar Perturbation}
In stability analysis within the context of modified gravity, the focus is on understanding how small deviations from a homogeneous and isotropic Universe influence its overall stability and evolution. Typically, cosmological models assume a background Universe that is homogeneous and isotropic. However, real Universe is not perfectly homogeneous and small fluctuations in density and other quantities inevitably arise. These fluctuations can evolve over time, either growing or diminishing, depending on the underlying cosmological model and the perturbation.

In this context, the investigation focuses on how small perturbations in the Hubble parameter $H(t)$ and the matter density $\rho(t)$ impact the evolution of the Universe. The Hubble parameter $H(t)$, which governs the rate of expansion of the Universe, is perturbed by a small term $\delta$, such that
$H(t) \rightarrow H(t)(1 + \delta)$. Similarly, the matter density $ \rho(t)$, which describes the distribution of matter in the Universe, is perturbed by a small deviation $\delta_m$ by $\rho(t) \rightarrow \rho(t)(1 + \delta_m)$. The perturbations $\delta$ and $\delta_m$ are assumed to be small in magnitude, representing first-order perturbations from the homogeneous background. These perturbations can be thought of as small fluctuations in the Hubble parameter and the matter density around their background. By analyzing these perturbations, the stability of the cosmological model in the presence of small fluctuations can be assess. Specifically, the goal is to determine whether these perturbations will grow or decay over time, which provides important information about the late-time behavior of the Universe. If the perturbations grow uncontrollably, the model may indicate an unstable or unrealistic cosmological evolution. Conversely, if the perturbations diminish, the system is considered stable and the Universe will tend to return to its homogeneous state.

In modified gravity theories, the presence of such small fluctuations is crucial. Modified gravity can lead to a very different evolution of perturbations compared to standard GR, depending on its specific form. Modified gravity theories, for instance, may predict that perturbations in the Hubble parameter or the matter density will behave in a manner inconsistent with observations \cite{Clifton_2012_513_1, Koyama_2016_79_046902, Sharma_2022_934_13}. Therefore, investigating the stability of these perturbations is not only important for understanding the behavior of the Universe in modified gravity but also for ensuring that these theories are consistent with current observational data. If perturbations in the Hubble parameter or matter density behave in a way that aligns with observations, it would lend support to the validity of the modified gravity theory. Conversely, if these perturbations grow too large or evolve in an unphysical manner, it may indicate that the theory needs to be adjusted or abandoned.

Overall, the stability analysis of small perturbations provides a powerful tool to test the robustness of cosmological models, especially in modified gravity frameworks and ensures that they align with the observable Universe. By systematically examining how small deviations evolve, one can gain insights into the dynamical behavior of the Universe and assess whether these deviations are consistent with a stable and observable cosmological evolution.

The subsequent chapters apply the symmetric teleparallel gravity or $f(Q)$ gravity theory previously discussed to address specific challenges mentioned above.
\chapter{Accelerating cosmological models in $f(Q)$ gravity and the phase space analysis} 

\label{Chapter2} 

\lhead{Chapter 2. \emph{Accelerating cosmological models in $f(Q)$ gravity and the phase space analysis}} 

\vspace{10 cm}
* The work, in this chapter, is covered by the following publication: \\

\textbf{S. A. Narawade}, Shashank P. Singh and B. Mishra, ``Accelerating cosmological models in $f(Q)$ gravity and the phase space analysis", \textit{Phys. of the Dark Universe} \textbf{42} (2023) 101282.
 
\clearpage
  
\section{Introduction}
Before we consider the $f(Q)$ gravity as a viable alternative to GR, one of the crucial issue of $f(Q)$ gravity to be addressed, i.e. the stability of its cosmological models. One among several stability analysis is the dynamical stability analysis \cite{Khyllep_2023_107_044022} that studies the behavior of a model under small perturbations. To be specific, it aims at to find the model coming back to the original state or evolve into a different solution. This analysis further provides an accurate prediction on the behaviour of the model pertaining to the physical scenario and thereby resulted in a robust theoretical framework. One can refer some articles on dynamical system analysis in modified theories of gravity \cite{Bonanno_2012_14_025008, Khyllep_2021_103_103521, Agrawal_2024_84_56, Duchaniya_2023_83_27, Samaddar_2023_83_283,Duchaniya_2024_43_101402, Pati_2023_83_445}.

This chapter focuses mainly on the history of the evolution of the Universe using one of the most universally accepted methodology, namely dynamic system analysis. In this chapter, we explored the different phases of the Universe such as radiation, matter and the current accelerating phase. As part of this study, the Friedmann equations are used in framing the autonomous dynamical system. In addition, the well-motivated forms of $f(Q)$ gravity are considered as well in framing the autonomous dynamical system. In addition, we discussed the critical points in relation to the deceleration and EoS parameter. We also analyzed the phase portraits for both models in order to examine their stability. The result may provide some crucial insights into the viability of $f(Q)$ gravity and its possible contribution in understanding the structure and evolution of the Universe. This chapter is structured as, in section \ref{sec2.2}, the formulation of $f(Q)$ gravity with some adjustment and its corresponding field equations are introduced. Section \ref{sec2.3} presents Hubble parameter for the dynamical analysis, along with two cosmological models, each with a specific form of $f(Q)$. In section \ref{sec2.4}, a dynamical system analysis is conducted to identify the critical points and explore their behavior. Finally, the results and conclusion are discussed in section \ref{sec2.5}.

\section{Modified field equations}\label{sec2.2}
The field equations for $F(Q)= Q+f(Q)$ in perfect fluid from Eqs. \eqref{fqr} and \eqref{fqp} can be obtained as \cite{Atayde_2021_104_064052},
\begin{eqnarray} 
\rho &=& \frac{1}{2}(Q+2Qf_{Q}-f)~,\label{eq2.1}\\
p &=& \frac{1}{2}(f-Q-2Qf_{Q})-2\dot{H}(2Qf_{QQ}+f_{Q}+1)~.\label{eq2.2}
\end{eqnarray}
We denote $f(Q)=f$ and $f_Q=\frac{\partial f}{\partial Q}$ and also we consider that the Universe is filled with dust and radiation fluids. Hence, 
\begin{eqnarray*}
\rho=\rho_{m}+\rho_{r},~~~~~~~~~~~~~p=p_{r}=\frac{1}{3}\rho_{r},
\end{eqnarray*}
with $\rho_{m}$ and $\rho_{r}$ represents the energy density for the matter and radiation phase respectively. Then from Eqs. \eqref{eq2.1} and \eqref{eq2.2}, we get
\begin{eqnarray}
3H^{2} &=& \rho_{total} = \rho_{r} + \rho_{m} + \rho_{DE}~,\label{eq2.3}\\
2\dot{H} + 3H^{2} &=& -p_{total} = -p_{r} - p_{m} - p_{DE}~,\label{eq2.4}
\end{eqnarray}
where $p_m$, $p_r$ respectively denotes the pressure of matter and radiation phase; $\rho_{DE}$ and $p_{DE}$ be the energy density and pressure of DE phase, which can be expressed as,
\begin{eqnarray*}
\rho_{DE} &=& \frac{f}{2} - Qf_{Q},\\
p_{DE} &=& 2\dot{H}(2Qf_{QQ} + f_{Q}) + Qf_{Q}-\frac{f}{2}.
\end{eqnarray*}
To note, the above two equations satisfies the conservation equation of the energy momentum tensor, which can be expressed as, $\dot{\rho}+3H(\rho + p) = 0$. Now, the total EoS parameter and the EoS parameter due to DE can be obtained respectively as,
\begin{eqnarray}
\omega_{total} = \frac{p_{total}}{\rho_{total}} &=& -1+\frac{\Omega_{m} + \frac{4}{3}\Omega_{r}}{2Qf_{QQ}+f_{Q}+1}~,\label{eq2.5}\\[5pt]
\omega_{DE}= \frac{p_{DE}}{\rho_{DE}} &=& -1 + \frac{4\dot{H}(2Qf_{QQ}+f_{Q})}{f-2Qf_{Q}}~.\label{eq2.6}
\end{eqnarray}
The density parameter pertaining to pressure-less matter, radiation are given in \eqref{densityparameter} and for DE it is represented as, 
\begin{equation}\label{eq2.7}
\Omega_{DE} = \frac{\rho_{DE}}{3H^{2}}~.
\end{equation}
The EoS parameter describes the present state of the Universe. A bunch of cosmological observations recently constrained the current value of the EoS parameter to be, $\omega=-1.29_{-0.12}^{+0.15}$ \cite{Valentino_2016_761_242}, $\omega=-1.3$ \cite{Vagnozzi_2020_102_023518}, Supernovae Cosmology Project, $\omega = -1.035_{-0.059}^{+0.055}$ \cite{Amanullah_2010_716_712}; WAMP+CMB, $\omega = -1.079_{-0.089}^{+0.090}$ \cite{Hinshaw_2013_208_19}; Plank 2018, $\omega = -1.03\pm 0.03$ \cite{Aghanim_2020_641_A1} or $\omega=-1.33_{-0.42}^{+0.31}$ \cite{Valentino_2021_23_404}.\\
These field equations provide the foundation for studying the dynamics of the Universe within the context of $f(Q)$ gravity. In the next section, we will explore Hubble parameter, deceleration parameter and jerk parameter which help us understand the cosmic evolution and along with the dynamic behaviour of the specific $f(Q)$ models.

\section{The $f(Q)$ gravity models}\label{sec2.3}
\begin{figure}[ht]
    \centering
    \begin{subfigure}{0.5\textwidth}
        \centering
        \includegraphics[width=75mm]{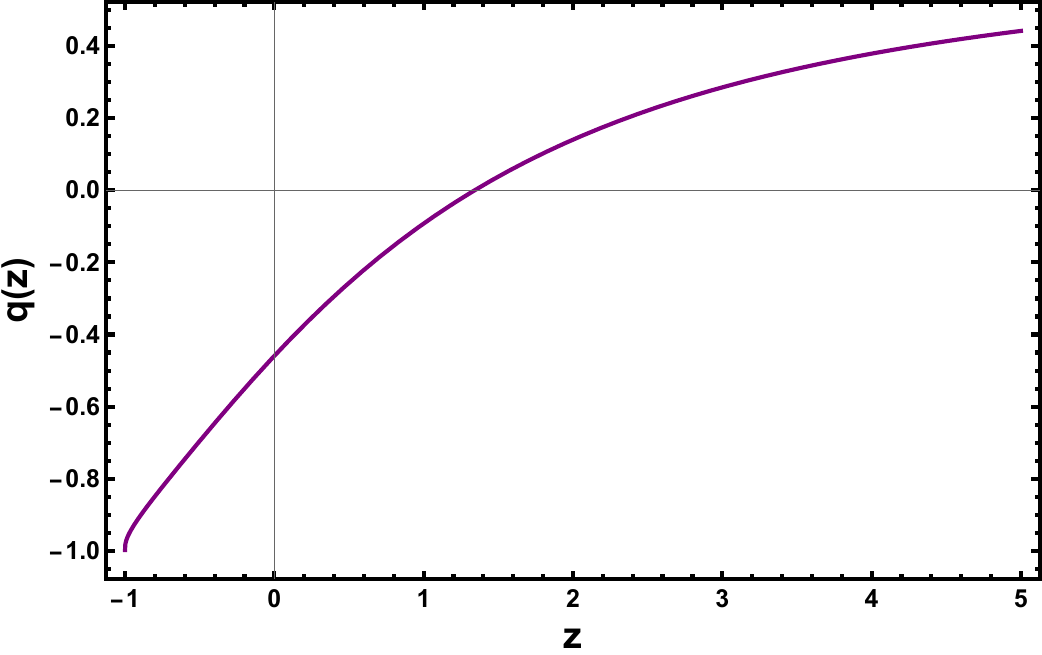}
        \caption{Evolution of deceleration parameter.}
        \label{q2.1}
    \end{subfigure}%
    \hfill
    \begin{subfigure}{0.5\textwidth}
        \centering
        \includegraphics[width=75mm]{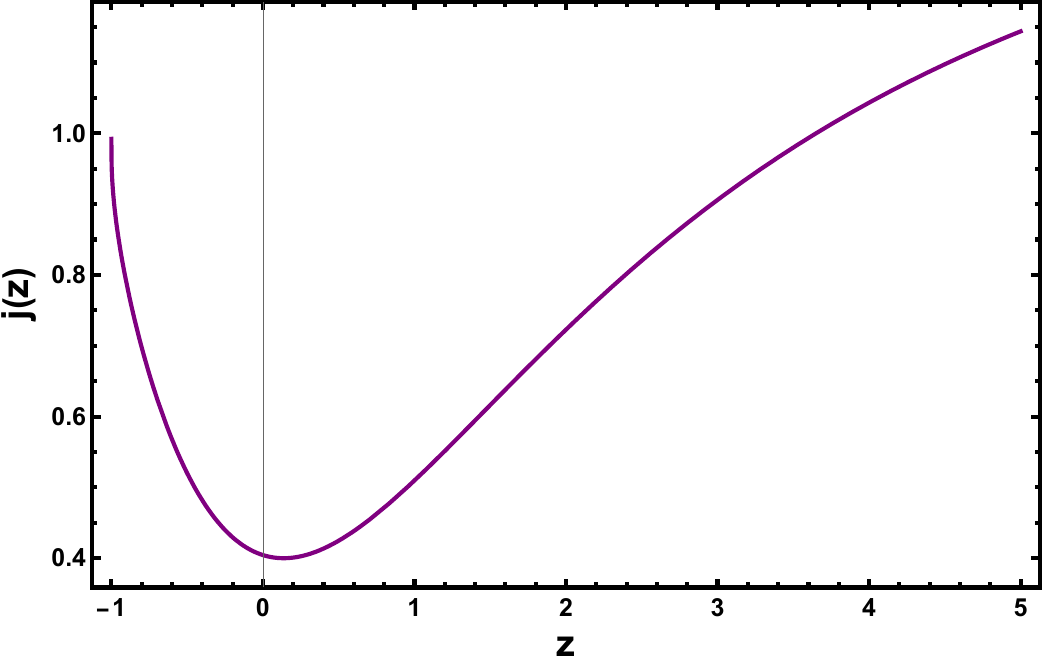}
        \caption{Evolution of jerk parameter.}
        \label{j2.1}
    \end{subfigure}%
    \caption{Evolution of cosmographic parameters as a function of redshift.}
    \label{fig2.1}
\end{figure}

To solve the above system and to analyse the behaviour of the dynamical parameters, we consider some form of Hubble parameter $H$ and the function $f(Q)$. The Hubble parameter, $H=\xi+\frac{\eta}{t}$ corresponds to the scale factor, $a(t)=e^{\xi t}t^{\eta}$, known as the hybrid scale factor \cite{Mishra_2015_30_1550175, Mishra_2018_33_1850052, Mishra_2019_79_34, Tripathy_2021_30_2140005}. Subsequently, the non-metricity scalar becomes, $Q=6H^2=6\left(\xi+\frac{\eta}{t}\right)^2$. This form of Hubble parameter further provides a time varying deceleration parameter $\left(q=-1+\frac{\eta}{(\xi t+\eta)^2}\right)$, which can simulate a cosmic transition from early deceleration to late-time acceleration. The deceleration parameter $q\approx -1+\frac{1}{\eta}$ when $t\rightarrow0$ and $q\approx-1$ when $t\rightarrow \infty$. To realise the positive deceleration at early Universe for the transient Universe, the scale factor parameter $\eta$ should be $0<\eta<1$. The transition can occur at $t=-\frac{\eta}{\xi}\pm\frac{\sqrt{\eta}}{\xi}$. We will restrict to the positivity of the second term and ignore the negativity. This is because the negativity of the second term would provide negative time, which may lead to unphysical situation at the Big Bang scenario. In that case also the parameter $\eta$ restricted to $0<\eta<1$, since the cosmic transit may have occurred at, $t=\frac{-\eta+\sqrt{\eta}}{\xi}$. The jerk parameter, $j=1-\frac{3\eta}{(\eta+\xi t)^2}+\frac{2\eta}{(\eta+\xi t)^3}$. We have used the parameter values for model analysis as $\xi = 0.965$ and $\eta = 0.60$ \cite{Mishra_2015_30_1550175}. The deceleration parameter shows a transient behaviour with the transition occurs at $z=1.34$. It reduces from early time to late-time and at present its value is, $q_0 = -0.46$ [Fig. \ref{q2.1}]. The evolution of jerk parameter remains entirely in the range $0<j<1$. It reduces from early time and attained the minimum value at present and eventually converges to $1$ at the late-time [Fig. \ref{j2.1}].

\subsection{Log-square-root model}
We first consider the logarithmic form of $f(Q)$ \cite{Anagnostopoulos_2023_83_58} as, 
\begin{equation}\label{eq2.8}
f(Q)=nQ_{0}\sqrt{\frac{Q}{\lambda_{1}Q_{0}}}\ln{\frac{\lambda_{1}Q_{0}}{Q}}~,
\end{equation}
where $n$ and $\lambda_{1}>0$ are free parameters; $Q_{0}=6H_{0}^{2}$, where $H_{0}=70.7~kms^{-1}Mpc^{-1}$ \cite{Hotokezaka_2019_3_940} represents the present value of $H$. For $n=0$, one can recover the GR equivalent model.

\begin{figure}[ht]
    \centering
    \begin{subfigure}{0.5\textwidth}
        \centering
        \includegraphics[width=75mm]{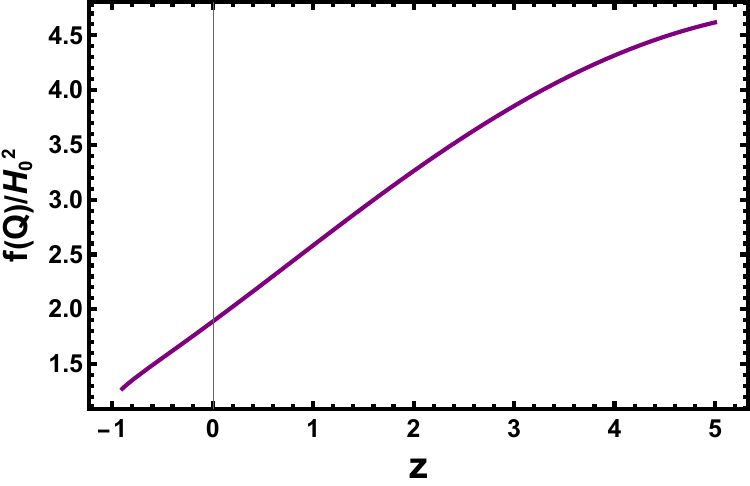}
        \caption{Evolution of $\frac{f(Q)}{H_{0}^2}$ in redshift.}
        \label{sign12.2}
    \end{subfigure}%
    \hfill
    \begin{subfigure}{0.5\textwidth}
        \centering
        \includegraphics[width=77mm]{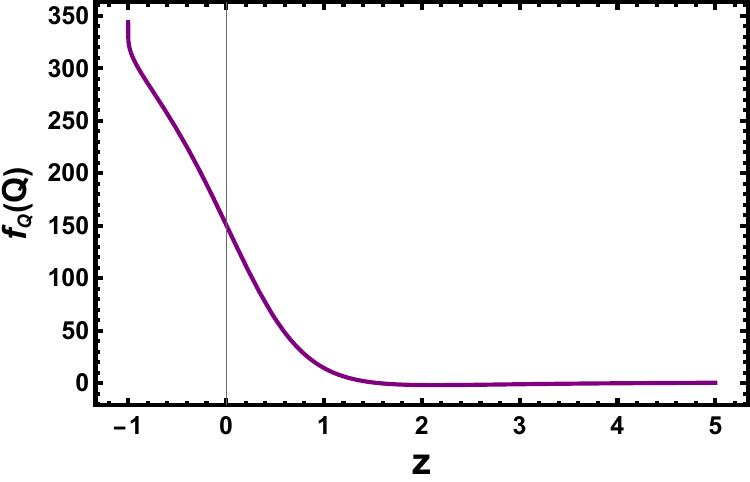}
        \caption{Evolution of $\frac{df(Q)}{dQ}$ in redshift.}
        \label{sign112.2}
    \end{subfigure}%
    \caption{Evolution of log-square-root model as a function of redshift.}
    \label{fig2.2}
\end{figure}

At the outset, we have studied the evolutionary behaviour of the function $f(Q)$ by plotting the graphs $\frac{f(Q)}{H_{0}^2}$ vs $z$ and $f_Q$ vs $z$ where $z$ is redshift \cite{Frusciante_2021_103_044021}. We wish to analyse $\frac{f(Q)}{H_{0}^2}$ and $f_Q$ as functions of redshift since the Hubble parameter is related to the scale factor as $H=\frac{\dot{a}}{a}$. One can see from Fig. \ref{sign12.2} as time passes, $\frac{f(Q)}{H_{0}^2}$ shows a decreasing behavior and gradually vanishes. Whereas, $f_Q$ [Fig. \ref{sign112.2}] starts with a lower positive value, increases over time and remains positive throughout.
Substituting Eq. \eqref{eq2.8} into Eqs. \eqref{eq2.1}, \eqref{eq2.2} and \eqref{eq2.5}, we obtain
\begin{eqnarray}
p_{total} &=&\frac{\lambda_{1}\Omega_{r}(z+1)^{4}\sqrt{\frac{H^{2}(z)}{\lambda_{1}Q_{0}}}-3\sqrt{6}H^{2}(z)n+\sqrt{6}H(z)n(z+1)H_{z}(z)}{3\lambda_{1}\sqrt{\frac{H^{2}(z)}{\lambda_{1}Q_{0}}}}~,\label{eq2.9}\\[10pt]
\rho_{total} &=& H(z)Q_{0}n\sqrt{\frac{6}{\lambda_{1}Q_{0}}}+\Omega_{r}(z+1)^{4}+\Omega_{m}(z+1)^{3}~, \label{eq2.10}\\[10pt]
\omega_{total} &=& \frac{\lambda_{1}\Omega_{r}(z+1)^{4}\sqrt{\frac{H^2(z)}{\lambda_{1}Q_{0}}}-3\sqrt{6}H^2(z)n+\sqrt{6}H(z)n(z+1)H_{z}(z)}{3\sqrt{\frac{H^2(z)}{\lambda_{1}Q_{0}}}\left(\sqrt{6}\lambda_{1}Q_{0}n\sqrt{\frac{H^2(z)}{\lambda_{1}Q_{0}}}+\lambda_{1}(z+1)^{3} (\Omega_{m}+\Omega_{r}z+\Omega_{r})\right)}~,\label{eq2.11}
\end{eqnarray}
\begin{figure}[H]
    \centering
    \begin{subfigure}{0.5\textwidth}
        \centering
        \includegraphics[width=76mm]{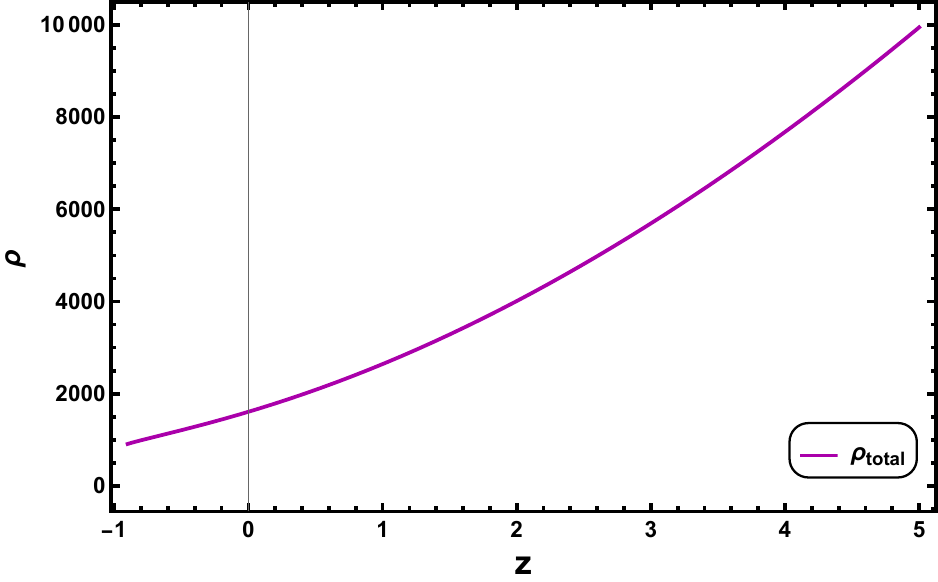}
        \caption{Evolution of energy density in redshift.}
        \label{rho2.3}
    \end{subfigure}%
    \hfill
    \begin{subfigure}{0.5\textwidth}
        \centering
        \includegraphics[width=75mm]{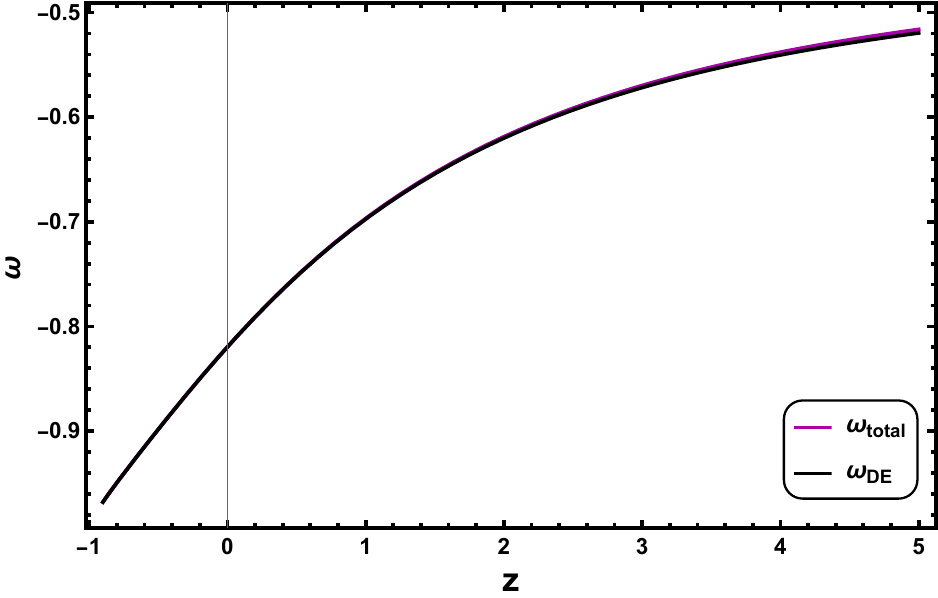}
        \caption{Evolution of EoS parameter in redshift.}
        \label{omega2.3}
    \end{subfigure}%
    \caption{The behavior of the dynamical parameter for log-square-root model in redshift.}
    \label{fig2.3}
\end{figure}
where $H_{z}$ denotes the derivative of $H$ with respect to $z$. The parameters $\xi$, $\eta$, $n$, $Q_{0}$ and $\lambda_{1}$ determine the evolutionary behavior of total energy density and EoS parameters. The model parameter has been chosen in such a way that a positive energy density can be obtained Fig. \ref{rho2.3}. The total energy density is showing decreasing behavior from early epoch to late epoch and remains positive throughout. The total EoS parameter shows quintessence behavior at present epoch, whereas it converges to $\Lambda$CDM at late epoch [Fig. \ref{omega2.3}]. At $z=0$, the value of total EoS parameter observed to be $\approx -0.82$. The behaviour of DE EoS parameter remains almost same with that of total EoS parameter. The parameter scheme for the plots is $\xi = 0.965$, $\eta = 0.60$, $n = 1$ $Q_{0} = 29990$, $\lambda_{1}=0.35$, $\Omega_{m}=0.3$ and $\Omega_{r}=0.00001$.

\subsection{Exponential model}
As a second model, we consider an exponential function of $f(Q)$ \cite{Anagnostopoulos_2021_822_136634} as,
\begin{equation}\label{eq2.12}
f(Q)=Qe^{\frac{\mu\lambda_{2}}{Q}}-Q~,
\end{equation}
where $\lambda_{2}$ is the free parameter. In the cosmological framework, the model gives rise to a scenario without $\Lambda$CDM as a limit, having the same number of free parameters as $\Lambda$CDM. In this model, in certain time period of cosmic history, the term $\frac{\mu}{Q}$ decreases, which ends up making the model as polynomial one.
\begin{figure}[ht]
    \centering
    \begin{subfigure}{0.5\textwidth}
        \centering
        \includegraphics[width=75mm]{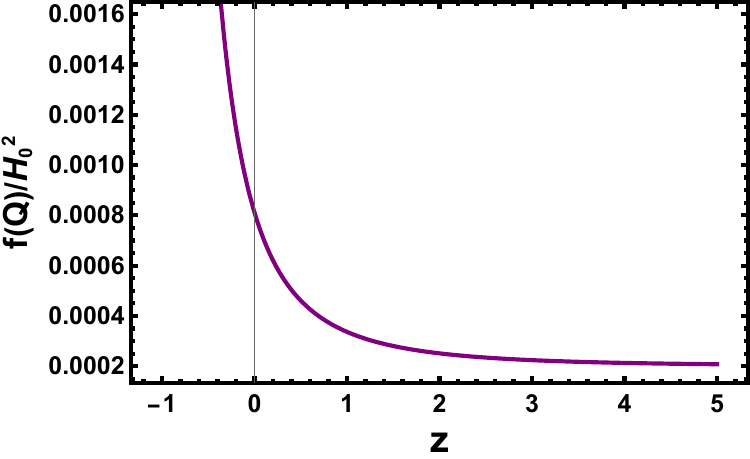}
        \caption{Evolution of $\frac{f(Q)}{H_{0}^2}$ in redshift.}
        \label{sign22.4}
    \end{subfigure}%
    \hfill
    \begin{subfigure}{0.5\textwidth}
        \centering
        \includegraphics[width=75mm]{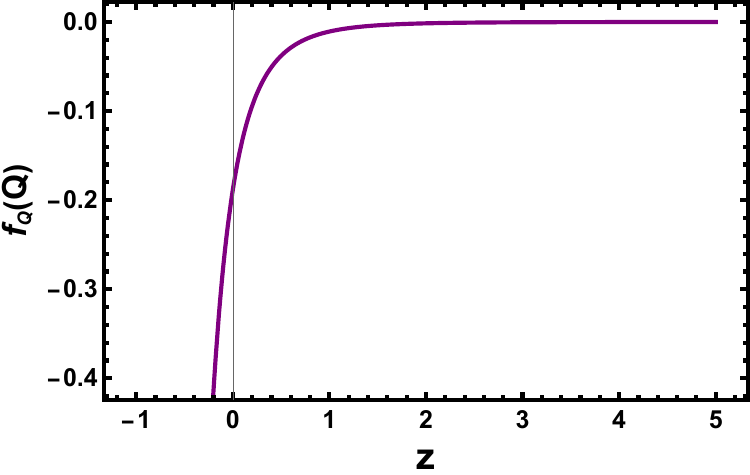}
        \caption{Evolution of $\frac{df(Q)}{dQ}$ in redshift.}
        \label{sign212.4}
    \end{subfigure}%
    \caption{Evolution of exponential model as a function of redshift.}
    \label{fig2.4}
\end{figure}
We can see the behavior of $\frac{f(Q)}{H_{0}^2}$ and $f_Q$ in Fig. \ref{fig2.4}. In this model also, the normalized non-metricity function $\frac{f(Q)}{H_{0}^2}$ increases with cosmic time due to exponential form of model [Fig. \ref{sign22.4}]. However, the functional $f_Q$ becomes a decreasing function of the redshift [Fig. \ref{sign212.4}]. The motivation behind these plots is that the non-metricity scalar $Q$ at the FLRW background assumes $6H^{2}$. To analyse the dynamic behavior of the model, the dynamical parameters can be derived from Eqs. \eqref{eq2.1}, \eqref{eq2.2} and \eqref{eq2.5} by incorporating Eq. \eqref{eq2.12} as,
\begin{align}
 \rho_{total} &= \frac{1}{6}e^{\frac{\lambda_{2}\mu}{6H^2(z)}} \left(\frac{\lambda_{2}\mu}{H^2(z)}+3\right)+3H^2(z)+(z+1)^{3}(\Omega_{m}+\Omega_{r}z+\Omega_{r})~,\label{eq2.13}\\[5pt]
p_{total} &= -3H^2(z)+2H(z)(z+1)H_{z}(z)+\frac{1}{3}\Omega_{r}(z+1)^4~,\nonumber\\
&-\frac{e^{\frac{\lambda_{2}\mu}{6H^2(z)}} \left(9H^3(z)\left(3H^2(z)+\lambda_{2}\mu\right)+\lambda_{2}\mu(z+1)H_{z}(z)\left(9H^2(z)+\lambda_{2}\mu\right)\right)}{54H^5(z)}~,\label{eq2.14}\\[10pt]
\omega_{total} &= \frac{-3H^2(z)+2H(z)(z+1)H_{z}(z)+\frac{1}{3}\Omega_{r}(z+1)^4}{\frac{1}{6}e^{\frac{\lambda_{2}\mu}{6H^2(z)}} \left(\frac{\lambda_{2}\mu}{H^2(z)}+3\right)+3H^2(z)+(z+1)^{3}(\Omega_{m}+\Omega_{r}z+\Omega_{r})} \nonumber\\
&-\frac{e^{\frac{\lambda_{2}\mu}{6H^2(z)}} \left(9H^3(z)\left(3H^2(z)+\lambda_{2}\mu\right)+\lambda_{2}\mu(z+1)H_{z}(z)\left(9H^2(z)+\lambda_{2}\mu\right)\right)}{54H^5(z)\left(\frac{1}{6}e^{\frac{\lambda_{2}\mu}{6H^2(z)}} \left(\frac{\lambda_{2}\mu}{H^2(z)}+3\right)+3H^2(z)+(z+1)^{3}(\Omega_{m}+\Omega_{r}z+\Omega_{r})\right)}.\label{eq2.15}
\end{align}

\begin{figure}[ht]
    \centering
    \begin{subfigure}{0.5\textwidth}
        \centering
        \includegraphics[width=75mm]{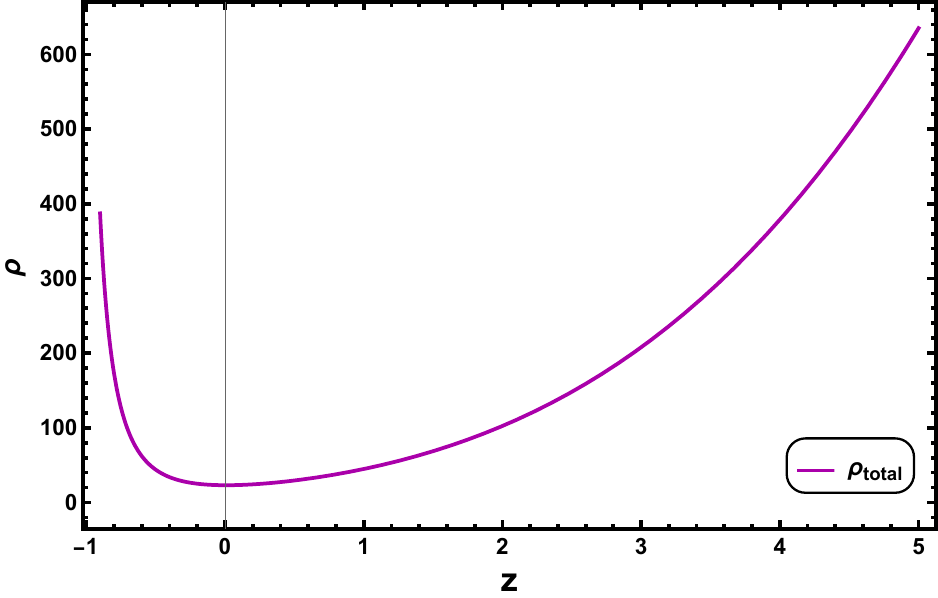}
        \caption{Evolution of energy density in redshift.}
        \label{rho2.5}
    \end{subfigure}%
    \hfill
    \begin{subfigure}{0.5\textwidth}
        \centering
        \includegraphics[width=75mm]{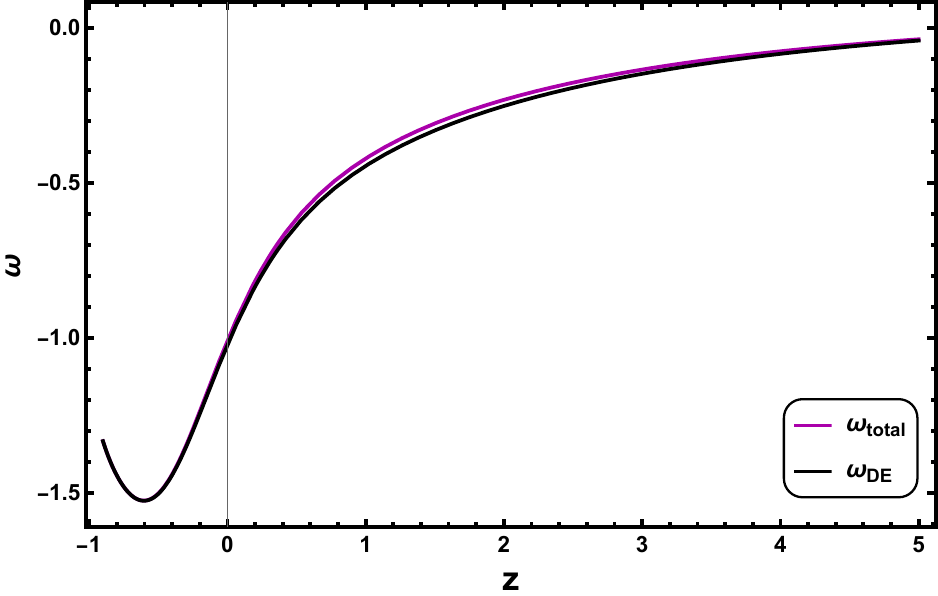}
        \caption{Evolution of EoS parameter in redshift.}
        \label{omega2.5}
    \end{subfigure}%
    \caption{The behavior of the dynamical parameter for exponential model in redshift.}
    \label{fig2.5}
\end{figure}
Fig. \ref{rho2.5} shows the evolutionary behavior of total energy density that remains positive throughout the evolution and at present time it attains the minimum value. The total EoS parameter remains entirely in the negative domain and decreases from early time. The present value has been obtained as, $\omega_{total}=-1.01$ [Fig. \ref{omega2.5}]. At present the DE EoS parameter remains alike to that of total EoS parameter though there was a marginal change noticed at early time. The parameter scheme for the plots is $\xi = 0.965$, $\eta = 0.60$, $\mu = 6.5$, $\lambda_{2} = 6.5$, $\Omega_{m}=0.3$ and $\Omega_{r}=0.00001$.\\
The evolutionary behaviors of the dynamical parameters for both the log-square-root model and exponential model have been established and now we proceed to investigate the phase space analysis of these models to further explore their stability and the potential for cosmological solutions.

\section{Phase space analysis}\label{sec2.4}
\begin{figure}[ht]
    \centering
    \begin{subfigure}{0.5\textwidth}
        \centering
        \includegraphics[scale=0.4]{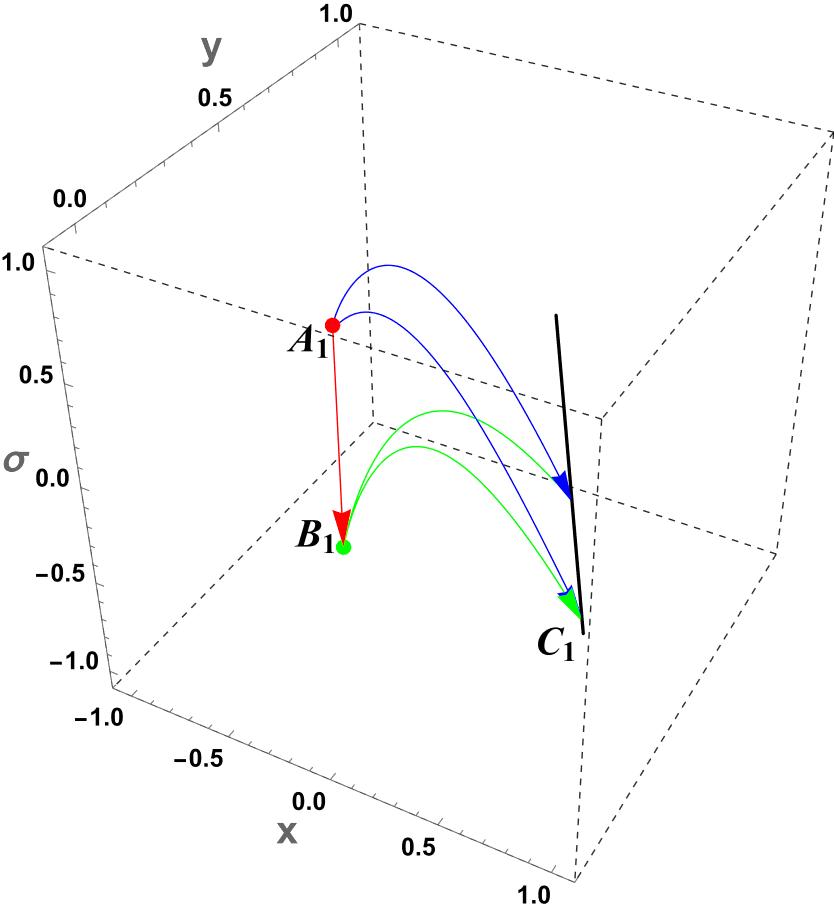}
        \caption{Phase-space trajectories on the $x$-$y$-$\sigma$ plane.}
        \label{3D2.6}
    \end{subfigure}%
    \hfill
    \begin{subfigure}{0.5\textwidth}
        \centering
        \includegraphics[scale=0.6]{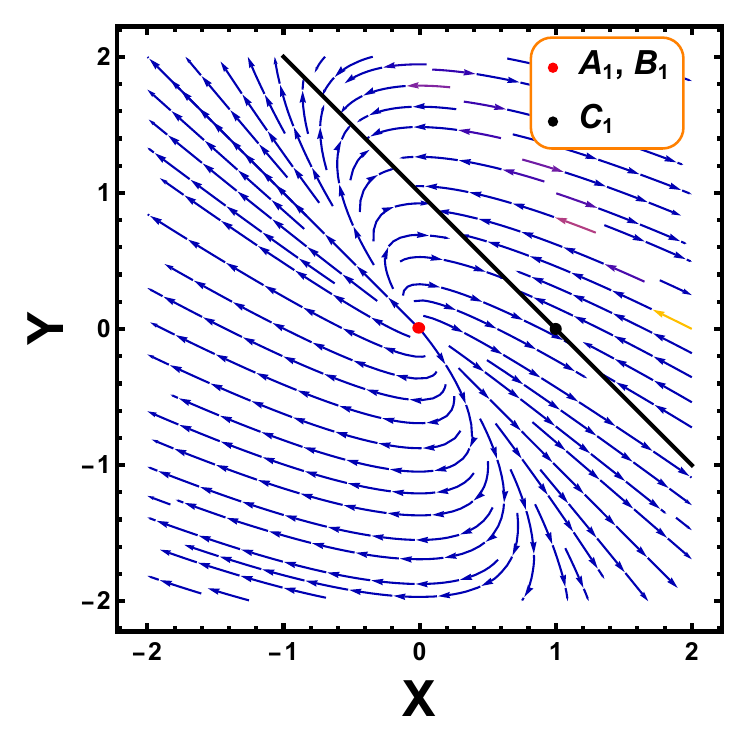}
        \caption{Phase-space trajectories on the $x$-$y$ plane.}
        \label{2D2.6}
    \end{subfigure}%
    \caption{The phase-space portrait for log-square-root model.}
    \label{fig2.6}
\end{figure}

Any group of elements that evolves over time is a dynamical system, whether they are real or even artificial. The dynamical system is based on differential equations associated with time derivatives. So, it is unlike that there exists any universal theory of dynamical systems. The evolution rule that governs the dynamical system should therefore be analyzed in various ways to find its characteristics \cite{Abraham_1992, Katok_1995, Strogatz_2018}. In order to probe the evolutionary dynamics of the theory, we use dynamical systems instead of solving the non-linear differential equations that describes the majority of cosmological models. The stability can be analysed in various methods, some of them are Jacobi stability, Kosambi-Cartan-Chern (KCC) theory or Lyapunov methods. We shall use the Jacobi stability analysis in this problem. We shall perform the dynamical system analysis of the background equations of the two models and will focus on its stability \cite{Bahamonde_2018_775_1, D'Agostino_2018_98_124013}. To do so, we consider 3-dimensionless parameters, $x$, $y$ and $\sigma$, which may give detailed idea about DE and transform the field equations in terms of the dynamical variables as,
\begin{eqnarray*}
x=\frac{f}{6H^2}~,\hspace{1.5cm} y=-2f_{Q}~,\hspace{1.5cm} \sigma=\frac{\rho_r}{3H^2}~.
\end{eqnarray*}
Subsequently, Eq. \eqref{eq2.7} reduces to,
\begin{eqnarray*}
\Omega_r = \sigma~,\quad\quad
\Omega_{DE} = x+y~,\quad\quad
\Omega_m = 1-x-y-\sigma~,\quad\quad
\frac{\dot{H}}{H^2} = \frac{-(3-3x-3y+\sigma)}{2(2Qf_{QQ}+f_Q+1)}~.
\end{eqnarray*}
Here, prime (~$\prime$~) represents differentiation with respect to the number of e-folds of the Universe, $N=ln a$. Then we can differentiate $x$, $y$ and $\sigma$ with respect to $N$ to obtain,
\begin{eqnarray}
x^{\prime} &=& -\frac{\dot{H}}{H^2}(2x+y) ~~~= \frac{(3-3x-3y+\sigma)(2x+y)}{2(2Qf_{QQ}+f_Q+1)},\label{eq2.16} \\[5pt]
y^{\prime} &=&- \frac{\dot{H}}{H^2}(4Qf_{QQ}) ~~= (3-3x-3 y+\sigma)\left[1+\frac{(y-2)}{2(2 Qf_{QQ}+f_Q+1)}\right], \label{eq2.17}\\[5pt]
\sigma^{\prime} &=& -\sigma\left[4+2\frac{\dot{H}}{H^2}\right]  ~= \sigma\left[\frac{(3-3x-3y+\sigma)}{(2Qf_{QQ}+f_Q+1)}-4\right]. \label{eq2.18}
\end{eqnarray}
Now, we can redefine $\omega_{total}$ and $\omega_{DE}$ as
\begin{eqnarray*}
\omega_{total} &=& -1-\frac{2}{3}\frac{\dot{H}}{H^2}~,\quad\quad\quad\quad
\omega_{DE} = -1-\frac{1}{3(x+y)}\left[y^{'}+\frac{\dot{H}}{H^2}y\right].
\end{eqnarray*}
For $f(Q)=nQ_{0}\sqrt{\frac{Q}{\lambda_{1}Q_{0}}}~ln\left(\frac{\lambda_{1}Q_{0}}{Q}\right)$, one can express Eqs. \eqref{eq2.16}-\eqref{eq2.18} in terms of dynamical variables as,
\begin{eqnarray}
 x^{\prime}& =& \frac{(2x+y)(3-3x-3y+\sigma)}{(2-x-y)},\label{eq2.19}\\[5pt]
 y^{\prime} &=& \frac{x(3x+3y-3-\sigma)}{(2-x-y)}, \label{eq2.20}\\[5pt]
 \sigma^{\prime} &=& \frac{2\sigma(\sigma-x-y-1)}{(2-x-y)}.\label{eq2.21}
\end{eqnarray}
Moreover, the total EoS parameter and EoS parameter for DE can be written in dynamical variables as,
\begin{eqnarray*}
\omega_{total} &=& -1-\frac{2}{3}\left( \frac{3x+3y-\sigma-3}{2-x-y}\right),\quad\quad\quad\quad
\omega_{DE} = -\frac{3-\sigma}{3(2-x-y)}~.
\end{eqnarray*}
\begin{table}[ht]
\renewcommand\arraystretch{1.5}
    \centering
    \addtolength{\tabcolsep}{-6pt}
    {\small
    \begin{tabular}{|c|c|c|c|c|c|c|c|c|c|}
        \hline 
 ~~Name~~ & ~~Point/Curve~~ & ~~$\Omega_{m}~~$ & ~~$\Omega_{r}$~~ & ~~$\Omega_{DE}$~~ & ~~$q$~~ & ~~$\omega_{total}$~~ & ~~$\omega_{DE}$~~& ~~Phase of Universe~~ & ~~Stability~~ \\ [0.2cm]
\hline \hline
$A_{1}$ & (0, 0, 1) & 0 & 1 & 0 & 1 & $\frac{1}{3}$ & $-\frac{1}{3}$ & \begin{tabular}{@{}c@{}}Radiation \\dominated\end{tabular} & Unstable Node\\
\hline
$B_{1}$ & (0, 0, 0) & 1 & 0 & 0 & $\frac{1}{2}$ & 0 & $-\frac{1}{2}$ & \begin{tabular}{@{}c@{}}Matter \\dominated\end{tabular} &  ~~Unstable Saddle~~\\
\hline
$C_{1}$ & ($x$, $1-x$, 0) & 0 & 0 & 1 & -1 & -1 & -1 & \begin{tabular}{@{}c@{}}Dark Energy \\ dominated\end{tabular} & Stable Node\\
\hline
\end{tabular}
\caption{Critical Points and the corresponding cosmology for log-square-root model.}
\label{table2.1}
}
\end{table}
Table \ref{table2.1} provides the critical points and the cosmological behaviour at these points. The details description of each critical point has been narrated below. In Fig. \ref{fig2.6}, the $2D$ and $3D$ phase portrait have been given to understand the stability of these points.
\begin{itemize}
 \item \textbf{Critical point $A_1$ (0, 0, 1) :} The
    corresponding EoS parameter and deceleration parameter is $\omega_{total}=\frac{1}{3}$ and $q=1$ respectively. This behaviour of the critical point leads to the decelerating phase
    of the Universe. The density parameters are, $\Omega_{DE}=0$, $\Omega_{m}=0$ and $\Omega_{r}=1$. This critical point is an unstable node because it contains all positive eigenvalues of the Jacobian matrix,
    \begin{eqnarray*}
    \{2,~2,~1\}.
    \end{eqnarray*}
    
    \item \textbf{Critical point $B_1$ ($0$, $0$, $0$) :} The critical point leads to the decelerating phase of the Universe, since the EoS parameter corresponding to this critical point is $\omega_{total}=0$ and deceleration parameter is $q=\frac{1}{2}$. The corresponding density parameter are $\Omega_{DE}=1$, $\Omega_{m}=0$ and $\Omega_{r}=0$. The critical point shows unstable saddle behaviour. The eigenvalues for corresponding critical point shows positive and negative signature as shown below:
    \begin{equation*}
    \left\{\frac{3}{2},~~\frac{3}{2},~-1\right\}. 
    \end{equation*}

    \item \textbf{Curve of critical points $C_1$ ($x$, $1-x$, $0$) :} At this point, $\Omega_{DE}=1$, $\Omega_{m}=0$ and $\Omega_{r}=0$, i.e. the Universe shows DE dominated phase. The accelerated DE dominated Universe is confirmed by the corresponding value of the EoS parameter ($\omega_{total}=-1$) and value of the deceleration parameter ($q=-1$). Jacobian matrices with critical points have negative real parts and zero eigenvalues. Further, there is only one vanishing eigenvalue and therefore the dimension of the set of eigenvalues equals its number. As a result, the set of eigenvalues is normally hyperbolic, the critical point associated with it cannot be a global attractor \cite{Aulbach_1984, Coley_1999_gr-qc/9910074}. This critical point, shows stable node behaviour. The corresponding eigenvalues are given below:
    \begin{equation*}
    \{-4,~-3,~~0\}.
    \end{equation*}
\end{itemize}

\begin{figure}[ht]
\centering
\includegraphics[width=8.5cm]{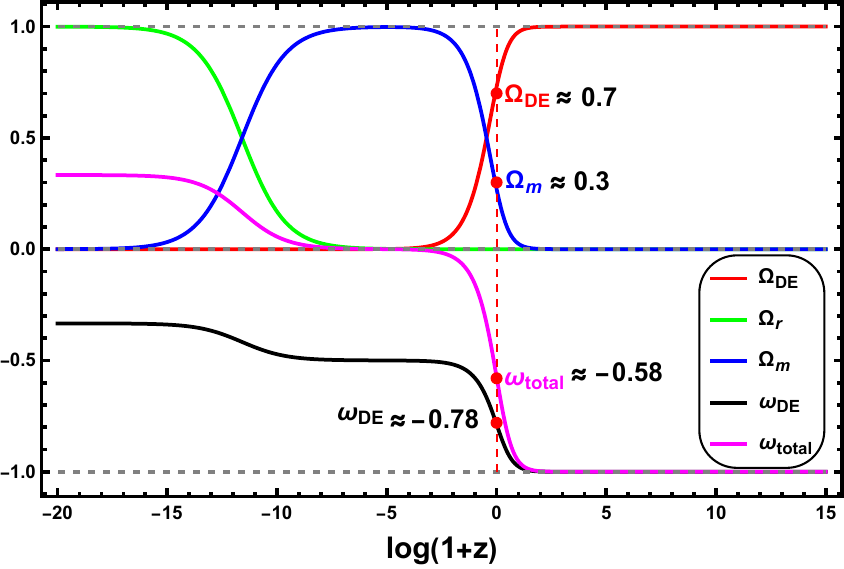}
\caption{Evolution of EoS Parameter and density parameters for log-square-root model. The vertical dashed red line denotes the present time.}\label{fig2.7}
\end{figure}

From the phase portrait [Fig. \ref{fig2.6}], we can see that the critical point $A_{1}$ is unstable node, where as curve \big[($x$, $1-x$, 0)\big] is stable node. The $B_{1}$ is unstable saddle point. Fig. \ref{3D2.6} describes the trajectories for critical points, where $A_{1}$ is the repeller, so it repels every trajectory and $C_{1}$ is the attractor, so it absorbs every trajectory coming towards it. The $B_{1}$ is saddle therefore, it absorbs the trajectories coming from $A_{1}$ and repel the trajectories originated from itself. Fig. \ref{2D2.6}, one can observe that the stability is not only specific to the single point but also in the entire curve ($x$, $1-x$, $0$). This kind of stability behaviour may be due to the fact that irrespective of the value of the dynamical variable $x$, it exhibits the stable behaviour. It may be due to the nature of the dynamical variable $x$. The $3D$ portrait shows the trajectory behaviour of the model starting from repeller point $A_{1}$ to the saddle point $B_{1}$ and then it is moving from $B_{1}$ to the stable curve $C_{1}$. Further, the evolution plot for log-square-root model has been given in Fig. \ref{fig2.7} utilizing the initial conditions $x=10^{-15}$, $y=10^{-6}$ and $\sigma=10^{-1}$. From the evolution curve, the present value of DE EoS parameter is $-0.80$ whereas the value obtained using the hybrid scale factor is $-0.82$. Hence in both the approaches, we can get similar value at present time and the Universe shows quintessence behaviour. At present the value of density parameters for DE and matter obtained as $\approx 0.7$ and $\approx 0.3$ respectively.

For, $f(Q)=Qe^{\frac{\mu\lambda_{2}}{Q}}-Q$, we can get the system of differential equations as,
\begin{eqnarray}
x^{\prime} &=& \frac{(3-3x-3y+\sigma)(2x+y)}{2-y+4(x+1)[\ln(x+1)]^2}, \label{eq2.22}\\[5pt]
y^{\prime} &=& \frac{4(x+1)(3-3x-3y+\sigma)[\ln(x+1)]^2}{2-y+4(x+1)[\ln(x+1)]^2}, \label{eq2.23}\\[5pt]
\sigma^{\prime} &=& 2\sigma\left[\frac{3-3x-3y+\sigma}{2-y+4(x+1)[\ln(x+1)]^2}-2\right]. \label{eq2.24}
\end{eqnarray}
On a similar note, we can obtain the EoS parameters as,

\begin{eqnarray*}
\omega_{total} &=& -1+ \frac{6(x+y-1)-2\sigma}{3\left(2-y+4(x+1)[\ln(x+1)]^2\right)}~,\\[10pt]
\omega_{DE} &=& \frac{-4(\sigma+3)[\ln(x+1)]^2-2 x\big(6 [\ln(x+1)]^2(x+1)+2\sigma[\ln(x+1)]^2+3\big)+y(\sigma-3)}{3(x+y)\left(2-y+4(x+1)[\ln(x+1)]^2\right)}~.
\end{eqnarray*}
\begin{table}[ht]
\renewcommand\arraystretch{1.5}
    \centering
    \addtolength{\tabcolsep}{-6pt}
    {\small
    \begin{tabular}{|c|c|c|c|c|c|c|c|c|c|}
        \hline 
 ~~Name~~ & ~~Point/Curve~~ & ~~$\Omega_{m}~~$ & ~~$\Omega_{r}$~~ & ~~$\Omega_{DE}$~~ & ~~$q$~~ & ~~$\omega_{total}$~~ & ~~$\omega_{DE}$~~& ~~ Phase of Universe~~&~~Stability~~ \\ [0.2cm]
\hline \hline
$A_{2}$ & (0, 0, 1) & 0 & 1 & 0 & 1 & $\frac{1}{3}$ & - & \begin{tabular}{@{}c@{}}Radiation \\dominated\end{tabular} & Unstable Node\\
\hline
$B_{2}$ & (0, 0, 0) & 1 & 0 & 0 & $\frac{1}{2}$ & 0 & - & \begin{tabular}{@{}c@{}}Matter \\dominated\end{tabular} & ~~Unstable Saddle~~\\
\hline
$C_{2}$ & ($x$, $1-x$, 0) & 0 & 0 & 1 & -1 & -1 & -1 & \begin{tabular}{@{}c@{}}Dark Energy \\ dominated\end{tabular} & Stable Node\\
\hline
    \end{tabular}
    \caption{Critical Points and the corresponding cosmology for exponential model.}
    \label{table2.2}
    }
\end{table}

\begin{figure}[ht]
    \centering
    \begin{subfigure}{0.5\textwidth}
        \centering
        \includegraphics[scale=0.4]{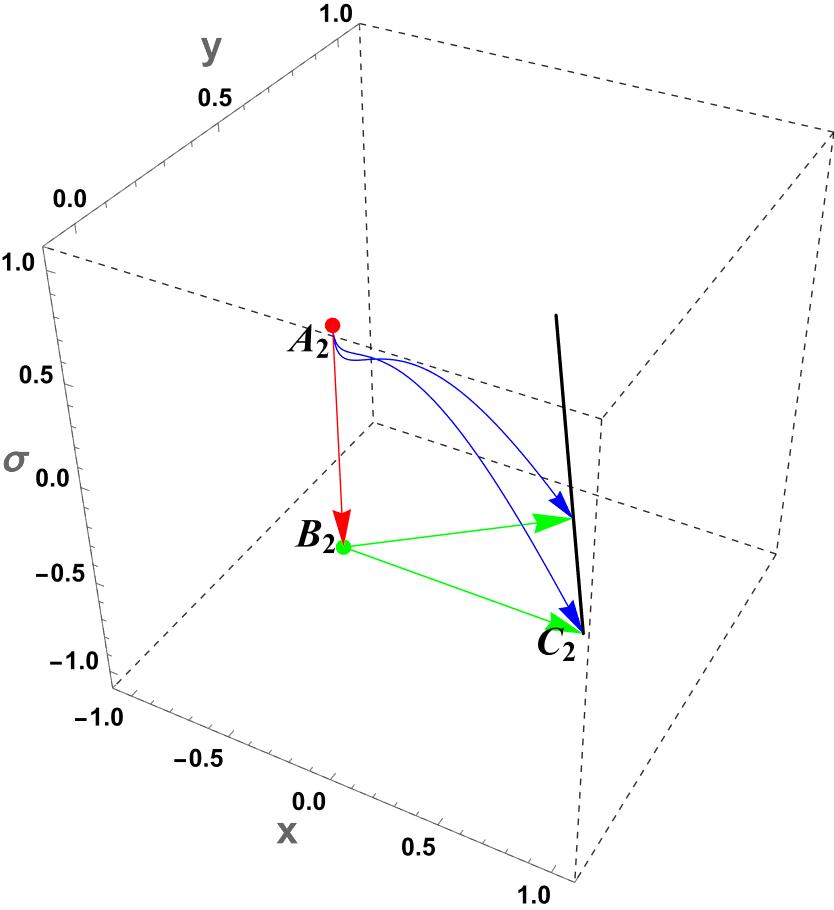}
        \caption{Phase-space trajectories on the $x$-$y$-$\sigma$ plane.}
        \label{3D2.8}
    \end{subfigure}%
    \hfill
    \begin{subfigure}{0.5\textwidth}
        \centering
        \includegraphics[scale=0.6]{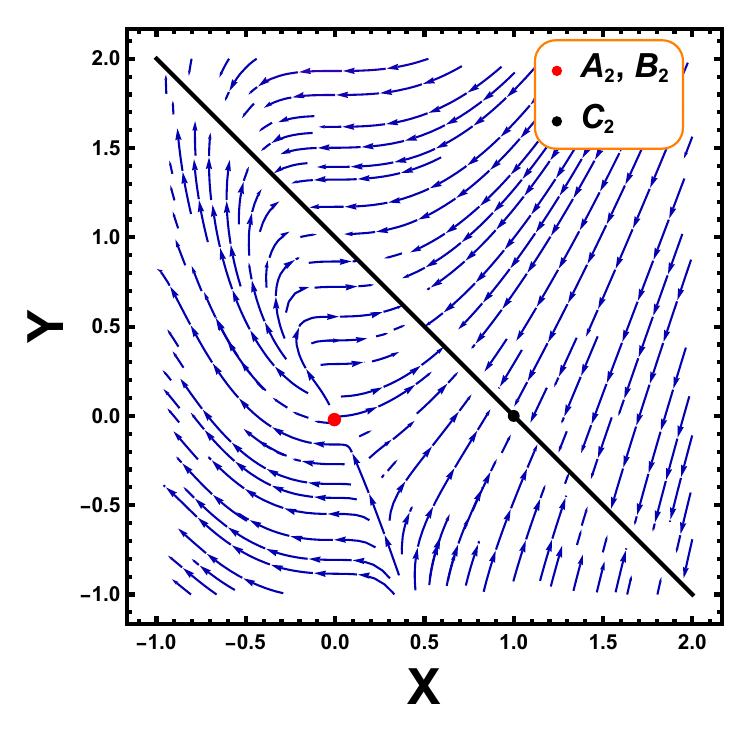}
        \caption{Phase-space trajectories on the $x$-$y$ plane.}
        \label{2D2.8}
    \end{subfigure}%
    \caption{The phase-space portrait for exponential model.}
    \label{fig2.8}
\end{figure}

Table \ref{table2.2} provides the critical points and the cosmological behaviour at these points. The details description of each critical point has been narrated below. In Fig. \ref{fig2.8}, the $2D$ and $3D$ phase portrait have been given to understand the stability of these points. 
\begin{itemize}
\item  \textbf{Critical point $A_2$ ($0$, $0$, $1$) :} The critical point leads to the decelerating phase of the Universe, since the EoS parameter and deceleration parameter corresponding to this critical point is $\omega_{total}=\frac{1}{3}$ and $q=1$ respectively. The corresponding density parameter are $\Omega_{DE}=0$, $\Omega_{m}=0$ and $\Omega_{r}=1$. This critical point shows unstable behaviour. The eigenvalues for the corresponding critical point shows positive signature as given below:  
\begin{equation*}
\left\{4,~1,~0\right\}. 
\end{equation*}

\item \textbf{Critical point $B_2$ ($0$, $0$, $0$) :} At this point, $\Omega_{DE}=0$, $\Omega_{m}=1$ and $\Omega_{r}=0$, i.e. the Universe shows matter dominated phase. The decelerated matter dominated Universe is confirmed by the corresponding value of the EoS parameter ($\omega_{total}=0$) and deceleration parameter $q=\frac{1}{2}$. Jacobian matrices with critical points have positive, negative real parts and zero eigenvalues. This critical point shows unstable saddle behaviour. The corresponding eigenvalues are given below:
    \begin{equation*}
    \{3,~-1,~~0\}.
    \end{equation*}

    \item \textbf{Curve of critical point $C_2$ ($x$, $1-x$, $0$) :} The corresponding EoS parameter is $\omega_{total}=-1$ and deceleration parameter is $q=-1$. This behaviour of the critical point leads to the accelerating phase of the Universe. Also, density parameters are $\Omega_{DE}=1$, $\Omega_{m}=0$ and $\Omega_{r}=0$. This critical point is a stable node because it contains negative real part and zero eigenvalues of the Jacobian matrix.
    \begin{eqnarray*}
    \{0,~-4,~-3\}.
    \end{eqnarray*}
\end{itemize}
\begin{figure}[ht]
\centering
\includegraphics[width=8.5cm]{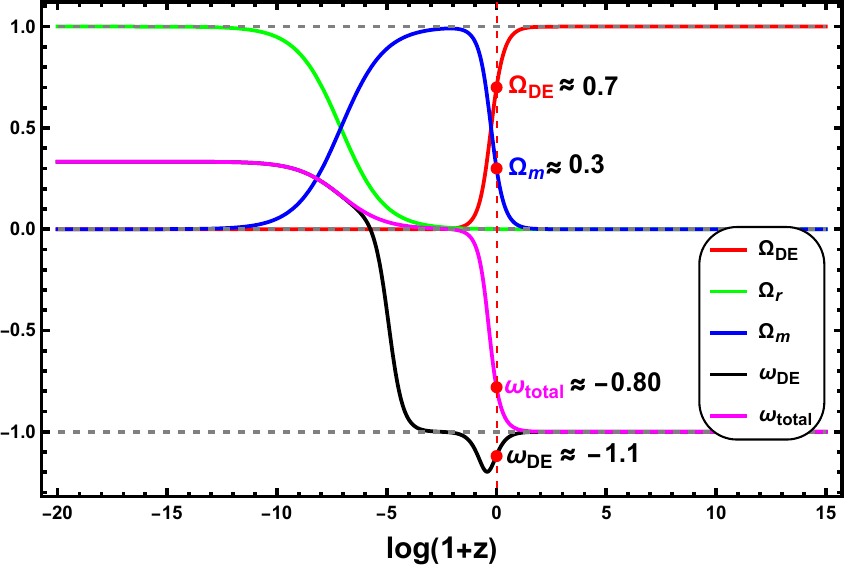}
\caption{Evolution of EoS parameter and density parameters for exponential model. The vertical dashed red line denotes the present time.}
\label{fig2.9}
\end{figure}
The phase portrait, which shows comparable trajectory plots, is an important tool in the study of dynamical systems. The stability of the models can be tested using the phase portrait. The phase portrait for system given in Eqs. \eqref{eq2.22}-\eqref{eq2.24} is shown in Fig. \ref{fig2.8}. The figure \ref{3D2.8} shows the trajectories in $x$-$y$-$\sigma$ ($3D$) plane and figure \ref{2D2.8} shows the trajectories in $x$-$y$ ($2D$) plane, since $\sigma=0$ is an invariant sub-manifold. Here, also the stability obtained along the curve $(x, 1-x, 0)$. This again shows the role of the dynamical variable $x$. Similar to log-square-root model, the $3D$ phase portrait for exponential model shows attracting behaviour at critical curve $C_{2}$. All the repelling trajectories going away from critical points $A_{2}$ and $B_{2}$ and moving towards the stable curve $C_{2}$. From the evolution plot [Fig. \ref{fig2.9}], the present value of $\omega_{DE}=-1.1$ which is almost same as that of the value obtained through the scale factor. The initial conditions are taken as $x=10^{-15}$, $y=10^{-6}$ and $\sigma=10^{-1}$.

\section{Results and Conclusion}\label{sec2.5}
In this chapter, we explored cosmological models within the framework of symmetric teleparallel gravity or $f(Q)$ gravity, by introducing two distinct models: a log-square-root model and an exponential model. Both models were examined using a time varying deceleration parameter to capture the transition of the Universe from early deceleration to late-time acceleration. A log-square-root model demonstrates a quintessence behavior at present, converging to $\Lambda$CDM at late-time, while exponential model shows a phantom behavior.\\
The dynamical system analysis revealed that both models exhibit unstable behavior during the radiation and matter phases but attain stability in the de Sitter phase, consistent with the desired features of an accelerating Universe. The phase portraits and critical points underscore the ability of the model to transition through different cosmological eras, with stable nodes indicating a late-time de Sitter phase. These findings support the potential of non-metricity gravity to produce stable cosmological models that align with the observed late-time cosmic acceleration, offering insights into the evolutionary history of the Universe. The study highlights the importance of the EoS parameter and the role of model parameters in constraining the behavior of dynamical parameters, confirming the stability of the models through comprehensive analysis.
\chapter{Phantom cosmological model with observational constraints in $f(Q)$ gravity} 

\label{Chapter3} 

\lhead{Chapter 3. \emph{Phantom cosmological model with observational constraints in $f(Q)$ gravity}} 

\vspace{10 cm}
* The work, in this chapter, is covered by the following publication: \\

\textbf{S. A. Narawade} and B. Mishra, ``Phantom cosmological model with observational constraints in $f(Q)$ gravity", \textit{Ann. der Phys.} \textbf{535} (2023) 2200626.
 
\clearpage
       
\section{Introduction}
In recent years, significant advancements have been made in cosmological models through the integration of observational data. Seikel et al. \cite{Seikel_2012_86_083001} introduced new consistency tests for the $\Lambda$CDM model, formulated in terms of the Hubble parameter $H(z)$. The acceleration of the expansion of the Universe has been studied by parametrize the EoS parameter of DE with analyses based on four distinct sets of Type Ia supernovae data \cite{Magana_2014_2014_017}. Wei et al. \cite{Wei_2007_654_139} tested ten cosmological models against the Hubble data set to validate their predictions. Farooq et al. \cite{Farooq_2017_835_26} constrained both spatially flat and curved dark energy models by utilizing $H(z)$ data from the redshift range $(0.07, 2.36)$. Mukherjee et al. \cite{Mukherjee_2016_93_043002} discussed a parametric reconstruction of the jerk parameter, which represents the third order time derivative of the scale factor in a dimensionless form. Additionally, for a bulk viscous anisotropic Universe, the Hubble parameter values obtained using various datasets Hubble, Pantheon and Hubble+Pantheon are $H_{0}=69.39\pm 1.54~kms^{-1}Mpc^{-1}$, $70.016\pm 1.65~kms^{-1}Mpc^{-1}$ and $69.36\pm 1.42~kms^{-1}Mpc^{-1}$ respectively \cite{Goswami_2021_69_2100007}.

Based on the analysis of dynamical stability in previous chapters, this chapter carried out an observational analysis to validate and assess the compatibility of theoretical models with observational data in $f(Q)$ gravity. To do that, we have obtained the Hubble parameter in redshift with some algebraic manipulation from the considered form of $f(Q)$. By performing statistical analysis using the MCMC method, we constrained the free parameters of the $f(Q)$ model, ensuring that it accurately reflects the observed behavior of the Universe. The incorporation of the BAO dataset in this analysis compare the constrained values of free parameter with cosmological observations. Additionally, the study of cosmographic parameters and the $Om(z)$ diagnostic provides further insight into the dynamics of cosmic acceleration, helping to evaluate the performance of the model and its potential to describe key observational features like cosmic expansion and dark energy. This chapter divided into seven sections. In section \ref{sec3.2}, we parametrized the Hubble parameter from the well motivated power-law $f(Q)$ model. The free model parameters are constrained through the MCMC technique and the observational datasets in section \ref{sec3.3}. The section \ref{sec3.4} validated the Hubble parameter and the constrained model parameters utilizing the BAO dataset. In section \ref{sec3.5} the cosmographic parameters introduced, while section \ref{sec3.6} discusses the $Om(z)$ diagnostic and the age of the Universe. Finally, the results and conclusions are presented in section \ref{sec3.7}.

\section{Power-law model}\label{sec3.2}
To obtain the cosmological parameters, a well-defined form of $f(Q)$ is needed, so that the analysis of the cosmological model can be performed. Capozziello et al. \cite{Capozziello_2022_832_137229} have assumed the Pade's approximation approach to compute a numerical reconstruction of the cosmological observables up to high redshifts. This can reduce the convergence issues associated with standard cosmographic methods and provides an effective method for describing cosmological observables up to high redshifts. To reconstruct $f(Q)$ through a numerical inversion procedure, utilizing the relation $Q = 6H^{2}$, the function that provides the best analytical match to the numerical results is given by
\begin{equation}\label{eq3.1}
f(Q) = \alpha + \beta Q^{n}~, 
\end{equation}
where $\alpha$, $\beta$ and $n$ are free model parameters, the parameter $n > 1$ is a real number responsible for the accelerating phase in the early Universe \cite{Capozziello_2022_37_101113}. There are several $f(Q)$ models which shares the same background evolution as in $\Lambda$CDM, while leaving precise and measurable effects on cosmological observable. But with the increase in redshift, the function $f(Q) = \alpha + \beta Q^{n}$ suggests small deviations from the $\Lambda$CDM model. For $\alpha = 0$, $\beta = 1$ and $n = 1$, one can recover the aforementioned class of theories with the same background evolution as in GR.

\subsection{$H(z)$ parameterization}
The model parameters present in Eq. \eqref{eq3.1} will regulate the dynamical behaviour of the model. The value of the model parameters $\alpha$, $\beta$ and $n$ to be chosen in such a way that the deceleration parameter attain the value, $q_{0} = -0.54$ \cite{Almada_2020_101_063516, Garza_2019_79_890,Akarsu_2019_79_846}. From the relationship between scale factor and redshift, $a(t)=\frac{1}{1+z}$, one can get, $H=\frac{-\dot{z}}{1+z}$. Now, from Eqs. \eqref{fqr} and \eqref{fqp}, we obtain
\begin{equation}\label{eq3.2}
\dot{H}= \frac{-(1+\omega)}{4}\frac{2Qf'-f}{2Qf''+f'}~.
\end{equation}
Subsequently,
\begin{equation}\label{eq3.3}
\frac{dH}{dz} = \frac{(1+\omega)}{4H(1+z)}\frac{2Qf'-f}{2Qf''+f'}~,
\end{equation}
Substituting Eq. \eqref{eq3.1} in Eq. \eqref{eq3.3}, we get
\begin{equation}\label{eq3.4}
H(z) = H_{0}\left[ \sqrt{\frac{\alpha + (1+z)^{3(1+\omega)}}{2\beta n 6^{n}-\beta 6^{n}}} \right]^{\frac{1}{n}},
\end{equation}
where $H_{0}$ is the present value of the Hubble parameter. Now, our aim is to put constraint on the parameters $\alpha$, $\beta$, $n$ and $\omega$ using the cosmological datasets. For dark energy dominated phase ($\omega$ is constant) and matter dominated phase ($\omega=0$), we obtain the following relation for the Hubble parameter, which is analogous to epsilon model given in \cite{Lemos_2018_483_4803}, 
\begin{equation}\label{eq3.5}
H(z)^{2} = H_{0}^{2}\left[ \left(1-\frac{\alpha + 1}{\beta (2n-1) 6^{n}}\right)(1+z)^{3} +  \frac{1}{\beta (2n-1) 6^{n}}(1+z)^{3(1+\omega)} + \frac{\alpha}{\beta (2n-1) 6^{n}}  \right]^{\frac{1}{n}}.
\end{equation}
The parametrization of $H(z)$ derived above will be used to place constraints on the model parameters through various observational datasets, which are discussed in the next section. These datasets will provide the required observational information to test the validity of the model.

\section{Observational datasets}\label{sec3.3}
\subsection{Hubble dataset}\label{3.3.1}
By estimating their differential evolution, early type galaxies offer measurements of the Hubble parameter. The process of compilation these observations is referred to as the cosmic chronometers method. The list of $32$ data points of Hubble parameter in the redshift range $0.07 \leq z \leq 1.965$ with errors (see \hyperref[Appendices]{Appendices}). By minimizing the chi-square value, we determine the mean values of the model parameters $\alpha$, $\beta$ and $n$. The Chi-square function can be given as, 
\begin{equation}\label{eq3.6}
\chi_{\mathrm{OHD}}^{2}(p_{s}) = \sum_{i=0}^{32}\frac{\big[H_{th}(z_{i}, p_{s}) - H_{obs}(z_{i})\big]^{2}}{\sigma_{H}^{2}(z_{i})}~,
\end{equation}
where $H_{th}(z_{i}, p_{s})$ represents the Hubble parameter with the model parameters, $H_{obs}(z_{i})$ represents the observed Hubble parameter values and $\sigma_{H}^{2}(z_{i})$ is the standard deviation.

\subsection{Pantheon+SH0ES dataset}\label{3.3.2}
The Pantheon+SH0ES sample dataset consists of 1701 light curves of 1550 distinct Type Ia Supernovae ranging in redshift from $z = 0.00122$ to $2.2613$ \cite{Brout_2022_938_110}. The model parameters are to be fitted by comparing the observed and theoretical value of the distance moduli. The distance moduli with nuisance parameter $\mu_{0}$ can be defined as,
\begin{equation}\label{eq3.7}
\mu(z, \theta) = 5log_{10}\big[d_{L}(z,\theta)\big] + \mu_{0}~,
\end{equation}
where $d_{L}$ is the dimensionless luminosity distance defined as,
\begin{equation}\label{eq3.8}
d_{L}(z) = (1+z)\int_{0}^{z}\frac{dz^*}{E(z^*)}~,\quad\quad\quad\quad\quad \left( E(z) = \frac{H(z)}{H_{0}}\right)
\end{equation}
where $z^{*}$ is variable change to define integration from $0$ to $z$. The $\chi^{2}$ is given by,
\begin{equation}\label{eq3.9}
\chi^{2}_{SN}(z,\theta) = \sum_{i=1}^{1701}\frac{\big[\mu(z_{i},\theta)_{th} - \mu(z_{i})_{obs}\big]^{2}}{\sigma^{2}_{\mu}(z_{i})}~,
 \end{equation}
 where $\sigma^{2}_{\mu}(z_{i})$ is the standard error in the observed value. In order to calculate $\chi^{2}_{SN}$, we use the fact that the SNe Ia dataset corresponds to redshifts below $2$ so that we can neglect the contribution from radiation in Einstein's equation.
 
 \subsection{MCMC analysis and Results}\label{sec3.3.3}
To obtain tight constraints on the parameters of the $f(Q)$ model, the MCMC analysis will be used to perform the test. This analysis will produce proficient fits of $\alpha$, $\beta$, $\omega$ and $n$ upon minimization of a total $\chi^{2}$. Also, this analysis will produce selection criteria, which will allow us to draw some conclusions. The panels on the diagonal in corner MCMC plot shows the $1D$ curve for each model parameter obtained by marginalizing over the other parameters, with a thick line curve to indicate the best fit value. The off diagonal panels show $2D$ projections of the posterior probability distributions for each pair of parameters, with contours to indicate $1\sigma$ (Blue) and $2\sigma$ (Light Blue) regions. 
\begin{figure}[H]
    \centering
    \begin{subfigure}{0.5\textwidth}
        \centering
        \includegraphics[width=75mm]{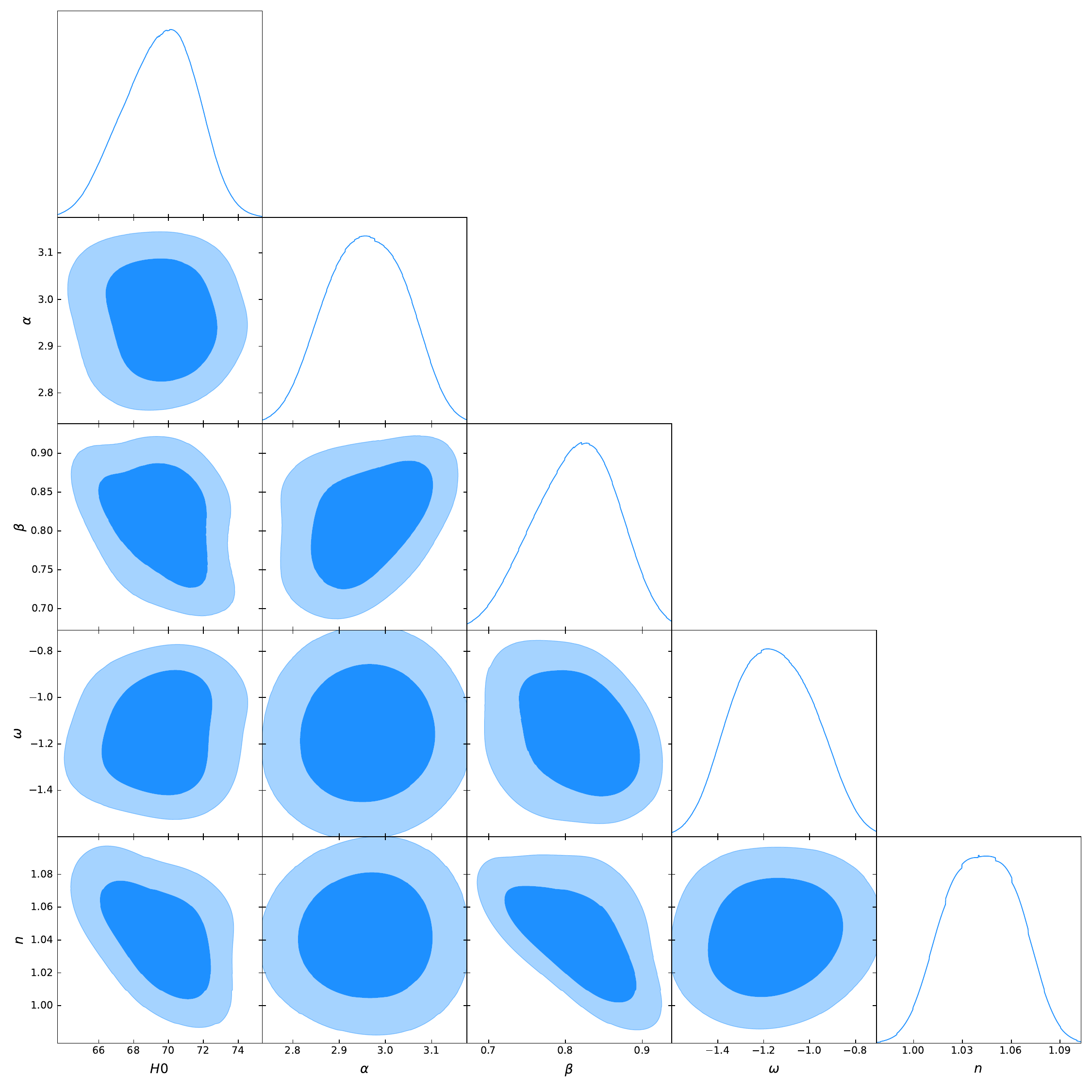}
        \caption{Contour plot obtained from $32$ data points of OHD dataset.}
        \label{H3.1}
    \end{subfigure}%
    \hfill
    \begin{subfigure}{0.5\textwidth}
        \centering
        \includegraphics[width=75mm]{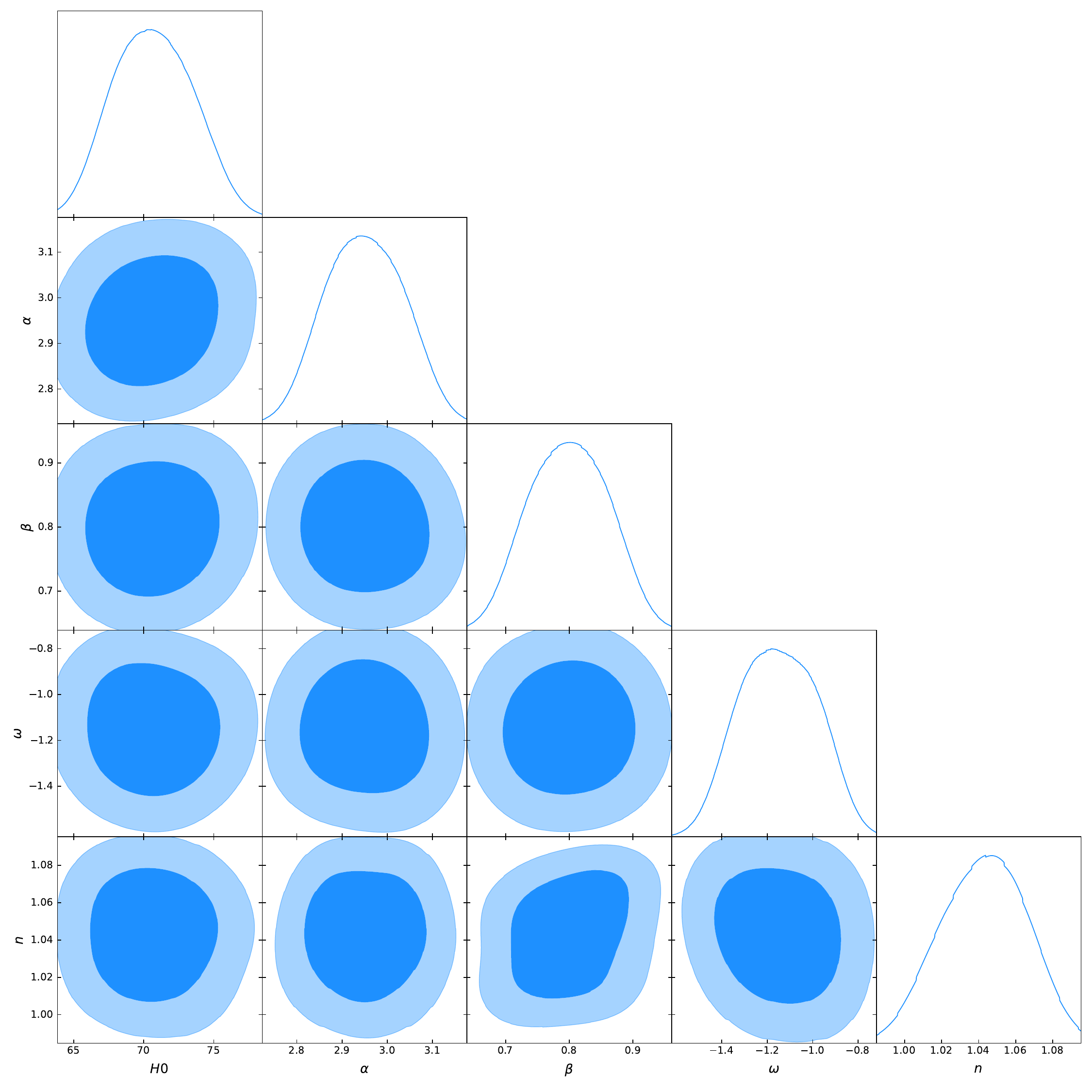}
        \caption{Contour plot obtained from Pantheon+SHOES dataset.}
        \label{P3.1}
    \end{subfigure}%
    \caption{MCMC contour plot obtained from the observational datasets for 1$\sigma$ and 2$\sigma$ confidence interval.}
    \label{fig3.1}
\end{figure}
The best fit values of $\alpha$, $\beta$, $\omega$ and $n$ are obtained from the triangle plot Fig. \ref{H3.1} through Hubble data and in triangle plot Fig. \ref{P3.1} using Pantheon+SHOES data with $1\sigma$ and $2\sigma$ confidence intervals. All the obtained values are listed in Table \ref{table3.1}. In Fig. \ref{EP3.2}, the curve for distance modulo has been given as the distance modulo can also expressed in Hubble parameter Eq. \eqref{eq3.5}. Here, the solid red line passed in the middle through the error bar plots [Fig. \ref{EP3.2}], where we used best fit values from Table \ref{table3.1} to plot the error bar plot. Also, in Fig. \ref{EH3.2}, we have shown the error bar plots of $H(z)$ and $H(z)/(1 + z)$ [Fig. \ref{EHZ3.2}] using the best fit values obtained in Table \ref{table3.1}. The dotted line represents the $\Lambda$CDM line and the solid red line represents the best fit curve for Hubble rate. It can be observed that in both the figures, the solid red line is traversing at the middle of the error bars. We have marginalized value of $H_{0}$ as $69.5^{+2.3}_{-1.9} kms^{-1}Mpc^{-1}$ and $70.7\pm2.7 kms^{-1}Mpc^{-1}$ respectively with Hubble data and Pantheon+SH0ES datasets. The detailed values of the parameters are given in Table \ref{table3.1}. For the subsequent study, we follow $H_0=70.7\pm2.7 kms^{-1}Mpc^{-1}$.
\begin{figure}[H]
    \centering
    \begin{subfigure}{\textwidth}
        \centering
        \includegraphics[width=106mm]{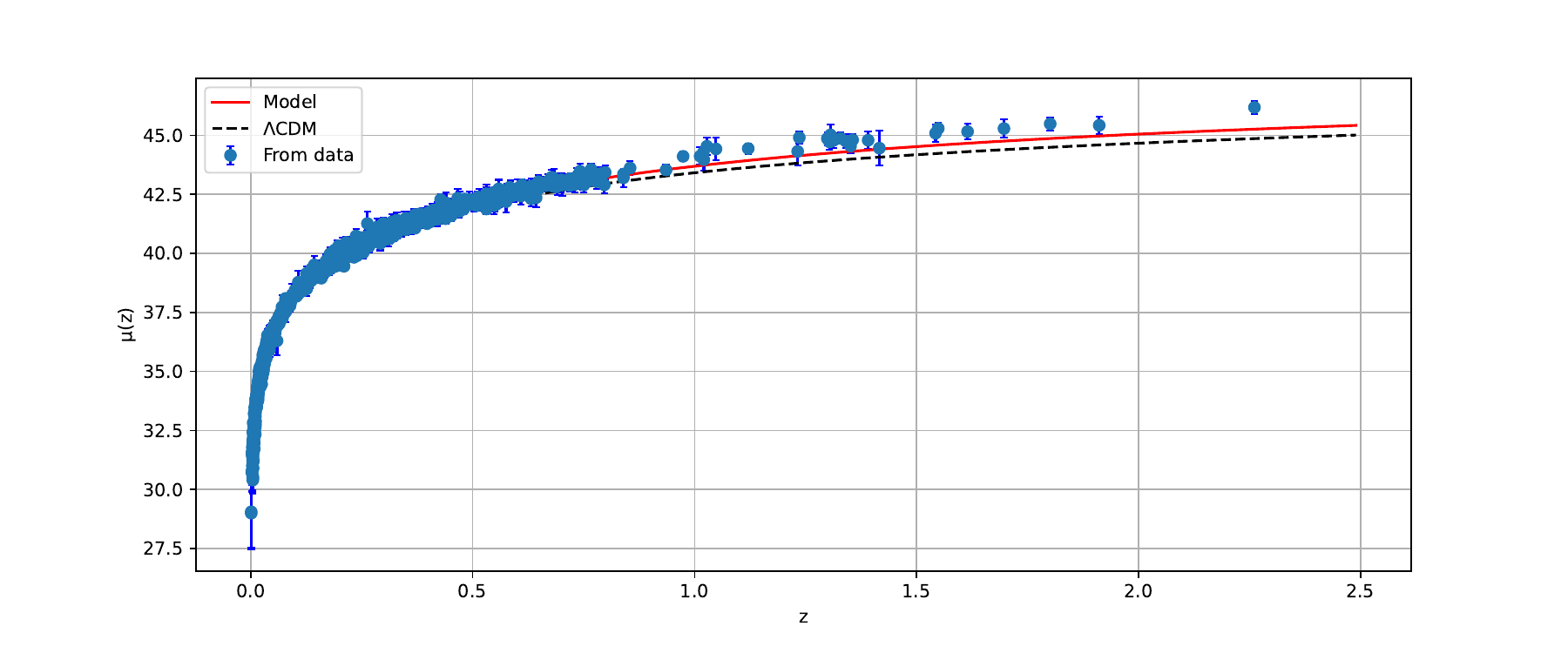}
        \caption{$\mu(z)$ in redshift for Pantheon+SH0ES dataset.}
        \label{EP3.2}
    \end{subfigure}%
    \hfill
    \begin{subfigure}{\textwidth}
        \centering
        \includegraphics[width=106mm]{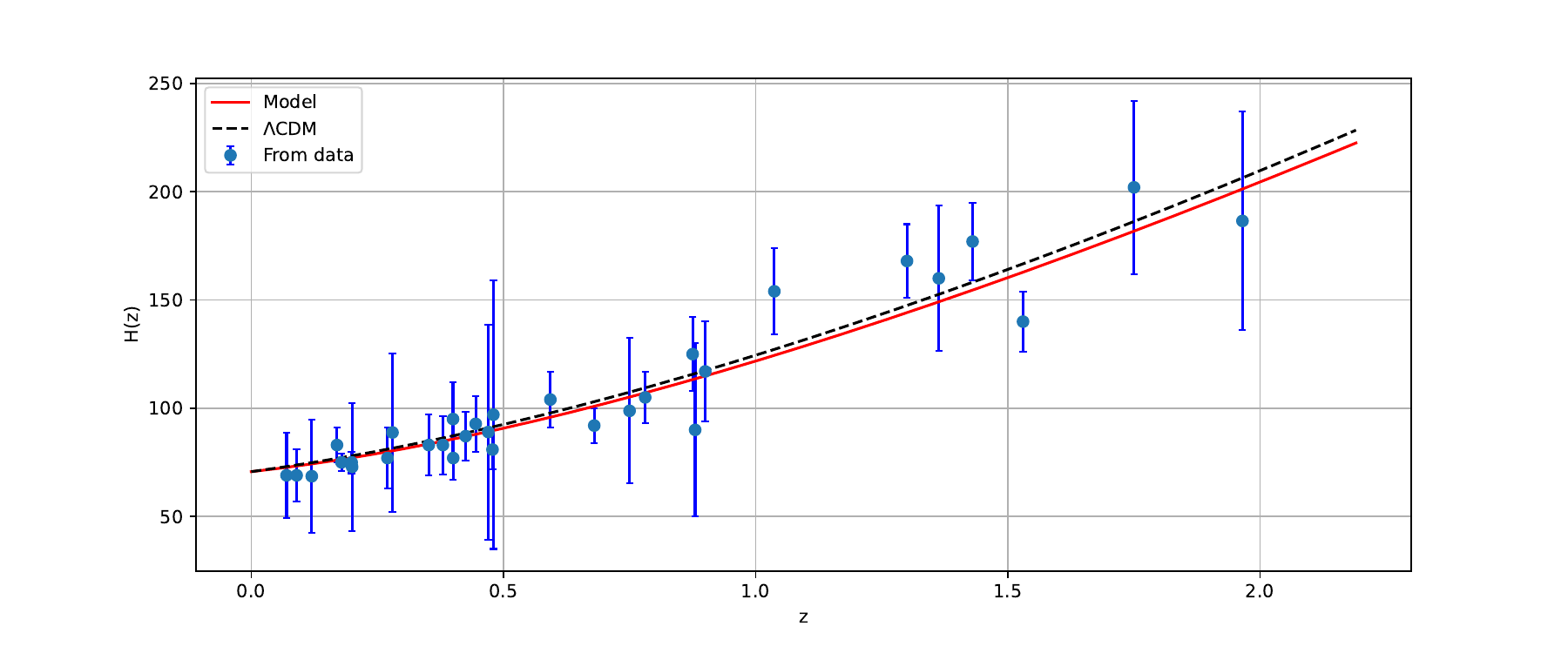}
        \caption{$H(z)$ in redshift for Hubble dataset.}
        \label{EH3.2}
    \end{subfigure}%
        \hfill
    \begin{subfigure}{\textwidth}
        \centering
        \includegraphics[width=106mm]{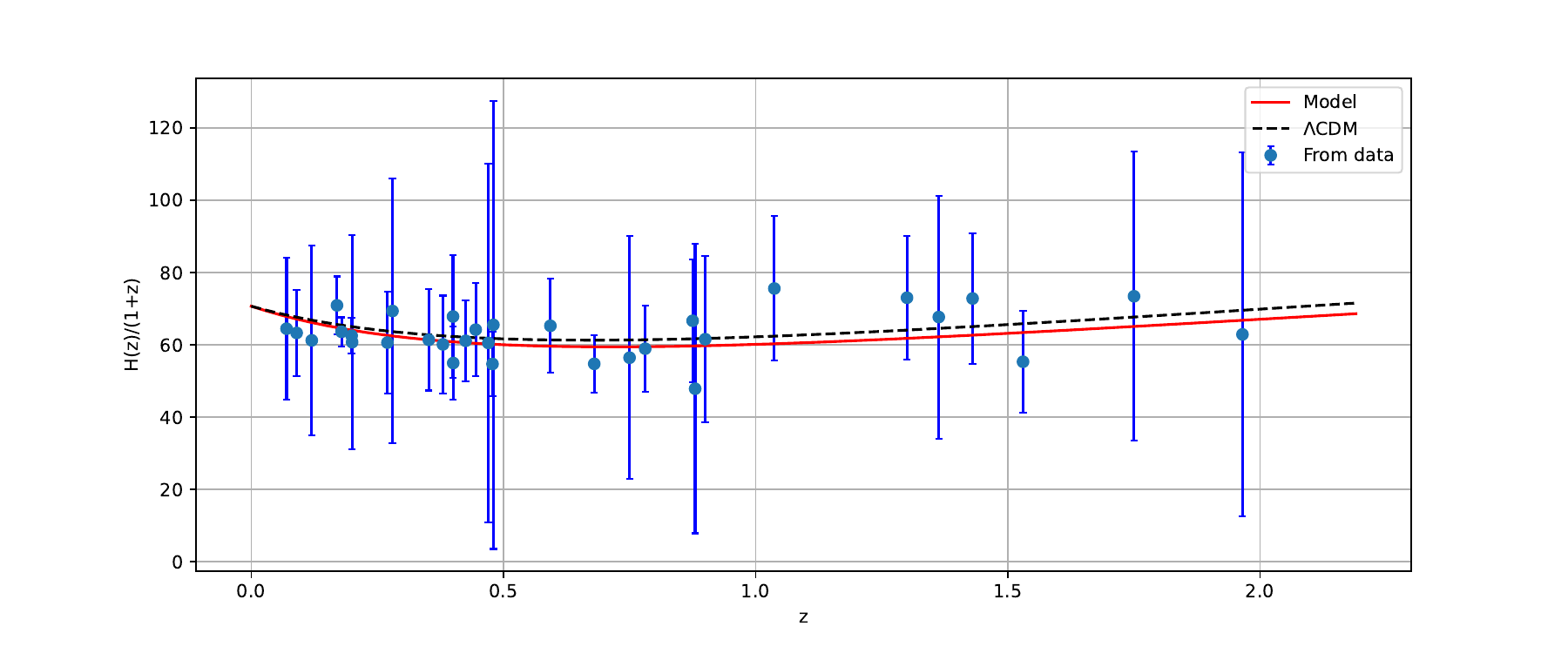}
        \caption{$H(z)/(1 + z)$ in redshift for Hubble dataset.}
        \label{EHZ3.2}
    \end{subfigure}%
    \caption{Error bar plot obtained from the observational datasets.}
    \label{fig3.2}
\end{figure}

\begin{table}[ht]
\renewcommand\arraystretch{1.5}
\centering
\begin{tabular}{|c|c|c|}
\hline
~Parameters~ &~~Hubble dataset~~&~~Pantheon+SH0ES dataset~~  \\ [0.1cm]
\hline\hline
$H_{0}$ & $69.5_{-1.9}^{+2.3}$ &  $70.7\pm 2.7$\\
\hline
$\omega$ & $-1.16\pm0.17$ &  $-1.15\pm 0.17$\\
\hline
$\alpha$ & $2.96\pm 0.082$ &  $2.951\pm 0.082$\\
\hline
$\beta$ & $0.813\pm 0.059$ &  $0.796\pm 0.0059$ \\
\hline
$n$ & $1.041 \pm 0.021$ &  $1.043\pm 0.022$\\
\hline
\end{tabular}
\caption{The marginalized constraining results of the parameters using Hubble and Pantheon+SH0ES data.}
    \label{table3.1}
\end{table}
The constraints derived from the Hubble and Pantheon+SH0ES datasets provide valuable insights into the model parameters. To further strengthen these results and test the consistency of the model, we now proceed to validate the parameters using BAO/CMB dataset, as outlined in the next section.

\section{BAO/CMB validation}\label{sec3.4}
In $\Lambda$CDM observations, the BAO standard ruler measurements are self consistent with CMB observations, as demonstrated by several galaxy surveys. As mentioned, our model has a similar background to $\Lambda$CDM but shows slight deviations at high redshift. So we can obtain stringent constraints on cosmological parameters by using this measurements. To map distance-redshift relationships, measuring the BAOs in large scale clustering patterns of galaxies is a promising technique. Also, BAO provides an independent way to measure the expansion rate of the Universe and also can describe the rate of change of expansion throughout the evolution history.\\

\subsection{BAO dataset}
The angular diameter distance through the clustering perpendicular to the line of sight can be measured using the BAO signals. Moreover, the expansion rate of the Universe $H(z)$ can be measured by the clustering along the line of sight. At the photon decoupling epoch, the comoving sound horizon can be defined as,
\begin{equation}\label{eq3.10}
r_{s}(z_{*}) = \frac{c}{\sqrt{3}}\int_{0}^{1/(1+z_{*})}\frac{d\tilde{z}}{\tilde{z}^{2}H(\tilde{z})\sqrt{1+\tilde{z}\left(3\Omega_{b_0}/4\Omega_{\gamma_0}\right)}}~,
\end{equation}
where, $\Omega_{b_0}$ and $\Omega_{\gamma_0}$ respectively represent the present value of the baryon and photon density parameter and $z_{*}$ is the redshift of photon decoupling. According to WMAP7 \cite{Jarosik_2011_192_14}, here we use, $z_{*} = 1091$. The expression for the comoving angular-diameter distance $\big[d_{A}(z_{*})\big]$ and the dilation scale \big[$D_{V}(z)$\big] are respectively,
\begin{eqnarray}
d_{A}(z_{*}) = \int_{0}^{z_{*}}\frac{d\tilde{z}}{H(\tilde{z})}~, \quad \quad \quad
D_{V}(z) = \left[\frac{(d_{A}(z))^{2}cz}{H(z)}\right]^{\frac{1}{3}}. \label{eq3.11}
\end{eqnarray}
The epoch at which baryons were released from photons called as drag epoch $(z_{d})$. At this epoch, the photon pressure is no longer able to avoid gravitational instability of the baryons. We use the value, $z_{d} = 1020$ \cite{Komatsu_2009_180_330}.

\subsection{CMB dataset}
The CMB is leftover radiation from the Big Bang or the time when the Universe starts evolution. In order to confront the dark energy models to CMB data, the distance priors method is more appropriate \cite{Wang_2007_76_103533, Wright_2007_664_633}. Using the CMB temperature power spectrum, this method measures two distance ratios:
\begin{itemize}
\item[(i)] The acoustic scale, measures the ratio of the angular diameter distance to the decoupling epoch. At decoupling epoch, it also measures the size of the comoving sound horizon. This first distance ratio can be expressed as,
\begin{equation*}
l_{A} = \pi\frac{d_{A}(z_{*})}{r_{s}(z_{*})}~.
\end{equation*}
\item[(ii)] The second one is at decoupling time, the ratio of angular diameter distance and the Hubble ratio, called the shift parameter. This can be expressed as,
\begin{equation*}
R = \sqrt{\Omega_{m}H_{0}^{2}}r(z_{*})~.
\end{equation*}
\end{itemize}
The acoustic scale is used to obtain the BAO/CMB constraints. Combining these results with the WMAP7-year \cite{Jarosik_2011_192_14} and WMAP9-year \cite{Bennett_2013_208_20} the value $l_{A} = 302.44\pm0.80$ and $l_{A} = 302.35\pm 0.65$ respectively. Percival et al. \cite{Percival_2010_401_2148} measured $\frac{r_{s}(z_{d})}{D_{V}(z)}$, at $z=0.2$ and $z =0.35$. The WiggleZ team \cite{Blake_2011_418_1707} obtained results at $z=0.44$, $z=0.60$ and $z=0.73$ and the 6dF Galaxy Survey also reported a new measurement of $\frac{r_{s}(z_{d})}{D_{V}(z)}$ at $z =0.106$ \cite{Beutler_2011_416_3017}. By using the WMAP 7 \cite{Jarosik_2011_192_14}, recommended values for $r_{s}(z_{d})$ and $r_{s}(z_{*})$ we get, $\frac{r_{s}(z_{d})}{r_{s}(z_{*})} = 1.045\pm0.016$. The BAO/CMB constraints $\frac{d_{A}(z_{*})}{D_{V}(z_{\text{BAO}})}$ also exhibited in Table \ref{table6.3}, along with the values for $\frac{d_{A}(z_{*})}{D_{V}(z_{\text{BAO}})}\frac{r_{s}(z_{d})}{r_{s}(z_{*})}$. The $\chi^{2}$ for the BAO/CMB can be written as,
\begin{eqnarray*}
    \chi_{\text{BAO/CMB}}^{2} &=& X^{T}C^{-1}X~,\quad \text{where} \quad
    X  = 
\renewcommand\arraystretch{1.2}
\begin{pmatrix}
\frac{d_{A}(z_{*})}{D_{V}(0.106)}-30.95 \\
\frac{d_{A}(z_{*})}{D_{V}(0.200)}-17.55 \\
\frac{d_{A}(z_{*})}{D_{V}(0.350)}-10.11 \\
\frac{d_{A}(z_{*})}{D_{V}(0.440)}-8.44 \\
\frac{d_{A}(z_{*})}{D_{V}(0.600)}-6.69 \\
\frac{d_{A}(z_{*})}{D_{V}(0.730)}-5.45
\end{pmatrix}~,
\end{eqnarray*}
and the inverse of covariance matrix $C$ is given by \cite{Giostri_2012_2012_027},
\begin{equation*}
 C^{-1} = 
\renewcommand\arraystretch{1.3}
\begin{pmatrix}
0.48435 & -0.101383 & -0.164945 & -0.0305703 & -0.097874 & -0.106738 \\
-0.101383 & 3.2882 & -2.45497 & -0.0787898 & -0.252254 & -0.2751 \\
-0.164945 & -2.45497 & 9.55916 & -0.128187  & -0.410404 & -0.447574  \\
-0.0305703 & -0.0787898 & -0.128187 & 2.78728 & -2.75632 & 1.16437 \\
-0.097874 & -0.252254 & -0.410404 & -2.75632 & 14.9245 & -7.32441 \\
0.106738 & -0.2751 & -0.447574 & 1.16437 & -7.32441 & 14.5022 \\
\end{pmatrix}~.
\end{equation*}

\begin{figure}[H]
    \centering
    \begin{subfigure}{0.5\textwidth}
        \centering
        \includegraphics[width=76.5mm]{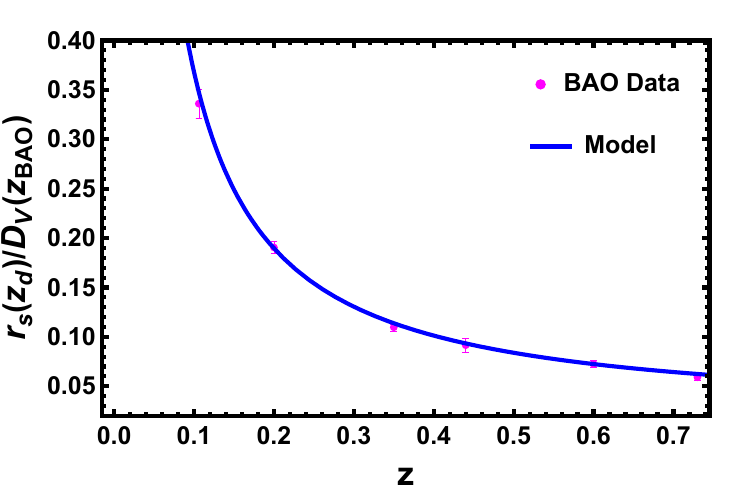}
        \caption{Plots of $\frac{r_{s}(z_{d})}{D_{V}(z_{\text{BAO}})}$ parameter versus $z$.}
        \label{BAOCMB3.4}
    \end{subfigure}%
    \hfill
    \begin{subfigure}{0.5\textwidth}
        \centering
        \includegraphics[width=75mm]{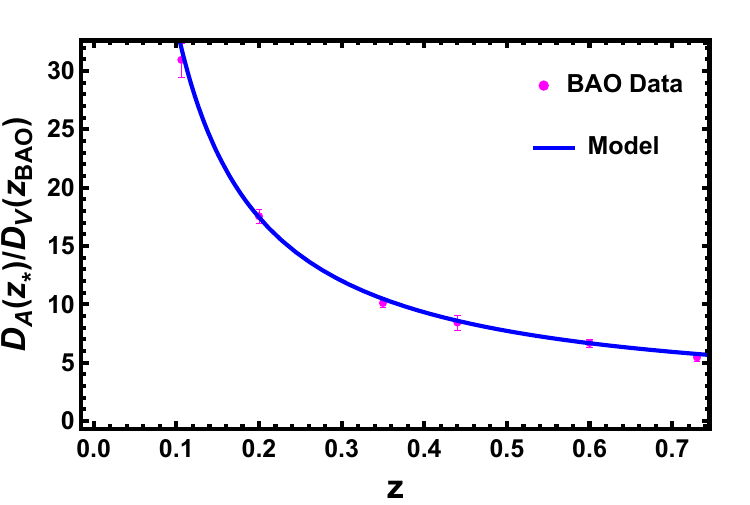}
        \caption{Plots of $\frac{d_{A}(z_{*})}{D_{V}(z_{\text{BAO}})}$ constraints versus $z$.}
        \label{BAOCMB13.4}
    \end{subfigure}%
    \caption{Error bar plot for the model utilizing BAO/CMB dataset.}
    \label{fig3.4}
\end{figure}
The plot for the distilled parameter and BAO/CMB constraints has been made using the $H_{0}=70.7kms^{-1}Mpc^{-1}$, $\omega = -1.15$, $\alpha = 2.95$, $\beta = 0.796$ and $n = 1.043$ that we obtained in Table \ref{table3.1}. The values which we have obtained in our results are best suited with the observational values of the BAO/CMB, as can be seen in Fig. \ref{fig3.4}.\\
The BAO/CMB datasets have validated the model parameters, ensuring their consistency with observational data. To gain a deeper understanding of the cosmological dynamics, we now proceed to explore the cosmographic parameters, which provide an alternative way to analyze the evolution of the Universe and test the predictions of the model, as discussed in the following section.

\section{Cosmographic parameters}\label{sec3.5}
 As explained in the cosmographic series \eqref{ATS}, the Hubble parameter is derived from the scale factor. The second and third derivative of the cosmographic series, the deceleration parameter ($q$) and jerk parameter ($j$) can be determined respectively. The Hubble parameter indicates the rate of the expansion of the Universe and it is a key element in characterizing the evolution of the Universe. For the expanding behaviour of the Universe, the Hubble parameter must be positive. The negative and positive sign of the deceleration parameter respectively gives the information on the accelerating and decelerating behaviour of the Universe. Now, using \eqref{eq3.5}, these parameters can be expressed in redshift as,
 \begin{eqnarray}
 q(z) &=& -1+\frac{(1+z)H_{z}(z)}{H(z)}~, \nonumber\\[5pt]
 j(z) &=& q(z)\big[1+2q(z)\big] + (1+z)q_{z}(z)~. \label{eq3.12}
 \end{eqnarray}
The interval on the value of $q$ ($q_0$ denotes the present value) describes the behaviour of the Universe as follow:
 \begin{itemize}
 \item[(i)] The Universe experiences expanding behaviour and undergoes deceleration phase for $q_{0}>0$. During this phase, one can obtain pressureless barotropic fluid or matter dominated Universe. However, the results from cosmological observations do not favor positive $q_{0}$. This situation would have occurred during early Universe.
 \item[(ii)] The expanding and accelerating Universe represents for $-1<q_{0}<0$, which is the present status of the Universe. 
 \item[(iii)] The entire cosmological energy budget is dominated by a de Sitter fluid for $q_{0} = -1$, representing a cosmic component with a constant energy density that remains unchanged as the Universe expands. This is the case of inflation during the very early Universe.
 \end{itemize}
From Eq. \eqref{eq3.12}, the present value of the jerk parameter can be, $j_{0} = 2q_{0}^{2}+q_{0}+ q_{z|_0}$. We wish to keep, $-1<q_{0}<-0.5$, which requires $2q_{0}^{2}+q_{0}>0$. Hence, if $q_{0}<-0.5$, then $j_{0}$ is linked to the sign of the variation of $q$. Accordingly the behaviour of this geometrical parameter can be interpreted as follows:
 \begin{itemize}
\item[(i)] when $j_{0}$ is negative, there is no change of the behaviour from the present phase to the accelerated phase. The dark energy influences early time dynamics without any change since the start of evolution.
\item[(ii)] when $j_{0}$ vanishes, the accelerating parameter tends smoothly to a precise value, without any change in its behavior. 
\item[(iii)] when $j_{0}$ is positive, there was a precise point during the evolution when the acceleration of the Universe began. The corresponding redshift can be referred as the transition redshift, at which the effect of dark energy becomes significant. As a consequence, it indicates the presence of further cosmological resources. In order to constrain the DE EoS, one would need to measure the transition redshift $z_{tr}$ directly. To note here, the slope of the Universe gets changed by changing the sign of $j_{0}$. 
\end{itemize}
\begin{figure}[ht]
\centering
\includegraphics[width=95mm]{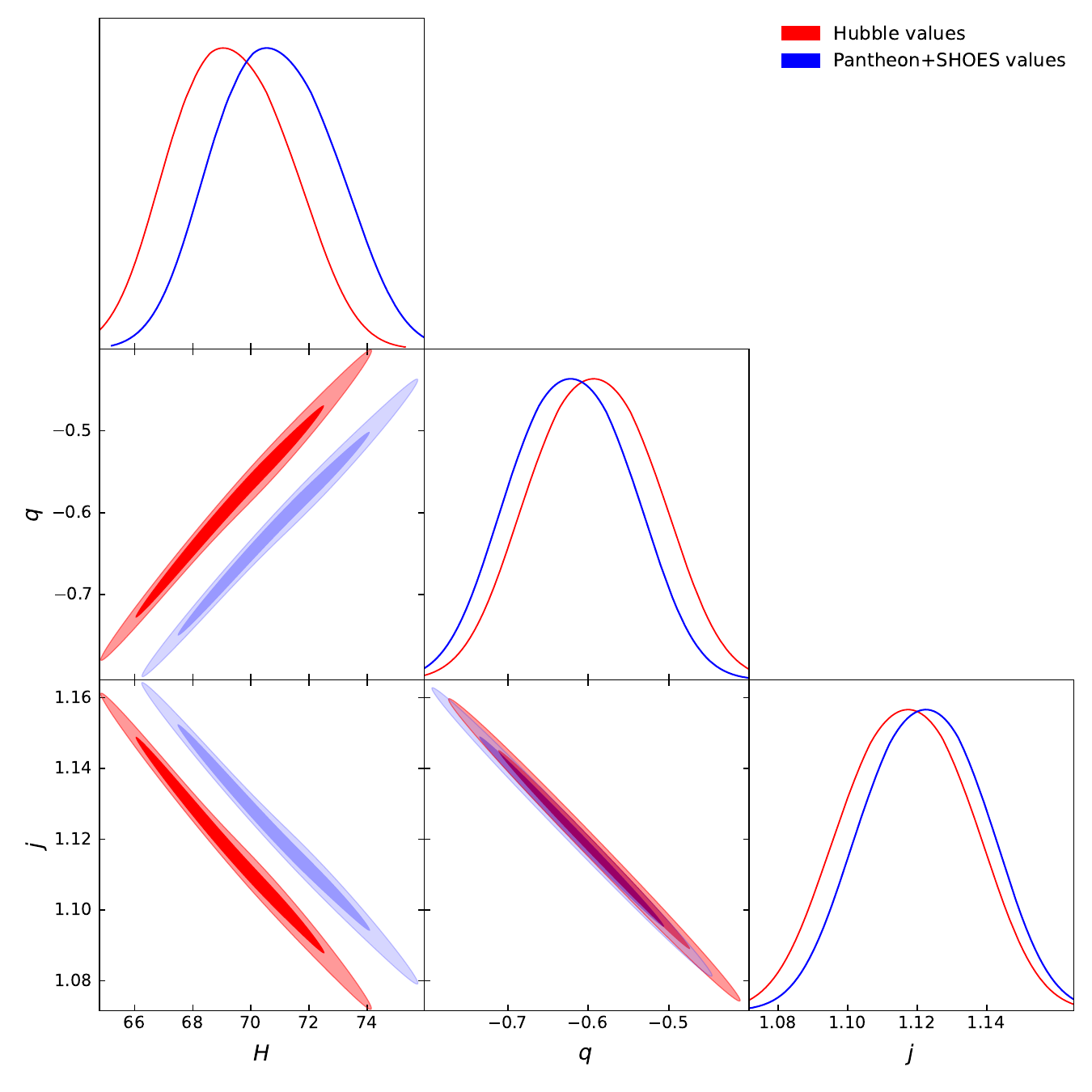} 
\caption{The marginalized constraints on the $H(z)$, $q(z)$ and $j(z)$ using redshift values.} \label{fig3.5}
 \end{figure}
In Fig. \ref{fig3.5}, we have marginalize the $z$-values for the present values of $H(z)$, $q(z)$ and $j(z)$. In Table \ref{table3.3} the best fit present value of $H(z)$, $q(z)$ and $j(z)$ are given, as obtained from the Hubble and Pantheon+SH0ES data using marginalized model parameter values.

\begin{table}[H]
\renewcommand\arraystretch{1.8}
\centering
\begin{tabular}{|c|c|c|}
\hline
~Parameters~ &~~Hubble dataset~~&~~Pantheon+SH0ES dataset~~  \\ [0.1cm]
\hline\hline
$H(z)$ & $69.5_{-1.9}^{+2.3}$ &  $70.7\pm 2.7$\\
\hline
$q(z)$ & $-0.59\pm 0.07$ &  $-0.61\pm 0.067$\\
\hline
$j(z)$ & $1.117\pm 0.02$ &  $1.122\pm 0.02$ \\
\hline
\end{tabular}
\caption{The marginalized constraining results of the cosmographic parameters using Hubble and Pantheon+SH0ES data.}\label{table3.3}
\end{table}

\begin{figure}[ht]
    \centering
    \begin{subfigure}{0.5\textwidth}
        \centering
        \includegraphics[width=75.5mm]{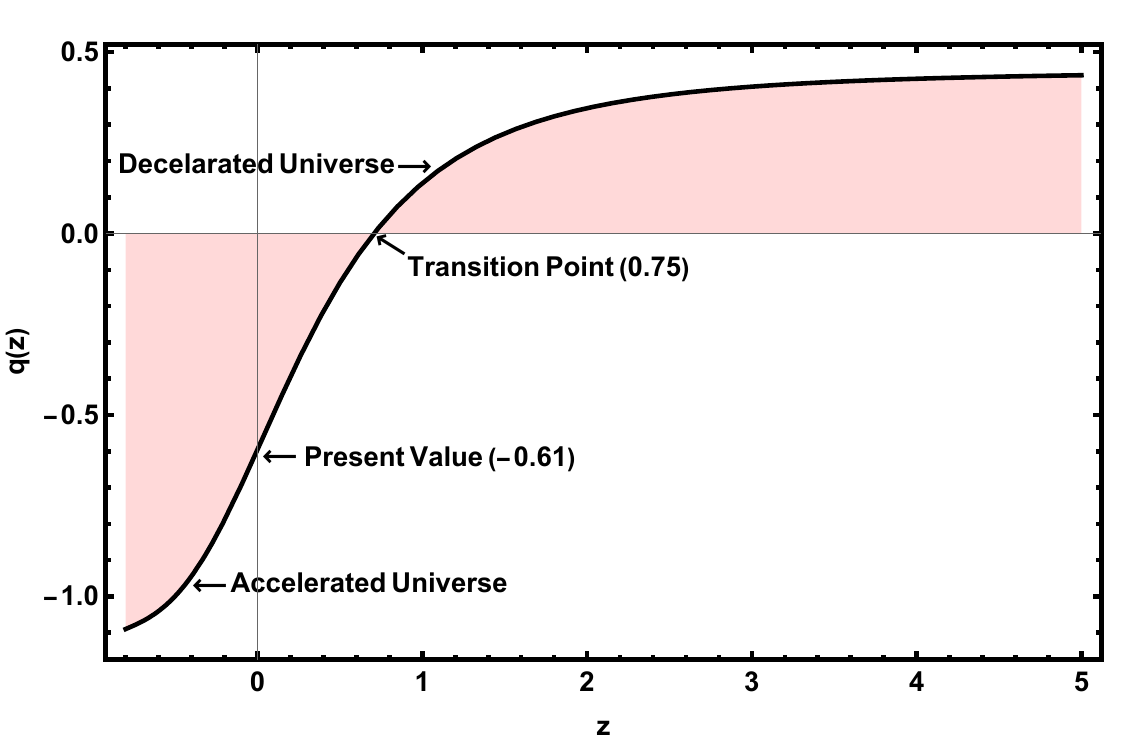}
        \caption{Evolution of deceleration parameter.}
        \label{q3.6}
    \end{subfigure}%
    \hfill
    \begin{subfigure}{0.5\textwidth}
        \centering
        \includegraphics[width=75mm]{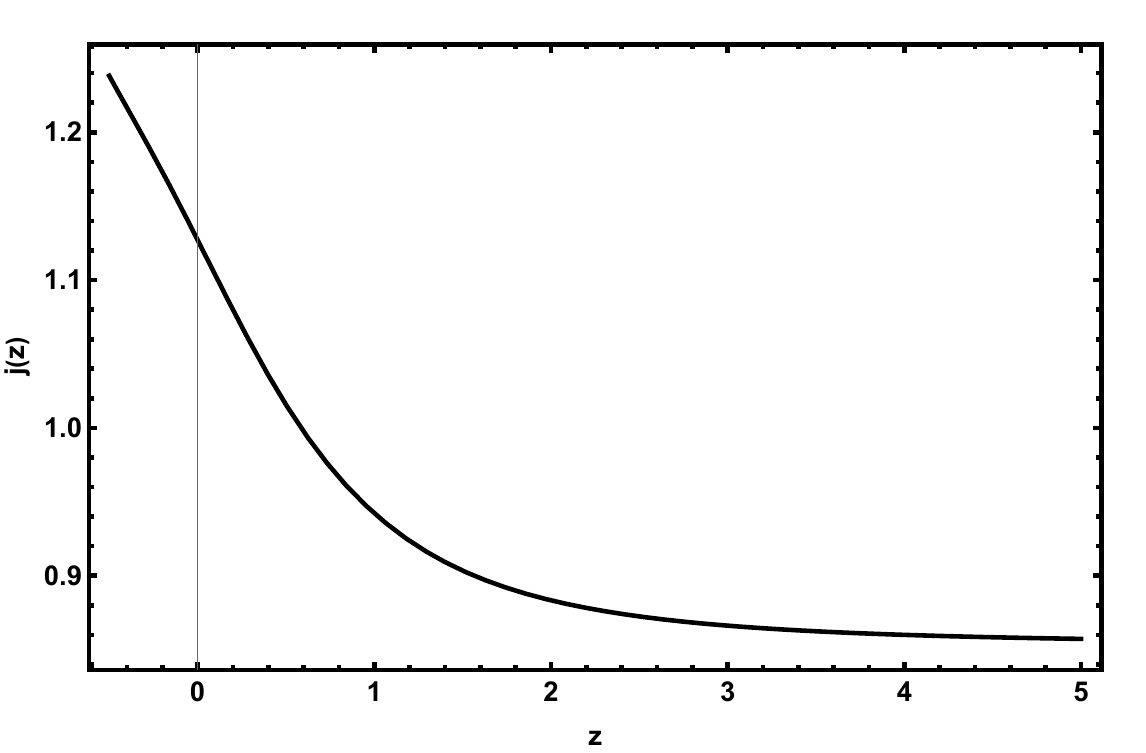}
        \caption{Evolution of jerk parameter.}
        \label{j3.6}
    \end{subfigure}%
    \caption{Evolution of cosmographic parameters as a function of redshift.}
    \label{fig3.6}
\end{figure}
The transition of the deceleration parameter from early deceleration to late-time acceleration is illustrated in Fig. \ref{q3.6}. This transition occurs at approximately $z_{t} \approx 0.75$, indicating a shift from a decelerating phase to an accelerating phase with the present value of the deceleration parameter being $q_{0} \approx -0.61$. Recently performed measurements have determined that the value of the deceleration parameter for the current cosmic epoch is within the range of $q_{0} = -0.528_{-0.088}^{+0.092}$ \cite{Gruber_2014_89_103506} and transition from deceleration to acceleration at $z_{t} = 0.60_{-0.12}^{+0.21}$ \cite{Yang_2020_2020_059,Capozziello_2015_91_124037}. From Fig. \ref{j3.6}, we can observed that $j>0$, which verify that there exists a transition time when the Universe modifies its expansion.\\
The next step is to test the validity of the model using observational tools like the $Om(z)$ diagnostic and the age of the Universe.

\section{Test for the validation}\label{sec3.6}
It is possible to verify the validity of any cosmological model through some theoretical and observational tests. For this discussion, we examine the $Om(z)$ cosmological test as a possible way to authenticate our derived model. To validate the model, we calculated the age of the Universe.

\subsection{$Om(z)$ diagnostic}
The $Om(z)$ diagnostic has been introduced as an alternative approach to test the accelerated expansion of the Universe with the phenomenological assumption, EoS, $p=\rho\omega$ filling the Universe with the perfect fluid. The $Om(z)$ diagnostic provides a null test to the $\Lambda$CDM model \cite{Sahni_2008_78_103502}. Also, there are evidences available in the literature on its sensitiveness with the EoS parameter \cite{Ding_2015_803_L22, Zheng_2016_825_17, Qi_2018_18_066}. The nature of $Om(z)$ slope differs between dark energy models because: the positive slope indicates the phantom phase $\omega < -1$ and the negative slope indicates the quintessence region $\omega > -1$. The $Om(z)$ diagnostic can be defined as, 
\begin{align}\label{eq3.13}
    Om(z) &= \frac{E^{2}(z) - 1}{(1+z)^{3}-1}~,\nonumber\\[10pt]
    Om(z) &= \frac{\left(\frac{\alpha }{\beta 6^{n}(2n-1)}+(z+1)^3 \left(-\frac{\alpha }{\beta 6^{n}(2n-1)}-\frac{1}{\beta 6^{n}(2n-1)}+1\right)+\frac{(z+1)^{3 (\omega +1)}}{\beta 6^{n}(2n-1)}\right)^{1/n}-1}{(z+1)^3-1}~.
\end{align}
In Fig. \ref{omz3.7}, $Om(z)$ shows the positive behavior and that confirms the phantom like behavior of model.

\subsection{Age of the Universe}
The age-redshift relationship determining the age of the Universe as a function of redshift, $t_{U}(z)$ is given by \cite{Vagnozzi_2022_36_27},
\begin{equation}\label{eq3.14}
 t_{U}(z) = \int_{z}^{\infty}\frac{d\tilde{z}}{(1+\tilde{z})H(\tilde{z})}~.
\end{equation}
 The age of the Universe at any redshift is inversely proportional to $H_{0} = H(z=0)$ as shown in Eq. \eqref{eq3.14}. Now, Using Eq. \eqref{eq3.5} the age of the Universe can be computed as,
\begin{align}\label{eq3.15}
 H_{0}(t-t_{0}) &= \int_{0}^{z}\frac{d\tilde{z}}{(1+\tilde{z})E(\tilde{z})}~, \nonumber\\[10pt]
    H_{0}t_{0} &= \lim_{z\to\infty}\int_{0}^{z}\frac{d\tilde{z}}{(1+\tilde{z})E(\tilde{z})}~.
\end{align}

\begin{figure}[ht]
    \centering
    \begin{subfigure}{0.5\textwidth}
        \centering
        \includegraphics[width=76mm]{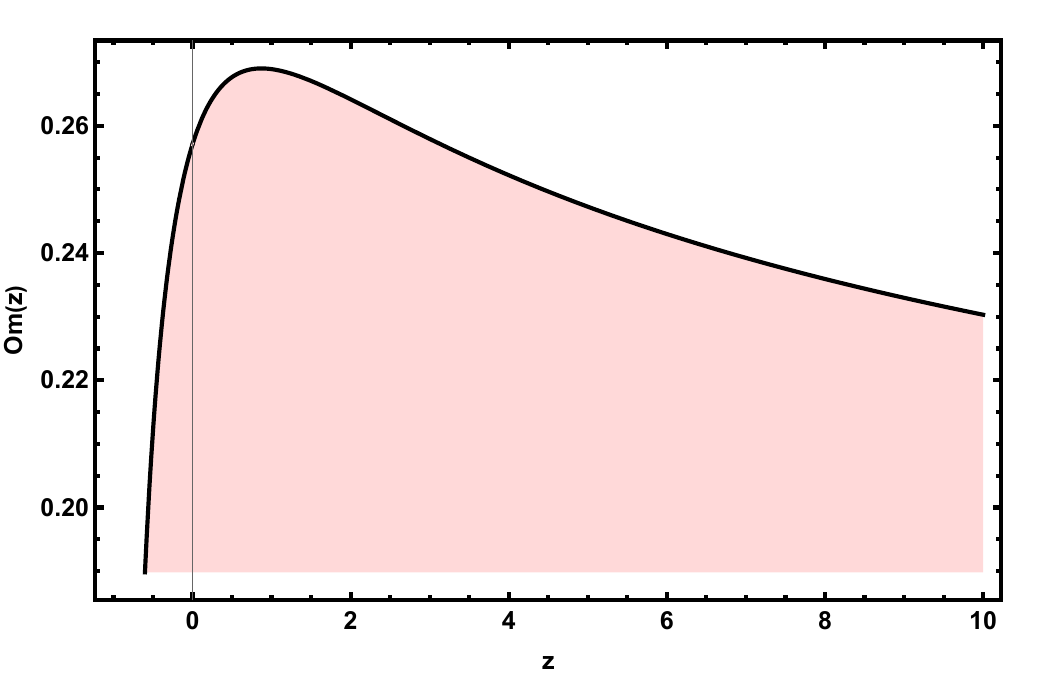}
        \caption{Behaviour of $Om(z)$ in redshift.}
        \label{omz3.7}
    \end{subfigure}%
    \hfill
    \begin{subfigure}{0.5\textwidth}
        \centering
        \includegraphics[width=75mm]{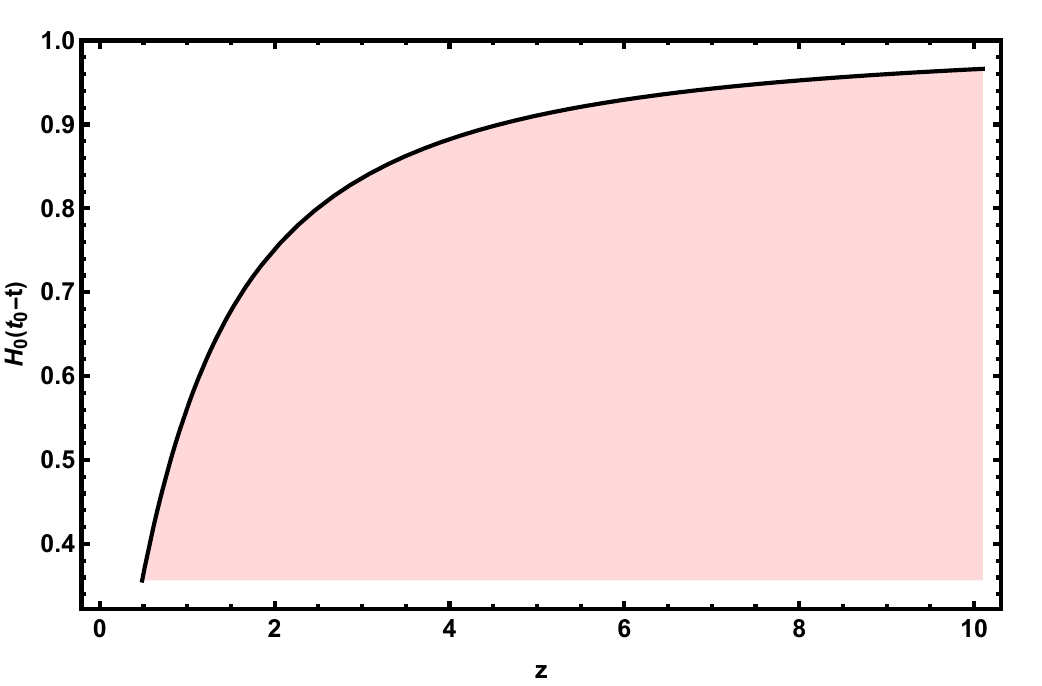}
        \caption{Behaviour of $H_{0}(t_{0}-t)$ in redshift.}
        \label{aou3.7}
    \end{subfigure}%
    \caption{Graphical representation of the test validation of the behavior of the model.}
    \label{fig3.7}
\end{figure}
The Fig. \ref{aou3.7} shows the behavior of time with redshift. It is found that $H_{0}(t_{0}-t)$ converges to $0.9791$ as $z\rightarrow\infty$. Utilizing this, we can calculate the current age of the Universe as $t_{0} = 0.9791H_{0}^{-1} \approx 13.85~~Gyrs$ which is quite close to the age calculated from the Planck result $t_{0} = 13.78 \pm 0.020~~Gyrs$. So, the results obtained in the model are in consistent with current data.

\section{Conclusion}\label{sec3.7}
In this chapter, we explored cosmological models by considering higher powers of non-metricity in the function $f(Q)$ and derived the expression for the Hubble parameter as a function of redshift. The $H(z)$ parameter was then constrained using the Hubble and the Pantheon+SH0ES datasets. We rebuild $H(z)$ and the distance modulus over $32$ data points within the redshift range $0.07\leq z\leq1.965$ using a $\chi^{2}$ minimization technique. Further analysis of the Pantheon+SHOES data, consisting of $1701$ SNe Ia apparent magnitude measurements, was conducted to obtain the best-fit values of model parameters and the EoS parameter through MCMC analysis. The results align well with the $\Lambda$CDM model, as shown by the error bar plots and the obtained parameter values listed in Table \ref{table3.1}. Using the BAO dataset, we have validated our results. We further constrained the expansion history $H(z)$ and the $Om(z)$ diagnostic analysis provided a null test for the $\Lambda$CDM model. The transition from deceleration to acceleration occurred at a redshift of approximately $z_{t} \approx 0.75$, with a deceleration parameter $q_{0} \approx -0.61$. The EoS parameter values derived from the Hubble and Pantheon+SH0ES datasets, $-1.16\pm0.17$ and $-1.15\pm0.17$ respectively suggest phantom behavior. Additionally, the age of the Universe was calculated based on the constrained Hubble parameter. Our models were compared with the concordance $\Lambda$CDM model through the cosmographic and $Om(z)$ parameters. The deviation from the $\Lambda$CDM model may indicate potential interactions between dark energy and dark matter, with our models favoring a phantom phase at this epoch.
\chapter{Insights into $f(Q)$ gravity: Modeling through the deceleration parameter} 

\label{Chapter4} 

\lhead{Chapter 4. \emph{Insights into $f(Q)$ gravity: Modeling through the deceleration parameter}} 

\vspace{10 cm}
* The work, in this chapter, is covered by the following publication: \\

\textbf{S. A. Narawade} and B. Mishra, ``Insights into $f(Q)$ gravity: Modeling through the deceleration parameter", \textit{J. High Energy Astrop.} \textbf{45} (2025) 409.
 
\clearpage
       
\section{Introduction} \label{sec1}
Building on the analysis of the $f(Q)$ model in comparison to observational datasets discussed in the previous chapter, we now explore into alternative cosmological frameworks via reconstruction method. This approach aims to enhance our understanding of the late-time acceleration of the Universe. By incorporating the reconstruction method, observational data can be directly integrated into cosmological models, enhancing our understanding of the Universe and improving the accuracy of future surveys. One of the key advantages of reconstruction is that it is independent of the specific gravity model used in cosmological studies. In this approach, two methods are used: the parametric reconstruction, which establishes a kinematic model with free parameters and constrains the parameters through statistical analysis of observational data. The second one is non-parametric reconstruction, which derives models directly from observational data by using statistical procedures. Various cosmological parameters, such as the Hubble parameter \cite{Roy_2022_36_101037, Lemos_2018_483_4803}, the deceleration parameter \cite{Capozziello_2022_36_101045, Mamon_2017_77_495} and the jerk parameter \cite{Mukherjee_2016_93_043002, Zhai_2013_727_8} have proven to be effective in exploring the accelerating Universe with parametric reconstruction. The parametrization of dynamical parameters such as pressure, energy density and EoS \cite{Mukherjee_2016_460_273, Pantazis_2016_93_103503} has been widely studied.

In this chapter, we began with the investigating the reconstruction of the deceleration parameter through the parametrization. Then the free parameters of the parameterized $q(z)$ are constrained using the observational datasets. Further, we have consider parametrized $H(z)$ obtained from $q(z)$ to reconstruct the $f(Q)$ models by applying the numerical approach with appropriate boundary conditions. Also, the statistical comparisons using AIC and BIC reveal a competitive performance with the $\Lambda$CDM model, positioning these $f(Q)$ models as promising alternatives for explaining DE and cosmic acceleration. This chapter aims to find viable alternatives to the standard $\Lambda$CDM model through the reconstruction of the deceleration parameter. It is therefore possible to replace the $\Lambda$CDM model effectively by reconstructing the $f(Q)$ gravity model using the deceleration parameter, providing an alternative explanation to the accelerating expansion of the Universe that does not rely on the cosmological constant. The chapter is organized as follows: The reconstruction of deceleration parameter are presented in section \ref{sec4.2}. The cosmological datasets are given in section \ref{sec4.3}. Our primary findings are presented in section \ref{sec4.4}, where we constrain the free parameters of both models based on the datasets. This section also features the reconstruction of the $f(Q)$ model, utilizing two forms of the deceleration parameter. Furthermore, we provide a comparative analysis of our results against the standard model of cosmology, employing the AIC and the BIC for evaluation. Finally, we summarize our main results in section \ref{sec4.5}.

\section{Reconstruction of $q(z)$}\label{sec4.2}
By combining the Taylor series for $a(t)$ given in Eq. \eqref{ATS} and the scale factor-redshift relation provided in Eq. \eqref{redshiftscalefactor}, we can rewrite Eq. \eqref{CP} as,
\begin{eqnarray}
  q(z)&&\equiv \frac{1}{2}\frac{(1+z)}{H(z)^{2}}\frac{d}{dz}\Big(H(z)^{2}\Big)-1~,\label{eq4.1}\\[5pt]
  j(z)&&\equiv -\left[\frac{1}{2}\frac{(1+z)^{2}}{H(z)^{2}}\frac{d^{2}}{dz^{2}}\Big(H(z)^{2}\Big)-\frac{(1+z)}{H(z)^{2}}\frac{d}{dz}\Big(H(z)^{2}\right)+1\Big].\label{eq4.2}
\end{eqnarray}
It is obvious that Eq. \eqref{eq4.1} lead to a homogeneous Euler equation when $q=constant$. Under this assumption, Eq. \eqref{eq4.1} can be solved as,
\begin{equation}\label{eq4.3}
  E^{2}(z)=\frac{H^{2}(z)}{H_{0}^{2}}=C_{1}(1+z)^{2(1+\nu)}~,
\end{equation}
where $C_{1}$ is an arbitrary constant and $H_{0}$ stands for the current value of a Hubble parameter. A definition of the $E(z)=\frac{H(z)}{H_{0}}$ can then be used to replace $E(z)$ in Eq. \eqref{eq4.1}. The constants in Eq. \eqref{eq4.1} and $H(z)$ can be determined without any fundamental change. In order to understand the physical meanings of the constants in Eq. \eqref{eq4.3}, one can refer to a particular cosmological model. From Eq. \eqref{eq4.3}, it is clear that if $q=\frac{1}{2}$, the above expression reduces to the Einstein-de Sitter solution, which is represented by \cite{Carneiro_2008_77_083504},
\begin{equation}\label{eq4.4}
  E(z)=\Big[C_{1}(1+z)^3\Big]^{\frac{1}{2}}.
\end{equation}
This solution describes a matter-dominated Universe with no cosmological constant, where the expansion rate is solely determined by the matter content. It is clear that $C_{1}$ becomes the matter density term $\Omega_{m_{0}}$. By evaluating deviation of $q(z)$ from $\frac{1}{2}$ or some other value in the past, we are able to determine how far it has strayed, but the calculation depends on the functional form of $q(z)$. The proposed reconstruction of deceleration parameter could be similar to the methods for studying the interaction between DE and DM based on phenomenological models \cite{Cai_2005_2005_002}, one possible proposal in reconstructing deceleration parameter is
\begin{equation}\label{eq4.5}
  q(z)=\frac{1}{2}+b\frac{g(z)}{E^{2}(z)}~,
\end{equation}
where $b$ is constant and needs to be constrained and $g(z)$ is an arbitrary function of redshift $z$. The different choice of $g(z)$ will lead to different reconstructions of $q(z)$. It is possible to solve Eq. \eqref{eq4.1} analytically under certain models of $g(z)$ with this kind of assumption. As a result, Eq. \eqref{eq4.1} becomes more symmetric by satisfying both sides of the equation which are composed of a constant term and a $E^{-2}$ term.

Using this method, we phenomenologically parameterize the $q(z)$ and solve the Euler equation. The first parametrization $f(z)=(1+z)^n$ simplifies modeling the dependence of the Hubble parameter on redshift. For $n = 4$, it represents the radiation-dominated era, while $n = 0$ corresponds to the standard $\Lambda$CDM model. Additionally, inspired by the reconstruction of the jerk parameter \cite{Zhai_2013_727_8}, we incorporate $f(z) = \ln(1 + z)$ as a secondary parametrization for our analysis.
\begin{eqnarray}
\bullet~\text{Model I} ~~~~~~ && q(z)=\frac{1}{2}+b\frac{(1+z)^n}{E^{2}(z)}~, \label{eq4.6} \\
\bullet~\text{Model II} ~~~~~ && q(z)=\frac{1}{2}+b\frac{\ln(1+z)}{E^{2}(z)}~. \label{eq4.7}
\end{eqnarray}
Substituting these equations into Eq. \eqref{eq4.1}, we can obtain the solutions of $E(z)$,
\begin{eqnarray}
&&\bullet~~\text{Power-law}~~~~~~~E^{2}(z)=C_1 (1+z)^3+\frac{2b(1+z)^n}{n-3} ~,\label{eq4.8} \\
&&\bullet~\text{Logarithmic}~~~~~E^{2}(z)=C_1 (1+z)^3-\frac{2b}{9}\Big[1+3\ln(1+z)\Big].  \label{eq4.9}
\end{eqnarray}
The coefficients $C_{1}$ and $b$ arise from the process of solving Eq. \eqref{eq4.1} which is a first order differential equation. Another constraint that $E(z=0)=1$ gives a relationship between the constants $C_{1}$ and $b$
\begin{eqnarray*}
&&\bullet~~\text{Power-law}~~~~~~~~C_{1}=1-\frac{2b}{n-3}~,\\
&&\bullet~~\text{Logarithmic}~~~~~~C_{1}=1+\frac{2b}{9}~.
\end{eqnarray*}
In the next section, the observational data has been explain to constrain the free parameters for each of the model listed above.

\section{Observational datasets}\label{sec4.3}
The model and cosmological parameters are constrained using the MCMC sampler by exploring the posteriors of the parameter space and varying them across a wide range of conservative priors discussed in section \ref{sec3.3.3}. The MCMC analysis below will utilize OHD, SNe Ia and BAO as our baseline dataset. With different combinations of observational datasets, we test the free parameter for the Hubble parameter derived from the deceleration parameter. We have used the \textit{emcee} package, which is available at Ref. \cite{Foreman-Mackey_2013_125_306}, to perform an MCMC analysis of every model and dataset combination. 

\subsection{Hubble dataset} 
The differential age method is widely used to estimate the expansion rate of the Universe at redshift $z$. The Hubble parameter can then be predicted using $(1+z)H(z) = -\frac{dz}{dt}$. The dataset includes $32$ measurements of the Hubble parameter, spanning a redshift range of $0.07 \leq z \leq 1.965$ \cite{Moresco_2022_25_6}. The mean value of the parameters is determined by minimizing the chi-square value given in section \ref{3.3.1}.

\subsection{Type Ia Supernovae compilation}
We used SNe Ia data as a baseline for MCMC analyses, with the chi-square formula outlined in section \ref{3.3.2}. Our analysis uses two SNe Ia datasets: Pantheon \cite{Scolnic_2018_859_101} and Pantheon$^+$ \cite{Scolnic_2022_938_113}, with the latter being an updated version of Pantheon (see \hyperref[Type Ia Supernovae]{1.6.2}).

\subsection{BAO dataset}
In this work, we have extended the BAO dataset by incorporating additional measurements from various surveys, including the BOSS DR11 quasar Lyman-alpha measurements at $z_{\mathrm{eff}} = 2.4$ \cite{Bourboux_2017_608_A130}, the SDSS Main Galaxy Sample at $z_{\mathrm{eff}} = 0.15$ \cite{Ross_2015_449_835} and the six-degree Field Galaxy Survey at $z_{\mathrm{eff}} = 0.106$ \cite{Beutler_2011_416_3017}. Furthermore, the $H(z)$ values and the angular diameter distances from the SDSS-IV eBOSS DR14 quasar survey at $z_{\mathrm{eff}} = \{0.98, 1.23, 1.52, 1.94\}$ \cite{Zhao_2018_482_3497} have been included. Also, the consensus BAO measurements for the Hubble parameter and comoving angular diameter distances from the SDSS-III BOSS DR12 at $z_{\mathrm{eff}} = \{0.38, 0.51, 0.61\}$ \cite{Alam_2017_470_2617} and the full covariance matrix associated with these two BAO datasets are taken into account. This extended dataset allows for a more comprehensive analysis, refining our cosmological constraints with the inclusion of measurements across a wider range of redshifts. By combining these new BAO data points, we enhance the precision of the constraints derived in section \ref{sec3.4}. 

The corresponding combination of parameter, utilizing the reported BAO results given by \cite{Briffa_2023_522_6024}, 
\begin{align*}
    \mathcal{G}(z_i) = \left(\frac{D_V(z_i)}{r_s(z_d)}\right)\left(\frac{r_s(z_d)}{D_V(z_i)}\right)D_H(z_i)\left(\frac{r_{s,\mathrm{fid}}(z_d)}{r_s(z_d)}D_M(z_i)\right)\left(\frac{r_s(z_d)}{r_{s,\mathrm{fid}}(z_d)}H(z_i)\right)\left(\frac{r_{s,\mathrm{fid}}(z_d)}{r_s(z_d)}D_A(z_i)\right)~,
\end{align*}
where, the angular diameter distance is $D_A(z)=(1+z)^{-2}D_L(z)$, the volume-average distance is $D_V(z) = \left[(1+z)^2D_A(z)^2 \frac{c z}{H(z)}\right]^{1/3}$, the comoving angular diameter distance is given by $D_M(z) = (1+z)D_A(z)$, the Hubble distance is $D_H(z) = \frac{c}{H(z)}$ and the comoving sound horizon at the end of the baryon drag epoch at redshift $z_d\approx 1059.94$ \cite{Aghanim_2020_641_A6} is calculated as,
\begin{align*}
    r_s(z)=~\int_z^\infty\frac{c_s(\tilde{z})}{H(\tilde{z})}\,\mathrm{d}z =~\frac{1}{\sqrt{3}}\int_0^{1/(1+z)}\frac{\mathrm{d}a}{a^2H(a)\sqrt{1+\left[3\Omega_{b,0}/(4\Omega_{\gamma,0})\right]a}}\,,
\end{align*}
where, we use the values $T_{0}=2.7255\,\mathrm{K}$ \cite{Fixsen_2009_707_916}, $\Omega_{b,0}=0.02242$ \cite{Aghanim_2020_641_A6} and a fiducial value of $r_{s,\mathrm{fid}}(z_d)=147.78\,\mathrm{Mpc}$.

The chi-square for the BAO data ($\chi^2_{\mathrm{BAO}}$) is determined by,
\begin{equation}\label{eq4.10}
\chi^2_{\text{BAO}}(\Theta) = \Delta G(z_i,\Theta)^T~ C_{\text{BAO}}^{-1}~\Delta G(z_i,\Theta)~,
\end{equation}
where $C_{\text{BAO}}$ is the covariance matrix of all the considered BAO observations and $\Delta G(z_i,\Theta) = G(z_i,\Theta)-G_{\text{obs}}(z_i)$.\\
These observational datasets offer essential constraints on the proposed Hubble function, providing valuable insights into their viability. In the next section, we will investigate the impact of these datasets on free parameter of the Hubble function and analyzing how these constrained free parameter of the Hubble parameter affect the $f(Q)$ models and refines the cosmological understanding.

\section{Impact on $f(Q)$ models}\label{sec4.4}
By considering the first Friedman equation \eqref{fqr}, we can reconstruct the $f(Q)$ function. Based on $Q$, the derivatives of the non-metricity tensor can be expressed as derivatives of the Hubble parameter with respect to redshift. One can find
\begin{equation*}
    f_{Q}=\frac{f'(z)}{12H(z)H'(z)}~,
\end{equation*}
where the prime denote derivative with respect to $z$. Now, the Eq. \eqref{fqr} can be re-written as,
\begin{eqnarray}\label{eq4.11}
    \frac{H'(z)}{H(z)}f'(z)-f(z) = 6H_{0}^{2}\Omega_{m_0}(1+z)^3~.
\end{eqnarray}
Based on Eq. \eqref{eq4.8} - \eqref{eq4.9} for both models and the results in Table \ref{table4.1} - \ref{table4.2} for each model, we can solve Eq. \eqref{eq4.11} numerically. To establish a boundary condition, we define the effective gravitational constant in $f(Q)$ gravity as $G_{eff}\equiv\frac{G}{f_Q}$ \cite{Saridakis_2021}. A natural requirement is that $G_{eff}$ coincident with Newton's constant at the present epoch, which implies that $f_Q=1$ when $z=0$. Incorporating this condition into Eq. \eqref{fqr} yields the following initial condition $f_0=6H_{0}^{2}\left(2-\Omega_{m_0}\right)$ \cite{Capozziello_2022_832_137229}.

\subsection{Power-law function}
\begin{figure}[ht]
    \centering
    \begin{subfigure}[b]{0.5\textwidth}
        \centering
        \includegraphics[width=75mm]{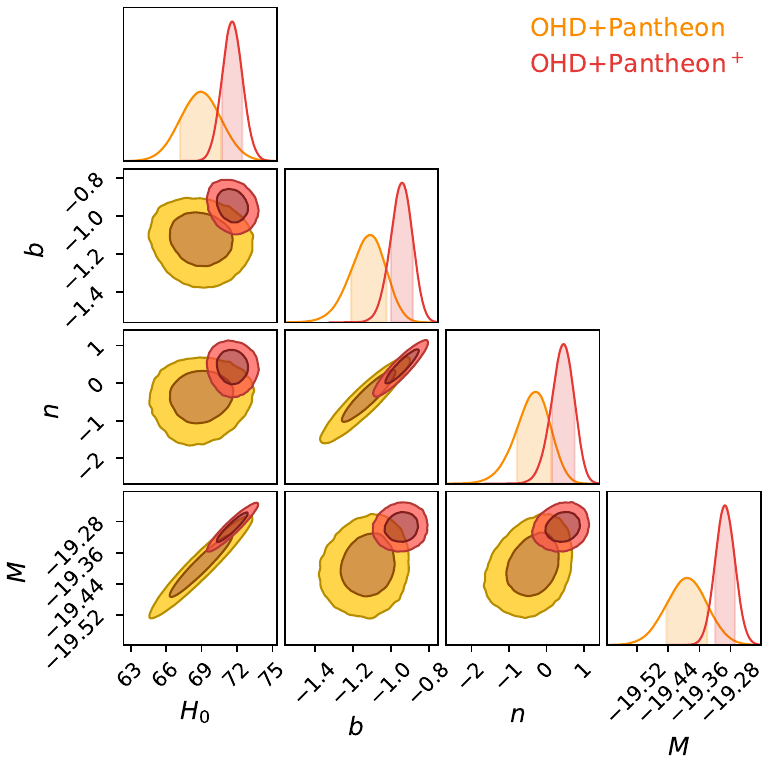}
        \caption{The yellow contours represent dataset combinations that include OHD+Pantheon, while the red contours show combinations that include the OHD+$\mathrm{Pantheon}^+$ datasets.}
        \label{Contour4.1}
    \end{subfigure}%
    \hfill\hfil
    \begin{subfigure}[b]{0.5\textwidth}
        \centering
        \includegraphics[width=75mm]{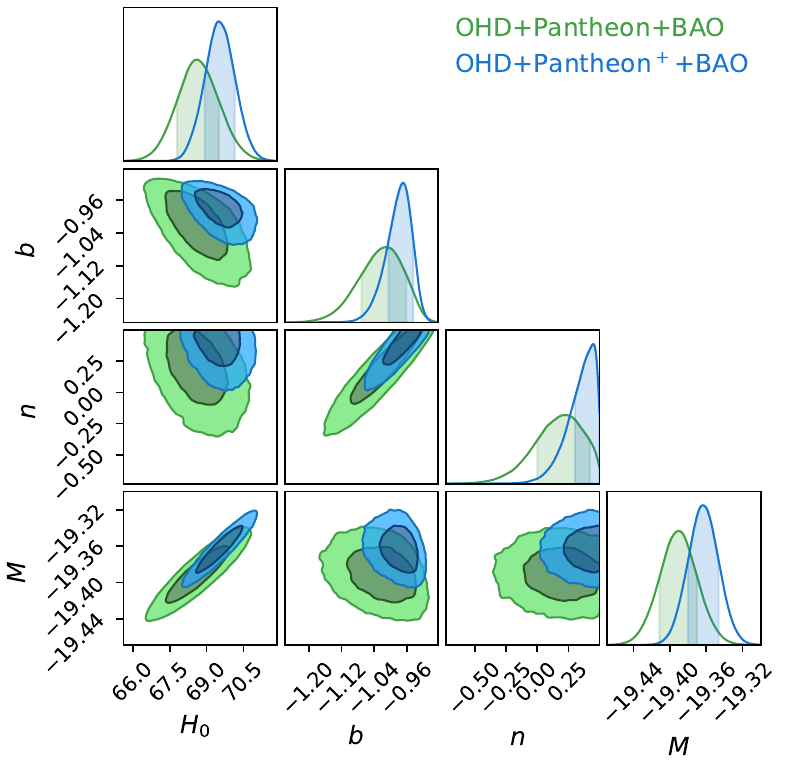}
        \caption{The green contours represent dataset combinations that include OHD+Pantheon+BAO, while the blue contours show combinations that include the OHD+$\mathrm{Pantheon}^+$+BAO datasets.}
        \label{Contour4.11}
    \end{subfigure}%
    \caption{Confidence contours and posteriors for the parameters $H_0$, $b$ and $n$ of the power-law function.}
    \label{fig4.1}
\end{figure}

The constraints on the free parameters for the power-law model has been displayed in Fig. \ref{fig4.1}. The figure \ref{Contour4.1} shows the posteriors as well as the confidence areas for various combination of OHD and SNe Ia datasets. It is clear from a deeper look at the posteriors that the parameters from the dataset combinations containing Pantheon$^+$ show tighter constraints, with the $H_0$ parameter demonstrating substantially improved precision. However, the contour plots for the OHD+Pantheon and OHD+Pantheon$^+$ dataset display a degeneracy between the $H_0$, $b$ and $n$. Notably, the OHD+$\mathrm{Pantheon}^+$ dataset combination exhibits a degeneracy between $H_0$ and the $\Omega_{m_0}$ parameter.

In Fig. \ref{Contour4.11} the contour plots for the combination of OHD, SNe Ia and BAO datasets are displayed. In the analysis of the OHD+Pantheon+BAO dataset combination, we observe that the constraints on the $H_0$, $b$ and $n$ parameters are more precise compared to those derived from the OHD+Pantheon dataset alone. The addition of BAO data helps break the degeneracy between these parameters, resulting in more robust constraints and tighter bounds on the $\Omega_{m_0}$ parameter. Also, the OHD+$\mathrm{Pantheon}^+$+BAO combination yields even more precise constraints, with excellent precision on $H_0$ while significantly minimizing the degeneracy between $H_0$, $b$ and $n$. As a result, the uncertainty associated with $\Omega_{m_0}$ is significantly reduced, making this combination the most precise across all parameters in the power-law model.

 \begin{table}[ht]
    \centering
    \renewcommand{\arraystretch}{1.8}
    \addtolength{\tabcolsep}{-1.0pt}
    {\small
    \begin{tabular}{|c|c|c|c|c|c|}
        \hline
		Dataset & $H_0$ & $b$ & $n$ & $\Omega_{m_{0}}$ & $M$\\ 
		\hline\hline
		OHD + Pantheon & $68.90^{+1.80}_{-1.70}$ & $-1.108^{+0.085}_{-0.101}$ & $-0.28^{+0.41}_{-0.50}$ & $0.324^{+0.036}_{-0.037}$ & $-19.391^{+0.051}_{-0.053}$ \\ \hline
		OHD + $\mathrm{Pantheon}^+$ & $71.62^{+0.84}_{-0.89}$ & $-0.941^{+0.055}_{-0.060}$ & $0.46^{+0.29}_{-0.31}$ & $0.26^{+0.05}_{-0.04}$ & $-19.294\pm 0.025$ \\ \hline
        OHD + Pantheon+BAO & $68.59^{+0.90}_{-0.80}$ & $-1.012^{+0.051}_{-0.059}$ & $0.22^{+0.20}_{-0.22}$ & $0.28^{+0.02}_{-0.03}$ & $-19.391\pm 0.021$ \\ \hline
		OHD + $\mathrm{Pantheon}^+$+BAO & $69.47^{+0.67}_{-0.55}$ & $-0.969^{+0.025}_{-0.036}$ & $0.46^{+0.04}_{-0.16}$ & $0.25^{+0.02}_{-0.01}$ & $-19.365\pm 0.020$ \\
		\hline
    \end{tabular}
    \caption{Constrained values of free parameters based on the different combinations of OHD, SNe Ia and BAO datasets for the power-law function.}
    \label{table4.1}
}
\end{table}

Table \ref{table4.1} displays the precise values for the cosmological and model parameters for power-law model, including the nuisance parameter $M$. For dataset combinations that contain $\mathrm{Pantheon}^+$, we found that the $H_0$ values are comparatively greater than the corresponding $H_0$ values that contain Pantheon data. According to the SH0ES team, $H_0 = 73.30 \pm 1.04 \,{\rm km\, s}^{-1} {\rm Mpc}^{-1}$ \cite{Riess_2022_934_L7} and $H_{0} = 73.24\pm 1.74\,{\rm km\, s}^{-1} {\rm Mpc}^{-1}$ \cite{Vagnozzi_2020_102_023518}, which is in accordance with the high value of $H_0$. The OHD+$\mathrm{Pantheon}^+$ yielded the highest values of $H_0$ as, $H_0 = 71.62^{+ 0.84}_{- 0.89} \,{\rm km\, s}^{-1} {\rm Mpc}^{-1}$. The $\Omega_{m_0}$ parameter interestingly reaches a minimal value in this scenario, suggesting that the majority of the energy of the Universe appears as an effective dark energy, consistent with the high value of $H_0$. For the dataset OHD + Pantheon + BAO and OHD + $\mathrm{Pantheon}^+$ + BAO, the constrained value of $H_0$ are marginally lower than those derived from the OHD + Pantheon and OHD + $\mathrm{Pantheon}^+$ combination. Specifically, the OHD + Pantheon + BAO dataset yields $H_0 = 68.59^{+0.90}_{-0.80} \,\mathrm{km\,s}^{-1}\,\mathrm{Mpc}^{-1}$, whereas the OHD + $\mathrm{Pantheon}^+$ + BAO configuration leads to $H_0 = 69.47^{+0.67}_{-0.55} \,\mathrm{km\,s}^{-1}\,\mathrm{Mpc}^{-1}$. Although these values are still in the high range compared to the CMB, which is approximately $67.4\pm0.5~\mathrm{km\,s}^{-1}\,\mathrm{Mpc}^{-1}$ based on Planck data \cite{Aghanim_2020_641_A6}, the inclusion of the BAO data imposes tighter constraints on the cosmological parameters, resulting in a slight reduction in the $H_0$ estimates. Additionally, the $\Omega_{m_0}$ parameter, while still on the lower end in both scenarios, shows a modest increase compared to the $\mathrm{Pantheon}^+$ analysis. This indicates that dark energy continues to be the dominant component of the Universe, while the inclusion of BAO data suggests a somewhat enhanced matter density contribution.

Using the Eq. \eqref{eq4.8} we can find
\begin{equation}\label{eq4.12}
    \frac{1}{H(z)}\frac{d}{dz}\left[H(z)\right] =\frac{3(1+z)^{3}(2b-n+3)-2bn(1+z)^{n}}{2\left(2b(1+z)^{n}+(1+z)^{3}(-2b+n-3)\right)}~.
\end{equation}
We determine $z(Q)$ by using Eq. \eqref{eq4.8} to invert $Q=6H^{2}$. The function $f(Q)$ is obtained by re-inserting $z(Q)$ into $f(z)$. Therefore, we find that the numerical solution is suitable for the function as,
\begin{equation}\label{eq4.13}
    f(Q) = Q+\alpha Q e^{\beta \frac{Q}{Q_{0}}}~,
\end{equation}
for the set of constant coefficients $(\alpha, ~\beta) = (2.460, ~0.191)$. \begin{figure}[ht]
    \centering
    \includegraphics[width=85mm]{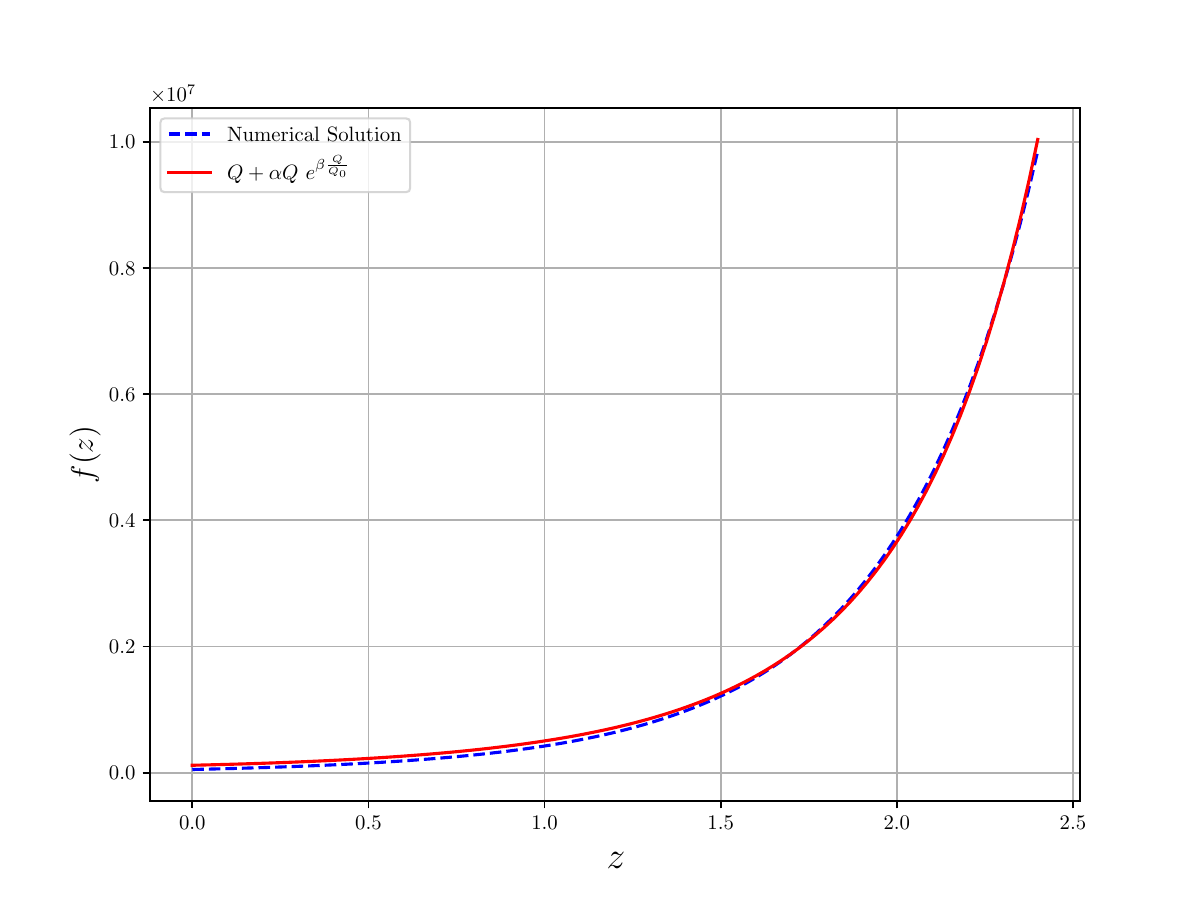}
    \caption{Reconstruction of $f(Q)$ model for power-law function. The best analytical matching (solid red) to the numerical solution (dashed blue).}
    \label{fig4.2}
\end{figure}
We display our results in Fig. \ref{fig4.2}. We note that the test function given in Eq. \eqref{eq4.13} recovers pure GR for $\alpha = 0$ and $\beta=0$. This reconstructed model has GR as a particular limit, it has the same number of free parameters as $\Lambda$CDM, but in a cosmological context it creates a scenario that does not have $\Lambda$CDM as a limit \cite{Anagnostopoulos_2021_822_136634}.

\subsection{Logarithmic function}
\begin{figure}[ht]
    \centering
    \begin{subfigure}[b]{0.5\textwidth}
        \centering
        \includegraphics[width=75mm]{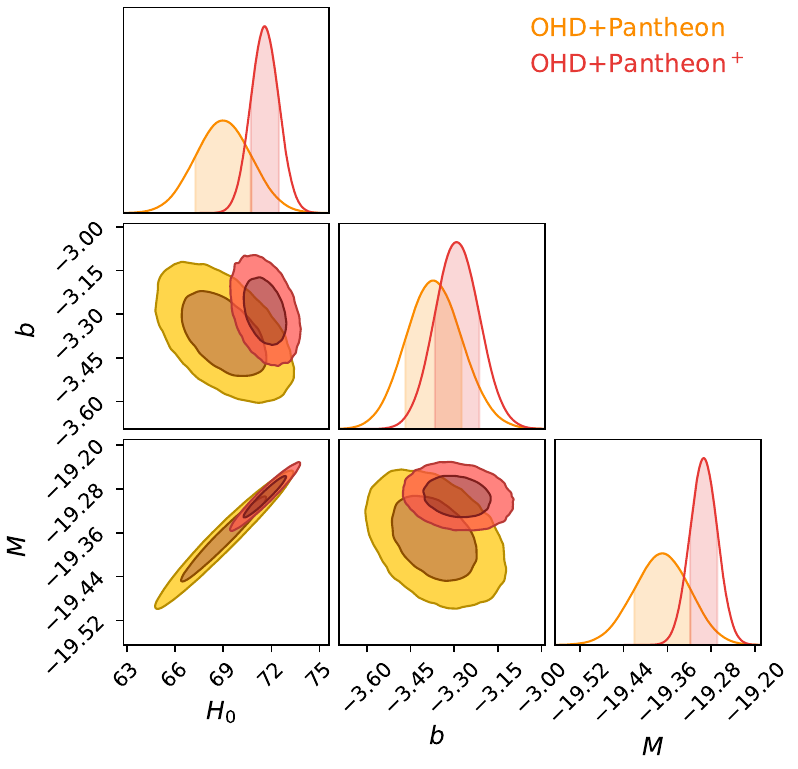}
        \caption{The yellow contours represent dataset combinations that include OHD+Pantheon, while the red contours show combinations that include the OHD+$\mathrm{Pantheon}^+$ datasets.}
        \label{Contour4.2}
    \end{subfigure}%
    \hfill\hfil
    \begin{subfigure}[b]{0.5\textwidth}
        \centering
        \includegraphics[width=75mm]{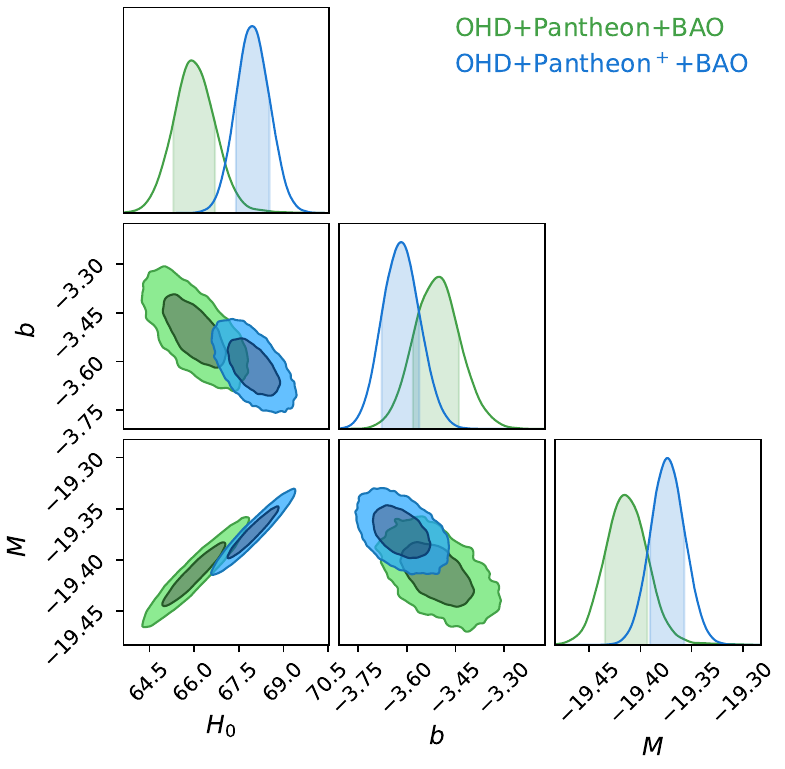}
        \caption{The green contours represent dataset combinations that include OHD+Pantheon+BAO, while the blue contours show combinations that include the OHD+$\mathrm{Pantheon}^+$+BAO datasets.}
        \label{Contour4.21}
    \end{subfigure}%
    \caption{Confidence contours and posteriors for the parameters $H_0$ and $b$ of the logarithmic function.}
    \label{fig4.3}
\end{figure}

Here, we have analyzed the observational constraints for logarithmic model given in Eq. \eqref{eq4.9}. The confidence levels and the posterior of the constrained parameters are shown in Fig. \ref{fig4.3}. The red contours show the combinations that consist of the OHD+Pantheon$^{+}$ samples, while the orange contours show the dataset combinations that include the OHD+Pantheon sample [Fig. \ref{Contour4.2}]. The precise numerical values of the parameters, including the nuisance parameter $M$, shown in Fig. \ref{fig4.3} are presented in Table \ref{table4.2}. These results indicate that the estimated $H_0$ values are comparable to those found in the power-law model. However, the inferred values of the matter density parameter $\Omega_{m_0}$ are somewhat lower than those of the power-law model since the logarithmic model is especially made to forecast an accelerating Universe at the late-time regime. In this case, the data constraints support a lower matter density to be consistent with the observed acceleration, while the model parameters allow for a more flexible description of the Universe.

Figure \ref{Contour4.21} displays the contour plots for the OHD+Pantheon and OHD+Pantheon$^{+}$ datasets, with the addition of BAO data. The inclusion of BAO data into the OHD+Pantheon and OHD+Pantheon$^{+}$ datasets offers additional constraints on the model parameters, resulting in minor adjustments to the estimated values. Specifically, for both dataset scenarios, the incorporation of BAO leads to a reduction in the estimated Hubble constant $H_0$, compared to analyses that exclude BAO. While the matter density parameter $\Omega_{m_0}$ remains nearly unchanged, but the BAO data helps refine the constraints on the other parameters, including the nuisance parameter. Overall, incorporating the BAO data enhances the precision of parameter estimates, thereby improving the capacity of the model to describe the late-time dynamics of the Universe. The BAO data incorporation yields the lowest value of $\Omega_{m_0}$ as $\Omega_{m_0} = 0.24$.
\begin{table}[ht]
    \centering
    \renewcommand{\arraystretch}{1.8}
    \begin{tabular}{|c|c|c|c|c|}
        \hline
		Dataset & $H_0$ & $b$ & $\Omega_{m_{0}}$ & $M$\\ 
		\hline\hline
		OHD + Pantheon & $69.00^{+1.80}_{-1.70}$ & $-3.372^{+0.098}_{-0.095}$ & $0.25\pm 0.02$ & $-19.367^{+0.049}_{-0.053}$ \\ \hline
		OHD + $\mathrm{Pantheon}^+$ & $71.57^{+0.90}_{-0.85}$ & $-3.293^{+0.080}_{-0.073}$ & $0.27\pm 0.02$ & $-19.294\pm 0.025$ \\ \hline
        OHD + Pantheon+BAO & $65.88^{+0.82}_{-0.58}$ & $-3.500^{+0.061}_{-0.081}$ & $0.24^{+0.01}_{-0.02}$ & $-19.416\pm 0.020$ \\ \hline
		OHD + $\mathrm{Pantheon}^+$+BAO & $67.98^{+0.55}_{-0.58}$ & $-3.617^{+0.056}_{-0.060}$ & $0.23^{+0.02}_{-0.02}$ & $-19.374\pm 0.017$ \\
		\hline
    \end{tabular}
    \caption{Constrained values of free parameters based on the different combinations of OHD, SNe Ia and BAO datasets for the logarithmic function.}
    \label{table4.2}
\end{table}

\begin{figure}[ht]
    \centering
    \includegraphics[width=85mm]{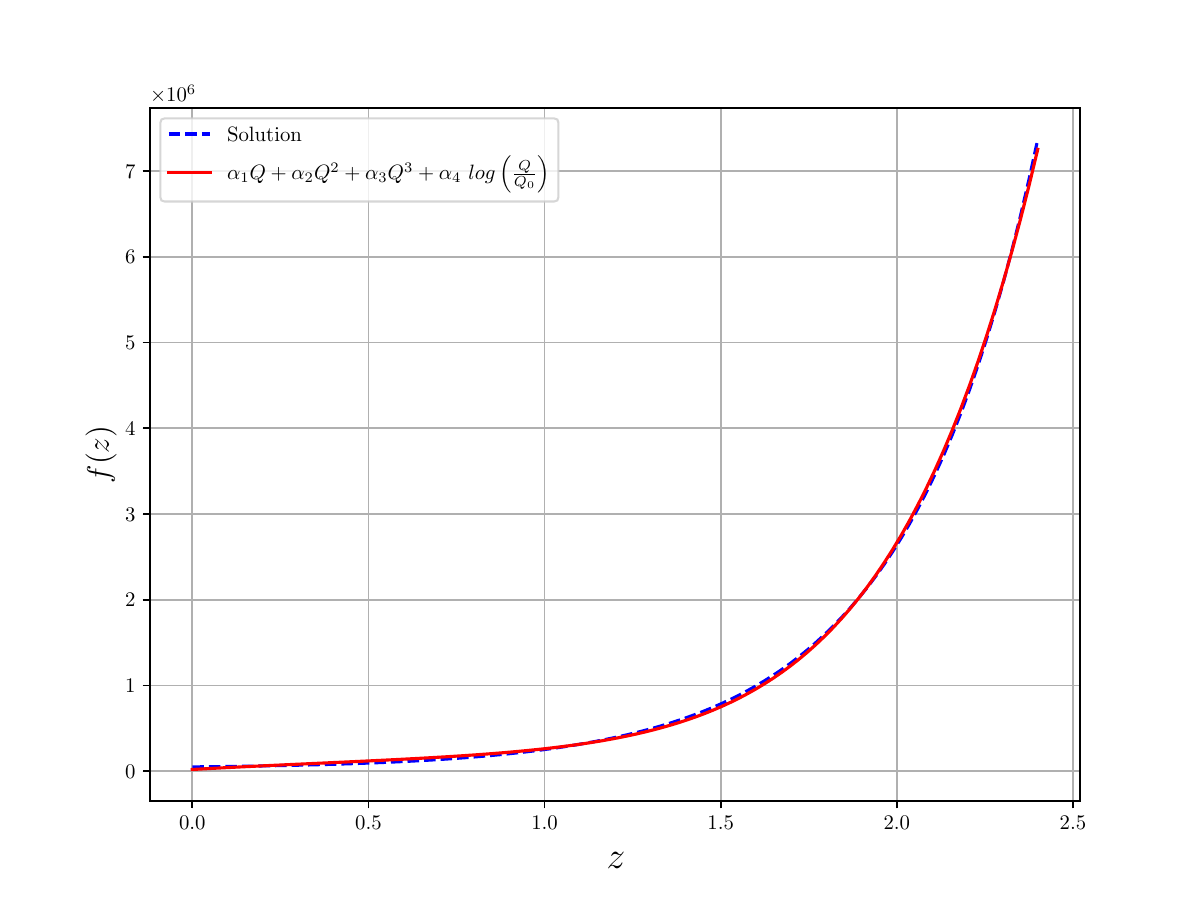}
    \caption{Reconstruction of $f(Q)$ model for logarithmic function. The best analytical match (solid red) to the numerical solution (dashed blue).}
    \label{fig4.4}
\end{figure}

We have reconstructed the $f(Q)$ for logarithmic function in a similar manner as power-law function. Subsequently, we obtain the most suitable $f(Q)$ function as,
\begin{equation}\label{eq4.15}
    f(Q) = \alpha_{1}Q+\alpha_{2}Q^{2}+\alpha_{3}Q^{3}+\alpha_{4} ~log\left(\frac{Q}{Q_0}\right),
\end{equation}
with the values of coefficients, $(\alpha_{1}, \alpha_{2}, \alpha_{3}, \alpha_{4}) = (1.196, ~-2.22\times10^{-5}, ~2.03\times10^{-10}, ~1.24\times10^{5})$. The results are shown in Fig. \ref{fig4.4}. We note that the test function [Eq. \eqref{eq4.15}] returns to GR when $(\alpha_{1}, ~\alpha_{2}, ~\alpha_{3}, ~\alpha_{4}) = (1,~0, ~0, ~0)$.
The Eq. \eqref{eq4.15} is presented as a novel approach for describing cosmic evolution, particularly focusing on avoiding early-time instabilities and phantom behaviors. In this model, the effective EoS parameter remains strictly non-phantom. As shown in \cite{Bamba_2011_2011_021}, it can be combined with an exponential and cross the phantom divide. Notably, this model passes the Big Bang Nucleosynthesis constraints trivially, meaning it does not interfere with the delicate balance required for early Universe nucleosynthesis, making it a stable and potentially viable model for both early and late Universe dynamics \cite{Anagnostopoulos_2023_83_58}.\\
The logarithmic model, offering a more flexible description of late-time acceleration and a lower matter density, serves as an alternative to the power-law function. In the next section, we will conduct a detailed comparison of both models, evaluating their cosmological implications and how well they fit the available observational data.

\subsection{Model comparison}
We assess the performance of each model and dataset by calculating its minimal $\chi^2_\mathrm{min}$ values, which are derived from the maximum likelihood $L_\mathrm{max}$ since $\chi^2_\mathrm{min} = -2 \ln L_\mathrm{max} $. Additionally, we use the Akaike Information Criteria (AIC), a measure of the goodness of fit and the complexity of the model based on the number of parameters $n$, to compare these models with the typical $\Lambda$CDM. The AIC is defined as \cite{Akaike_1974_19_716}, 
\begin{equation}\label{eq4.16}
    \mathrm{AIC} = \chi^2_\mathrm{min} + 2n \,.    
\end{equation}
When the AIC value is lower, the model fits the data more effectively while taking its complexity into account. Models with more parameters are penalized by the AIC, even if they improve the fit. This means that, provided the difference is significant, a model with a lower AIC is preferable to one with a higher value. As well as the AIC, we examine the Bayesian Information Criterion (BIC), which places a greater emphasis on model complexity than the AIC and is defined as \cite{Schwarz_1978_6_461},
\begin{equation}\label{eq4.17}
\mathrm{BIC} = \chi^2_\mathrm{min} + n \ln m \,,
\end{equation}
where $m$ is the sample size of the observational data combination. 

We determine the differences in AIC and BIC relative to the $\Lambda$CDM model as a reference, in order to assess the performance of different models with different dataset combinations. $\Delta \mathrm{AIC}=\Delta\chi^{2}_\mathrm{min}+2\Delta n$ and $\Delta\mathrm{BIC} = \Delta\chi^2_\mathrm{min} + \Delta n \ln m$, respectively represent the differences between AIC and BIC. Better performance is suggested by smaller $\Delta$AIC and $\Delta$BIC values, which show that the model utilizing the chosen dataset is more comparable to the $\Lambda$CDM model. The values for several statistical measures, including $\chi^2_{\mathrm{min}}$, AIC, $\Delta$AIC, BIC and $\Delta$BIC, are presented in Table \ref{table4.3} - \ref{table4.4} for both models that incorporate various combinations of OHD, SNe Ia and BAO data.

\begin{table}[H]
\centering
\renewcommand{\arraystretch}{1.8}
\addtolength{\tabcolsep}{-1.3pt}
{\small
\begin{tabular}{|c|c|c|c|c|c|c|c|c|c|c|}
\hline
    Model & \multicolumn{5}{c|}{OHD + Pantheon}& \multicolumn{5}{c|}{OHD+ $\mathrm{Pantheon}^+$}\\
    \cline{2-11}
     &$\chi^2_{\mathrm{min}}$ & AIC & $\Delta$AIC & BIC & $\Delta$BIC & $\chi^2_{\mathrm{min}}$ & AIC & $\Delta$AIC & BIC &$\Delta$BIC \\
     \hline\hline
    $\Lambda$CDM & 1041.16 & 1047.16 & 0 & 1050.63 & 0 & 1539.22 & 1545.22 & 0 &  1548.93 & 0   \\ \hline
    \textbf{Model I}  & 1040.76 & 1048.76 & 1.60 & 1052.90 & 2.27 & 1552.82 & 1560.82 & 15.60 &  1565.78 & 16.85 \\ \hline
    \textbf{Model II}   & 1043.90 & 1049.90 & 2.74 & 1053.00 & 2.37 & 1552.82 & 1558.82 & 13.60 &  1562.53 & 13.60 \\ \hline
\end{tabular}
    \caption{The comparison of the models with the $\Lambda$CDM model for OHD+Pantheon and OHD+$\mathrm{Pantheon}^+$ datasets.}
    \label{table4.3}
    }
\end{table}

\begin{table}[ht]
\centering
\renewcommand{\arraystretch}{1.8}
\addtolength{\tabcolsep}{-1.3pt}
{\small
\begin{tabular}{|c|c|c|c|c|c|c|c|c|c|c|}
\hline
    Model & \multicolumn{5}{c|}{OHD + Pantheon + BAO}& \multicolumn{5}{c|}{OHD+ $\mathrm{Pantheon}^+$ + BAO}\\
    \cline{2-11}
     &$\chi^2_{\mathrm{min}}$ & AIC & $\Delta$AIC & BIC & $\Delta$BIC & $\chi^2_{\mathrm{min}}$ & AIC & $\Delta$AIC & BIC &$\Delta$BIC \\
     \hline\hline
    $\Lambda$CDM & 1046.40 & 1052.40 & 0 & 1056.13 & 0 & 1576.10 & 1582.10 & 0 &  1585.83 & 0   \\ \hline
    \textbf{Model I} & 1045.26 & 1053.26 & 0.86 & 1057.42 & 1.29 & 1568.90 & 1576.90 & -5.2 &  1581.03 & -5.8 \\ \hline
    \textbf{Model II}   & 1063.31 & 1069.31 & 16.91 & 1073.03 & 16.90 & 1596.82 & 1602.82 & 20.72 & 1606.53 &  20.70\\ \hline
\end{tabular}
    \caption{The comparison of the models with the $\Lambda$CDM model for OHD+Pantheon+BAO and OHD+$\mathrm{Pantheon}^+$+BAO datasets.}\label{table4.4}
}
\end{table}

Several studies have explored the reconstruction of the $f(Q)$ model. For instance, the numerical approximation of the $f(Q)$ model to a polynomial form using the luminosity distance and P\'ade approximation was conducted by \cite{Capozziello_2022_832_137229}. Additionally, \cite{Singh_2023_41_101240} examined various modified gravity models for reconstructing the $f(Q)$ function. In contrast, our approach utilizes a parametrized deceleration parameter for reconstructing the $f(Q)$ models, resulting in exponential and logarithmic approximations. Notably, the exponential model has been rigorously tested against observational data in \cite{Anagnostopoulos_2021_822_136634}, while the logarithmic model has been studied in the context of Big Bang Nucleosynthesis in \cite{Anagnostopoulos_2023_83_58}.

\section{Conclusion}\label{sec4.5}
In this chapter, we studied the impact of the deceleration parameter on the modified $f(Q)$ gravity models. An important step in cosmological analysis is reconstructing a parametric form for $q(z)$. To validate the parametrization and ensure its predictive capability across all redshift ranges, it is essential to approach the process with careful consideration of both observational constraints and theoretical frameworks. Consequently, robust investigations of the dynamics of the Universe are enabled and modified gravity theories can be explored. A new parametrization for the deceleration parameter has been presented in this chapter, providing a more flexible way of studying the dynamics of the Universe. As a result of the utilization of observational datasets including OHD, SNe Ia and BAO, cosmological parameters were constrained and the proposed models were validated. With the use of Bayesian statistical inference techniques and MCMC methods, we were able to accurately analyze the data and draw meaningful conclusions from the analysis. Based on these fitting results, we analyze the kinematic behavior of the Universe. Our results demonstrate that the parameters of our models match the observational data well.

The model I, which follows a power-law formulation, demonstrates compatibility with the latest high values of $H_{0}$, aligning with recent observational values and indicating a dark energy-dominated Universe. Moreover the reconstructed $f(Q)$ model has GR as a particular limit, it has the same number of free parameters as $\Lambda$CDM, but in a cosmological context it creates a scenario that does not have $\Lambda$CDM as a limit. Whereas, the model II, featuring a logarithmic term, aligns with a slightly reduced matter density and accurately models the accelerated expansion. The reconstructed $f(Q)$ function for model II is presented as a novel approach for describing cosmic evolution, particularly focusing on avoiding early-time instabilities and phantom behaviors. In this model, the effective equation of state parameter remains strictly non-phantom. Also, this model passes the Big Bang Nucleosynthesis constraints trivially, meaning it does not interfere with the delicate balance required for early Universe nucleosynthesis, making it a stable and potentially viable model for both early and late Universe dynamics. Finally the comparative analysis against $\Lambda$CDM using AIC and BIC metrics shows that these models provide competitive fits, with Model II offering a slight edge in simplicity. Thus, the reconstructed $f(Q)$ functions from deceleration parameter in the $f(Q)$ gravity presents a robust framework for addressing dark energy and cosmic acceleration.
\chapter{Stable $f(Q)$ gravity model through trivial and non-trivial connection} 

\label{Chapter5} 

\lhead{Chapter 5. \emph{Stable $f(Q)$ gravity model through trivial and non-trivial connection}} 

\vspace{10 cm}
* The work, in this chapter, is covered by the following publications: \\

\textbf{S. A. Narawade}, Santosh V Lohakare and B. Mishra, ``Stable $f(Q)$ gravity model through non-trivial connection", \textit{Annals Phys.} \textbf{474} (2025) 169913.\\[2mm]
\textbf{S. A. Narawade}, S. H. Shekh, B. Mishra, Wompherdeiki Khyllep and Jibitesh Dutta, ``Modelling the Accelerating Universe with $f(Q)$ Gravity: Observational Consistency", \textit{Eur. Phys. J. C}, \textbf{84}, 773 (2024).

\clearpage

\section{Introduction}\label{Sec:I}
In symmetric teleparallel gravity, the choice of connection is crucial as it determines the nonmetricity scalar $Q$, this impacts the equations of motion and influences the cosmological dynamics. Various connections provide unique insights into the relationship between geometry and cosmic evolution. For example, Connection I corresponds to coincident gauge choices, ensures a more straightforward dynamic formulation. In contrast, Connection II and Connection III introduce more complex dynamical characteristics, allowing for a deeper investigation into cosmological phenomena. Recently, there has been considerable attention on $f(Q)$ gravity and its cosmological implications in several important studies, see \cite{Lazkoz_2019_100_104027, Lu_2019_79_530, Jimenez_2020_101_103507, Barros_2020_30_100616, Frusciante_2021_103_044021, Anagnostopoulos_2021_822_136634, Khyllep_2021_103_103521, Lin_2021_103_124001, D'Ambrosio_2022_105_024042, Capozziello_2022_82_865, Capozziello_2023_83_915, Jensko_2024_2407.17568}. It is worth noting that all of these studies were conducted with coincident gauges and line elements in Cartesian coordinates. Due to this specific choice, the covariant derivative is reduced to a partial derivative, which simplifies the calculations. On the other hand, the pressure and energy equations are identical to the $f(T)$ theory. In cosmological context, whether in flat \cite{Subramaniam_2023_71_2300038, Shabani_2023_83_535, Paliathanasis_2023_41_101255} or curved \cite{Dimakis_2022_106_043509, Heisenberg_2023_83_315, Shabani_2023_84_285, Subramaniam_2023_41_101243} Universe, the $f(Q)$ theory, which does not rely on coincident gauges, is garnering significant interest.

In the previous chapter, we analyzed the reconstruction of the gravity model $f(Q)$ with the help of the deceleration parameter within the framework of coincident gauge. Now, in this chapter, we explores the reconstruction of the $f(Q)$ gravity model using NEC within the framework of non-trivial connection. A covariant $f(Q)$ gravity within Connection-III \cite{Zhao_2022_82_303} and FLRW metric is utilized to reconstruct the $f(Q)$ model. The dynamic behavior of two models, particularly model within coincident gauge and the reconstructed model are thoroughly studied using the Hubble parameter and various observational datasets for the comparative analysis. Also the energy conditions are analyzed, because of the fundamental casual structure of space-time, the gravitation attraction can be characterized by the energy conditions. Moreover, these boundary conditions play a significant role in shaping the cosmic evolution of the Universe. Furthermore, the stability of the model is achieved through the scalar perturbation. To explore the potential impact of nonmetricity on the evolution of the Universe, we constructed an $f(Q)$ model that incorporates nontrivial connections. By comparing this model to the $\Lambda$CDM paradigm, the goal is to assess the role of nonmetricity in cosmic dynamics. The structure of the chapter is as follows: In section \ref{sec5.2}, we applying specific conditions along with connection-III, we reconstruct the $f(Q)$ model. The section \ref{sec5.3} discusses the Hubble parameter, along with the observational datasets and the results for parameter constrains. The dynamical behaviour of the two models, particularly model within coincident gauge and model reconstructed through non-coincident gauge are studied for the comparative analysis in section \ref{sec5.4} and also the stability of the model reconstructed through non-coincident gauge has bedd studied using scalar perturbation. We conclude with a general discussion of our results in section \ref{sec5.5}.

\section{Reconstruction of the $f(Q)$ model}\label{sec5.2}
The Connection I corresponds to coincident gauge choices, it simplifies the dynamical formulation. While Connection II and Connection III share a similar structure, the function $\gamma$ which acts as an additional degree of freedom, behaves differently in each case \cite{Guzman_2024_110_124013}. The $\frac{1}{a^2}$ factor in Connection III plays a crucial role by suppressing the contributions of $\gamma$ and $\dot{\gamma}$ at late times, ensuring a smoother cosmological evolution and reducing the risk of sudden singularities compared to Connection II. This scaling enables a clear distinction between early and late-time dynamics, making the model more stable and better aligned with observational data. Thus we will focus on model within coincident gauge and Connection III in this chapter. By employing $\gamma = \gamma_{1}a^{2}(t)$ \cite{Subramaniam_2023_71_2300038}, where $\gamma_{1}$ is arbitrary constant. We derive $Q=-6H^{2}+9\gamma_{1}H$, which leads to $\dot{Q} = -12H\dot{H}+9\gamma_{1}\dot{H}$. Consequently, Eqs. \eqref{connection3r} and \eqref{connection3p} are transformed as,
\begin{eqnarray}
    \rho &=& \frac{f}{2}+3H^{2}f_{Q}-\frac{1}{2}Qf_{Q}-\frac{3\gamma_{1}}{2}\dot{Q}f_{QQ}~,\label{eq5.1}\\
    p &=& -\frac{f}{2}-3H^{2}f_{Q}+\frac{1}{2}Qf_{Q}-2\dot{H}f_{Q}+\frac{(\gamma_{1}-4H)}{2}\dot{Q}f_{QQ}~.    \label{eq5.2}\\
\end{eqnarray}
From Eqs. \eqref{eq5.1} and \eqref{eq5.2}, we derive the expression
\begin{equation}\label{eq22}
    \rho + p = \left(-\gamma_{1}\dot{Q}f_{QQ}-2\dot{H}f_{Q}-2H\dot{Q}f_{QQ}\right)~.
\end{equation}
By applying the chain rule to $f_{Q}$ and considering the condition $p+\rho\rightarrow 0$, we obtain
\begin{eqnarray*}
    -\frac{d}{dt}\left(\gamma_{1}f_{Q}\right) &=&  \frac{d}{dt}\left(2Hf_{Q}\right),\\[5pt]
    \frac{df}{dH} &=& \frac{C(4H-3\gamma_{1})}{2H+\gamma_{1}}.
\end{eqnarray*}
Upon integrating both sides, we find
\begin{eqnarray*}
    f(H) &=& 4H-5\gamma_{1}\,\text{ln}(2H+\gamma_{1})~.
\end{eqnarray*}
By applying these results up to the $4^{th}$ order and substituting the value of $Q$, we can reconstruct $f(Q)$ as,
\begin{equation}\label{eq5.4}
    f(Q) = \alpha_{1}+\alpha_{2}Q+\alpha_{3}Q^{2}+\left(\beta_{1}+\beta_{2}Q\right)\sqrt{81\gamma_{1}^{2}-24Q}~.
\end{equation}
The reconstruction of the $f(Q)$ model provides a theoretical framework to explore the underlying dynamics of the Universe. In the following section, we will analyze observational data to estimate the free parameters of the Hubble parameter.

\section{Data analysis and parameter estimation}\label{sec5.3}
\begin{figure}[ht]
    \centering
    \begin{subfigure}{0.5\textwidth}
        \centering
        \includegraphics[width=75mm]{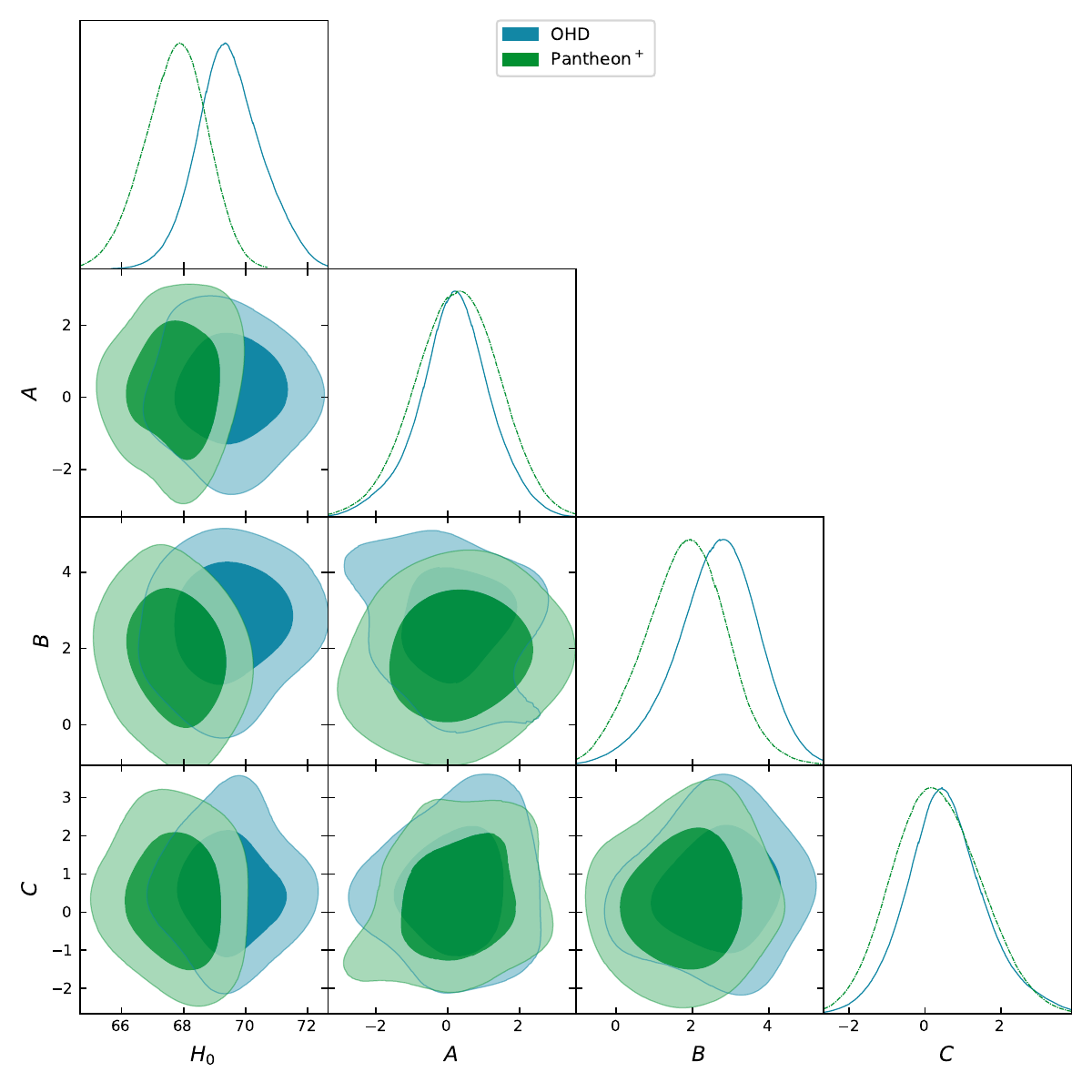}
        \caption{Contour plot obtained from OHD and Pantheon$^+$ dataset.}
        \label{HP}
    \end{subfigure}%
    \hfill
    \begin{subfigure}{0.5\textwidth}
        \centering
        \includegraphics[width=75mm]{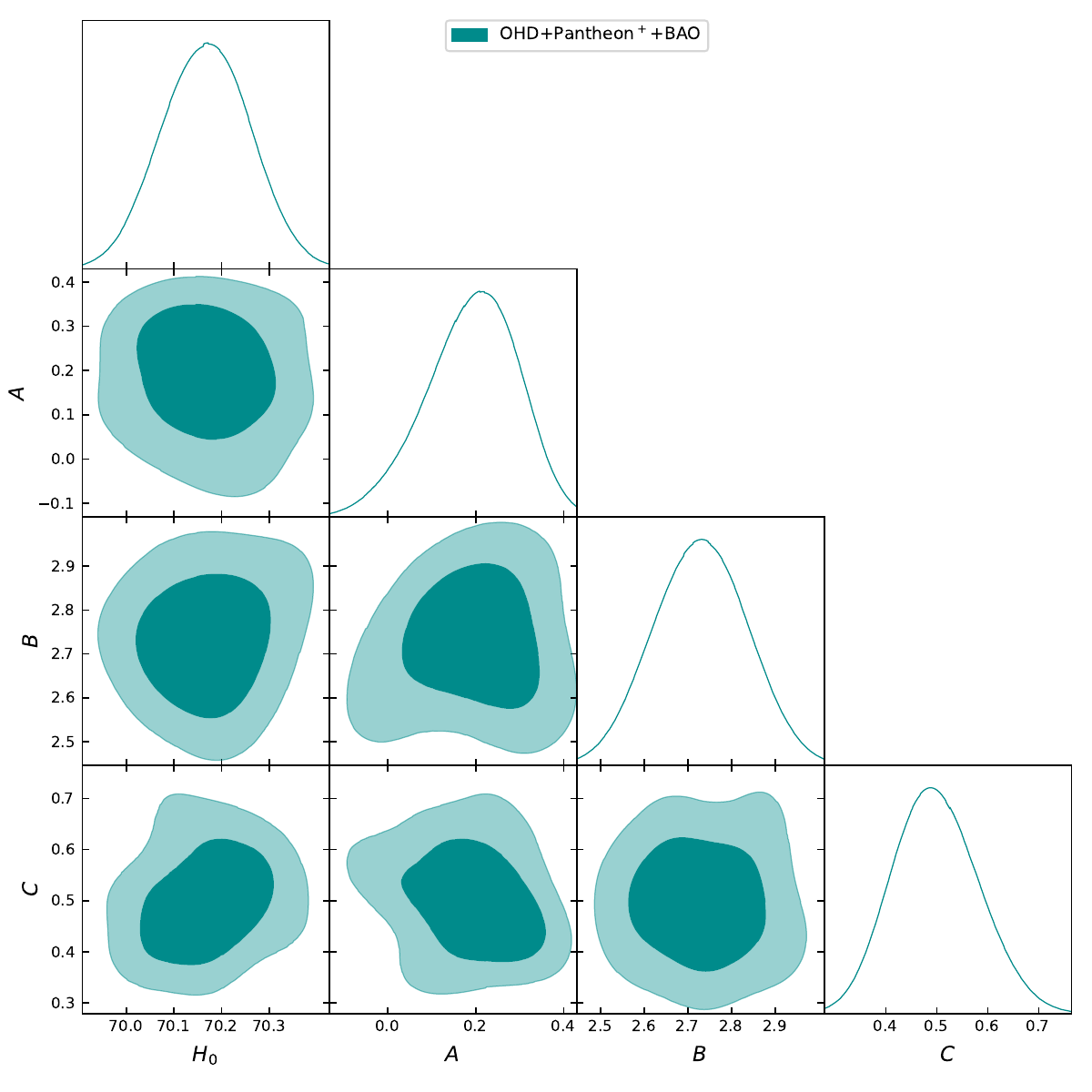}
        \caption{Contour plot obtained from $\text{OHD}+\text{Pantheon}^{+}+\text{BAO}$ dataset.}
        \label{HPB}
    \end{subfigure}%
    \caption{MCMC contour plot obtained from the observational datasets for 1$\sigma$ and 2$\sigma$ confidence interval.}
    \label{fig5.1}
\end{figure}

In this section, we analyze the behavior of the $f(Q)$ model by utilizing the functional form of the Hubble parameter given by $H(z)^{2} = H_{0}^2\left[A(1+z)^{B}+\sqrt{A^{2}(1+z)^{2B}+C}\right]$ \cite{Sahni_2003_2003_014}. We will conduct MCMC analysis using the \textit{emcee} package described in section \ref{sec3.3.3}, to constrain the free parameters 
$H_{0}$, $A$, $B$ and $C$ based on observational datasets. Our analysis will use Hubble data, Pantheon$^{+}$ data and BAO data as the baseline discussed in the section \ref{sec3.3} and \ref{sec4.3}.\\
The free parameters $H_{0}$, $A$, $B$ and $C$ are constrained using the OHD, Pantheon$^+$ and $\text{OHD}+\text{Pantheon}^{+}+\text{BAO}$ datasets [Fig. \ref{fig5.1}] and are given in Table \ref{table5.1}.\vspace{2mm}

 \begin{table}[ht]
		\begin{center}
  \renewcommand{\arraystretch}{1.8}
			\begin{tabular}{|c|c|c|c|c|}
				\hline
				Dataset & $H_{0}$ & $A$ & $B$ & $C$ \\[0.1cm] 
				\hline\hline  
				OHD & $69.49\pm 1.10$ & $0.19\pm 0.99$ & $2.66^{+1.10}_{-0.97}$ & $0.55^{+0.95}_{-1.10}$\\\hline
                Pantheon$^+$ & $69.5^{+2.10}_{-1.80}$ & $0.20^{+2.00}_{-2.20}$ & $2.70\pm 2.10$ & $0.5^{+2.20}_{-2.10}$\\\hline
                $\text{OHD}+\text{Pantheon}^{+}+\text{BAO}$ & $70.16^{+1.00}_{-0.90}$ & $0.19^{+0.15}_{-0.13}$ & $2.73\pm 0.10$ & $0.49^{+0.08}_{-1.00}$\\[0.1cm]
				\hline
			\end{tabular}
		\caption{Best-fit values of parameter space using OHD, Pantheon$^+$ and $\text{OHD}+\text{Pantheon}^{+}+\text{BAO}$ dataset.}\label{table5.1}
		\end{center}
	\end{table}
The estimated value of the Hubble constant $H_{0}$ from our analysis, using the OHD, Pantheon$^+$ and the combined $\text{OHD} + \text{Pantheon}^{+} + \text{BAO}$ datasets are respectively $69.49$~km/s/Mpc, $69.50$~km/s/Mpc and $70.16$~km/s/Mpc, are consistent with current local measurements, with a value around $70.4\pm1.4$~km/s/Mpc \cite{Jarosik_2011_192_14}. This result is in mild tension with the value inferred from the CMB, which is approximately $67.4$~km/s/Mpc based on Planck data \cite{Aghanim_2020_641_A6}. The slight variations observed in our estimates reflect the impact of different datasets and highlight the ongoing challenges in resolving the Hubble tension, a key issue in modern cosmology.

In this section, we have reconstructed the cosmological model within the $f(Q)$ gravity framework and constrained the free parameters for Hubble parameter using MCMC analysis using different observational datasets. These constrained parameters and the $H(z)$ will be utilized in the next section to study the dynamical aspects of this reconstructed model by analyzing the time evolution of energy density and the equation of state, along with testing stability under perturbations. This analysis is essential for understanding the physical validity and long-term behavior of the model, ensuring that the reconstructed model align with observational data and remain stable over cosmic evolution.

\section{Comparative analysis}\label{sec5.4}
In this section, we will analyze the dynamic behaviour of two different cosmological models utilizing the constrained free parameters for $H(z)$ from the various observational datasets. The equation for $\dot{H}$ is provided in the context of redshift $z$ as,
\begin{equation}\label{eq5.5}
    \dot{H} = -\frac{1}{2}ABH_{0}^2 (1+z)^B \left(1+\frac{A (1+z)^B}{\sqrt{A^2 (1+z)^{2 B}+C}}\right)~.
\end{equation}

\subsection{Model within coincident gauge}
\begin{figure}[H]
    \centering
    \begin{subfigure}{0.5\textwidth}
        \centering
        \includegraphics[width=76mm]{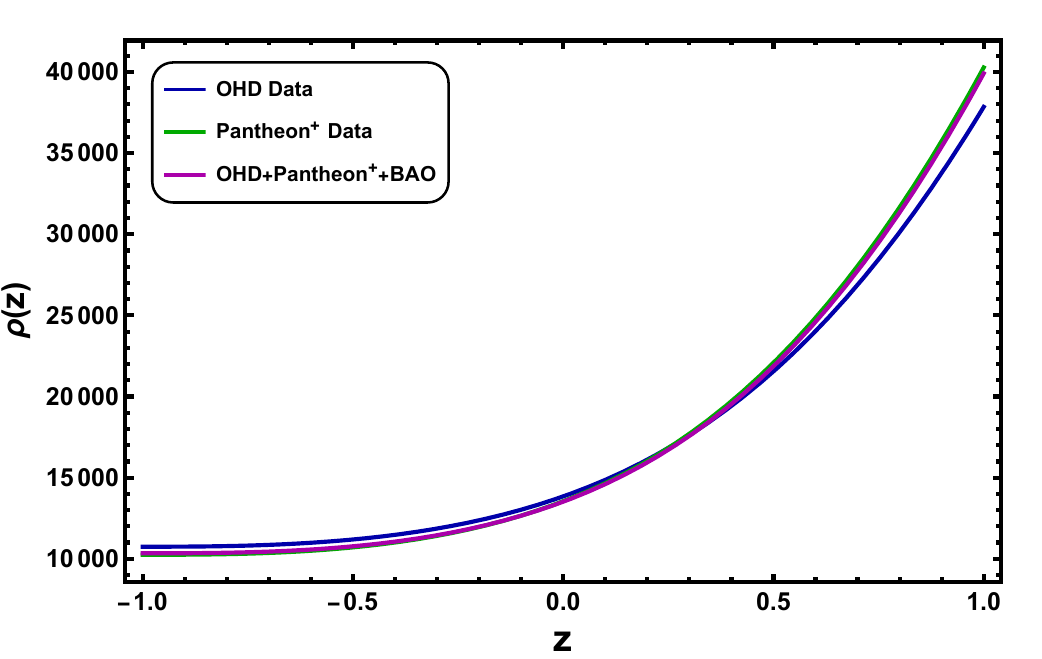}
        \caption{Evolution of energy density in redshift.}
        \label{rho11}
    \end{subfigure}%
    \hfill
    \begin{subfigure}{0.5\textwidth}
        \centering
        \includegraphics[width=75mm]{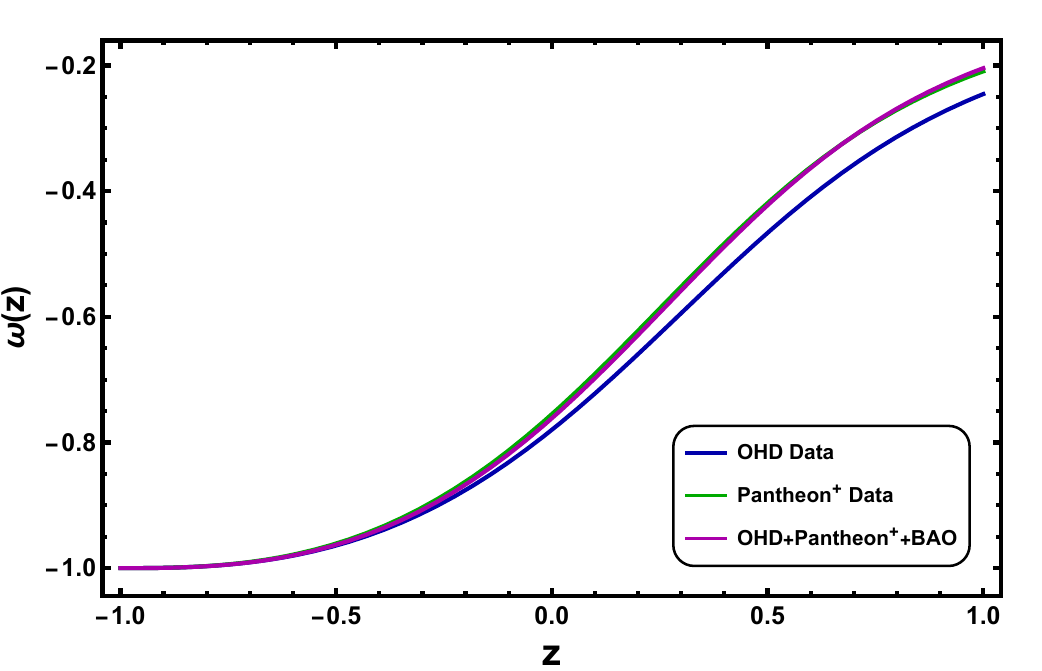}
        \caption{Evolution of EoS parameter in redshift.}
        \label{omega11}
    \end{subfigure}%
    \caption{The behavior of the dynamical parameters for model within coincident gauge.}
    \label{fig5.2}
\end{figure}

To study the background evolution of the dynamical parameter, we consider a well-motivated form $f(Q)$ given by \cite{Harko_2018_98_084043},
\begin{equation}\label{eq5.6}
f(Q)= \frac{\alpha Q}{Q_{0}} + \frac{\beta Q_{0}}{Q}~,
\end{equation}
where $Q_{0}=6H_{0}^{2}$ ($H_{0}$ is present Hubble value). This form is useful in describing the late time acceleration without invoking the DE component \cite{Jimenez_2020_101_103507}. Using the Hubble parameter and Eq. \eqref{eq5.6}, Eqs. \eqref{fqr} - \eqref{fqp} can be written as,
{\small
\begin{align}
\rho &= H(z)^2 \left(3-\frac{\alpha}{2H_{0}^2}\right)+\frac{3\beta H_{0}^2}{2 H(z)^2}+(1+z)^3 (\Omega_{m}+\Omega_{r}z+\Omega_{r})~,\label{eq5.7}\\[10pt]
p &= \frac{\alpha \left(3H(z)^6+2H(z)^4 H_{z}\right)+H_{0}^2 \left(-9\beta H(z)^{2}H_{0}^2+6\beta H_{0}^2 H_{z}-18 H(z)^6+2H(z)^4 \left(\Omega_{r}(1+z)^4-6H_{z}\right)\right)}{6H(z)^{4}H_{0}^2}~.\label{eq5.8}    
\end{align}
}
From Fig. \ref{fig5.2}, we can observe that the effect of the radiation term $\rho_{r}$ reflects here and also the energy density decreases from early times to late times but does not vanish. The total EoS parameter yields present values of $\omega_{0} \simeq -0.78$, $\omega_{0} \simeq -0.77$ and $\omega_{0} \simeq -0.76$ for the constrained free parameters from the OHD dataset, Pantheon$^{+}$ dataset and the combined OHD+Pantheon$^{+}$+BAO dataset, respectively [Fig. \ref{omega11}]. The energy density and EoS parameter are solely dependent on the values of the model coefficients, which govern the evolutionary behavior of the parameters. This model shows the quintessence behavior at present time \cite{Koussour_2022_31_2250115}.
In the study of cosmological models within modified gravity theories, examining energy conditions are crucial. Specifically, in the context of covariant $f(Q)$ gravity, we investigate the behavior of various energy conditions. The energy density must remain positive throughout cosmic evolution
\begin{figure}[H]
    \centering
    \begin{subfigure}{0.5\textwidth}
        \centering
        \includegraphics[width=75mm]{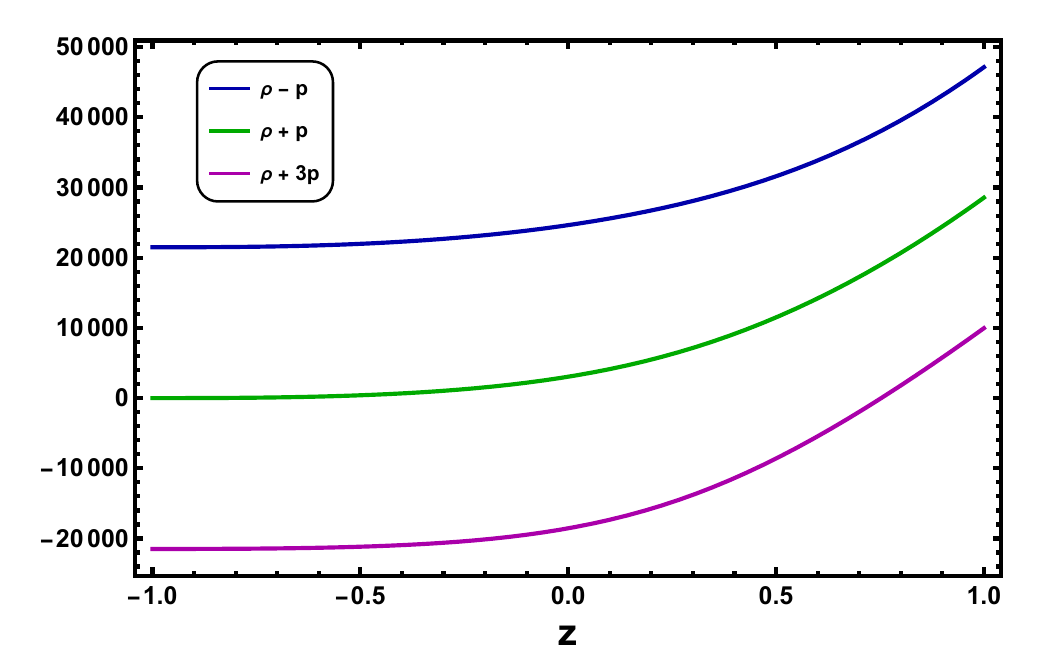}
        \caption{Energy conditions in redshift for OHD data.}
        \label{ec11}
    \end{subfigure}%
    \hfill
    \begin{subfigure}{0.5\textwidth}
        \centering
        \includegraphics[width=75mm]{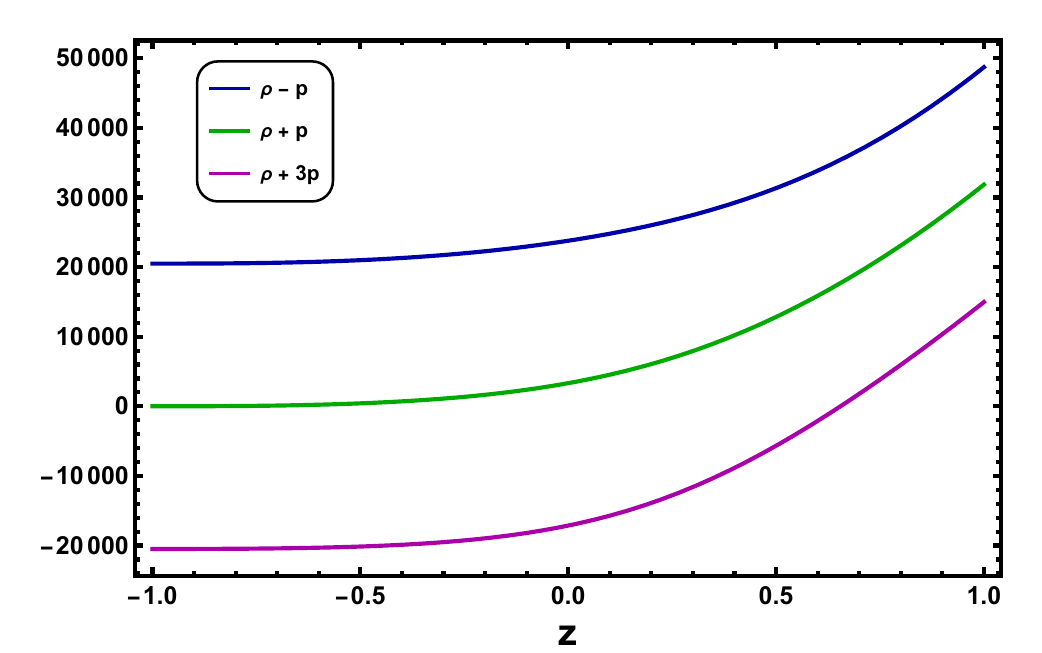}
        \caption{Energy conditions in redshift for Pantheon$^+$ data.}
        \label{ec12}
    \end{subfigure}%
    \hfill\vspace{5mm}
     \begin{subfigure}{\textwidth}
        \centering
        \includegraphics[width=75mm]{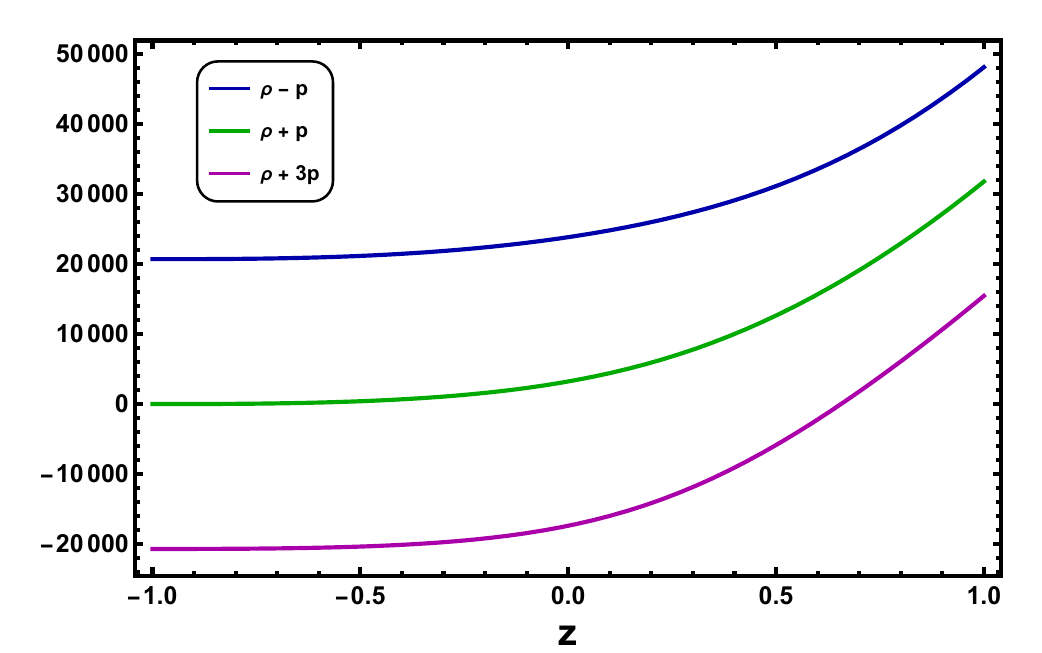}
        \caption{Energy conditions in redshift for $\text{OHD}+\text{Pantheon}^{+}+\text{BAO}$ data.}
        \label{ec13}
    \end{subfigure}%
    \caption{The behavior of energy conditions for the model within coincident gauge.}
    \label{fig5.3}
\end{figure}
and these energy conditions essentially serve as boundary conditions. The violation of the SEC due to DE suggests that these boundary conditions may not hold physical relevance. However, because of the fundamental causal structure of space-time, the gravitation attraction can be characterized by the energy conditions \cite{Capozziello_2019_28_1930016}. Moreover, these boundary conditions play a significant role in shaping the cosmic evolution of the Universe \cite{Carroll_2003_68_023509}.

The evolutionary behaviour of the energy conditions is depicted in Fig. \ref{fig5.3}, for the constrained values of the free parameters obtained from the OHD [Fig. \ref{ec11}], Pantheon$^+$ [Fig. \ref{ec12}] and $\text{OHD}+\text{Pantheon}^{+}+\text{BAO}$ [Fig. \ref{ec13}] datasets. We observed that during the early epoch, the NEC decreases and remains positive throughout but vanishes at the late epoch. The DEC remains positive throughout and does not violate. While the SEC does not violate at early times, it does violate at late times. These observations highlight the dynamic nature of the energy conditions across different cosmological epochs. The parameter scheme for the plots is $Q_{0}=69$, $\alpha= -0.0287$ and $\beta= 0.345$.

\subsection{Model reconstructed through non-coincident gauge}
Here, we will analyze the dynamical behavior of the Universe for reconstructed model. Using the Eqs. \eqref{eq5.1}, \eqref{eq5.2}, \eqref{eq5.4} and Hubble parameter one can obtain the expressions for the variables $\rho$ and $p$, which are derived from a field equations. These equations incorporate various parameters such as $\alpha_{1}$, $\alpha_{2}$, $\alpha_{3}$, $\beta_{1}$, $\beta_{2}$ and $\gamma_{1}$ along with dynamical quantities like the Hubble parameter 
$H$ and its derivative with respect to time $t$ (i.e. $\dot{H}$).
{\small
\begin{align*}
    \rho =& ~\frac{-\sqrt{3}}{2(27\gamma_{1}^{2}-8Q)^{3/2}}\bigg(648  \beta_{1} \gamma_{1}^{2}  H^2-729  \beta_{1} \gamma_{1}^{4}-4374  \beta_{2} \gamma_{1}^{4} H^2-192  \beta_{1} H^2 Q+3240  \beta_{2} \gamma_{1}^{2}  H^2 Q+32  \beta_{2} Q^3\\[10pt] 
    &+576  \beta_{1} H \dot{H}-576  \beta_{2} H^2 Q^2 +7776  \beta_{2} \gamma_{1}^{2}  H \dot{H}-1728  \beta_{2} H \dot{H} Q-32  \beta_{1} Q^2-108  \beta_{2} \gamma_{1}^{2}  Q^2-432  \beta_{1} \dot{H}\\[10pt]
    &+1296  \beta_{2} \dot{H} Q-5832  \beta_{2} \gamma_{1}^{2}  \dot{H}+324  \beta_{1} \gamma_{1}^{2}  Q \bigg)-\frac{1}{2(27\gamma_{1}^{2}-8Q)}\bigg(96 \alpha_{3} H^2 Q^2 +48 \alpha_{2} H^2 Q -162 \alpha_{2} \gamma_{1}^{2}  H^2 \\[10pt]
    & +576 \alpha_{3} H \dot{H} Q-324 \alpha_{3} \gamma_{1}^{2}  H^2 Q-8 \alpha_{3} Q^3 +27 \alpha_{3} \gamma_{1}^{2}  Q^2 -1944 \alpha_{3} \gamma_{1}^{2}  H \dot{H}  +8 \alpha_{1} Q -432\alpha_{3} \dot{H} Q -27 \alpha_{1} \gamma_{1}^{2}\\[10pt]
    &+1458 \alpha_{3} \gamma_{1}^{2}  \dot{H}   \bigg)~,\\[10pt]
    p =& ~48 \alpha_{3} H^2 \dot{H}-3 \alpha_{2} H^2-6 \alpha_{3} H^2 Q-48 \alpha_{3} H \dot{H}-2 \alpha_{2} \dot{H}+9 \alpha_{3} \dot{H}-4 \alpha_{3} \dot{H} Q-\frac{1}{2} (\beta_{1}+\beta_{2}Q)\sqrt{81 \gamma_{1} -24 Q}\\[10pt]
    &+\frac{\alpha_{3} Q^2}{2}-\frac{\alpha_{1}}{2}+\frac{\sqrt{3}}{(27 \gamma_{1}^{2} -8 Q)^{3/2}}\bigg(384  \beta_{1} H \dot{H}-384  \beta_{1} H^2 \dot{H}+1152  \beta_{2} H^2 \dot{H} Q-72  \beta_{1} \dot{H}+216  \beta_{2} \dot{H} Q\\[10pt]
    &+5184  \beta_{2} \gamma_{1}^{2} H \dot{H}-5184  \beta_{2} \gamma_{1}^{2} H^2 \dot{H}-1152  \beta_{2} H \dot{H} Q-972  \beta_{2} \gamma_{1}^{2} \dot{H}\bigg)+\frac{1}{\sqrt{9 \gamma_{1}^{2} -\frac{8 Q}{3}}}\bigg(12 \beta_{1} H^2+36 \beta_{2} H^2 Q\\[10pt]
    &+8 \beta_{1} \dot{H}-81 \beta_{2} \gamma_{1}^{2}  H^2-54 \beta_{2} \gamma_{1}^{2}  \dot{H}+24 \beta_{2} \dot{H} Q-6 \beta_{2} Q^2-2 \beta_{1} Q+\frac{27 \beta_{2} \gamma_{1}^{2}  Q}{2}\bigg)~.
\end{align*}
}
\begin{figure}[H]
    \centering
    \begin{subfigure}{0.5\textwidth}
        \centering
        \includegraphics[width=78.5mm]{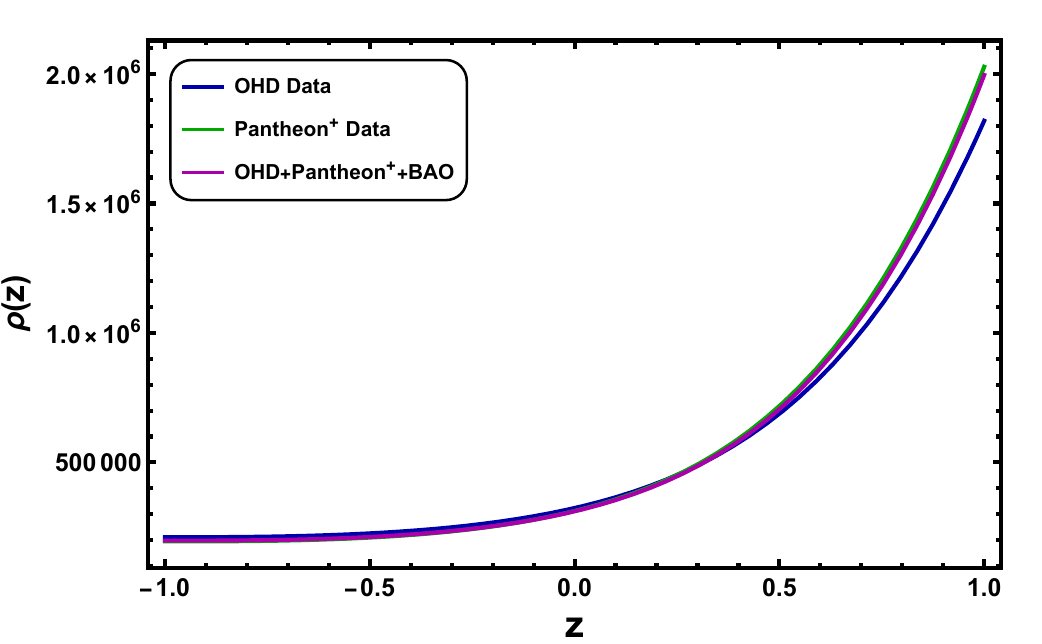}
        \caption{Evolution of energy density in redshift.}
        \label{rho1}
    \end{subfigure}%
    \hfill
    \begin{subfigure}{0.5\textwidth}
        \centering
        \includegraphics[width=75mm]{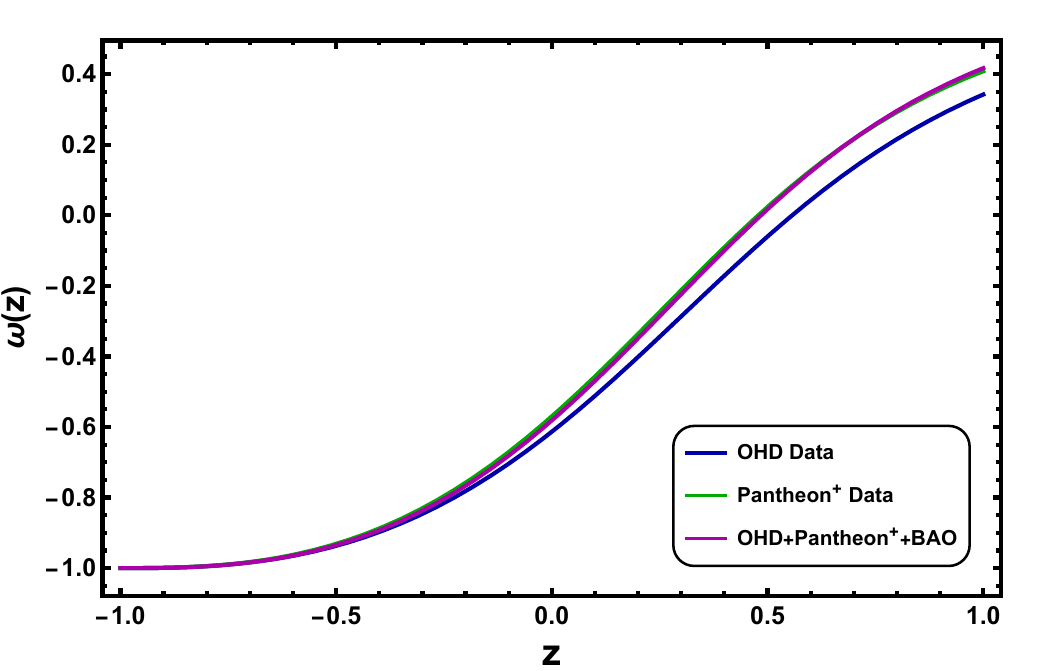}
        \caption{Evolution of EoS parameter in redshift.}
        \label{omega1}
    \end{subfigure}%
    \caption{The behavior of the dynamical parameters for the model reconstructed through non-coincident gauge.}
    \label{fig5.4}
\end{figure}
Fig. \ref{rho1} shows that while the energy density decreases over time, it never completely vanishes. The current values of the EoS parameter are $\omega_{0} \simeq -0.63$, $\omega_{0} \simeq -0.59$ and $\omega_{0} \simeq -0.60$ for the constrained free parameters from the OHD dataset, Pantheon$^{+}$ dataset and the combined $\text{OHD}+\text{Pantheon}^{+}+\text{BAO}$ dataset, respectively from Fig. \ref{omega1}. The energy density and EoS parameters are determined by the coefficients of the model, which influence their evolution. This model exhibits quintessence behavior at present.
\begin{figure}[ht]
    \centering
    \begin{subfigure}{0.5\textwidth}
        \centering
        \includegraphics[width=75mm]{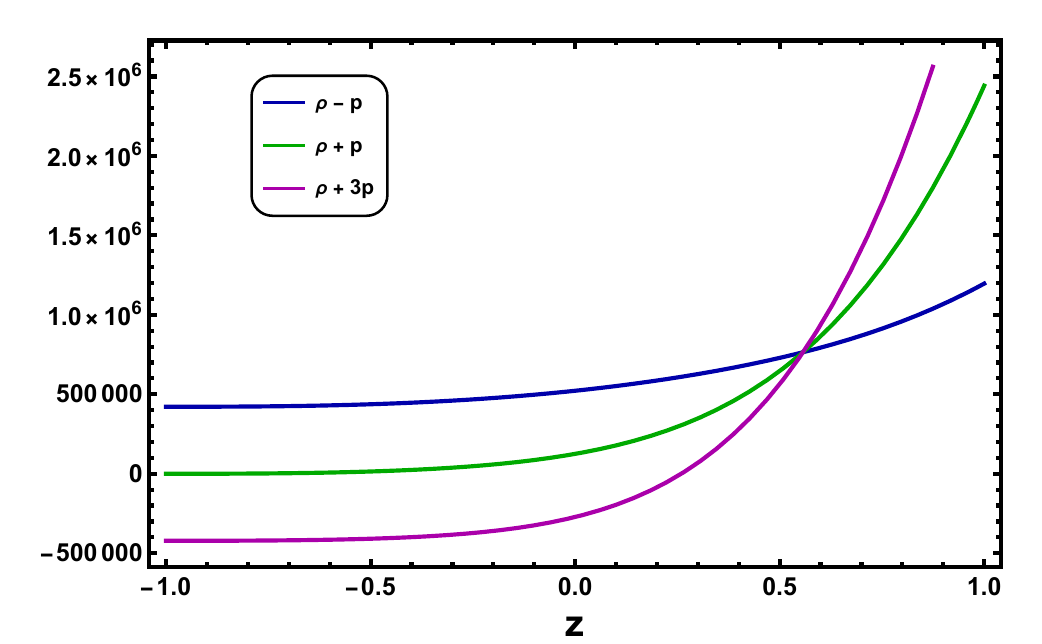}
        \caption{Energy conditions in redshift for OHD data.}
        \label{ec1}
    \end{subfigure}%
    \hfill
    \begin{subfigure}{0.5\textwidth}
        \centering
        \includegraphics[width=75mm]{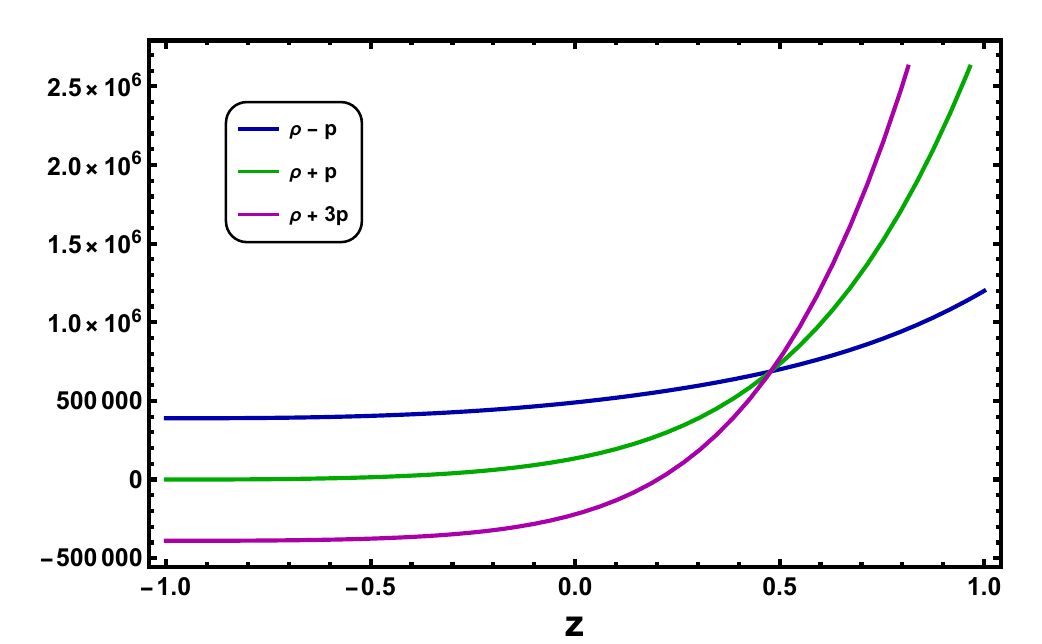}
        \caption{Energy conditions in redshift for Pantheon$^+$ data.}
        \label{ec2}
    \end{subfigure}%
    \hfill\vspace{5mm}
     \begin{subfigure}{\textwidth}
        \centering
        \includegraphics[width=75mm]{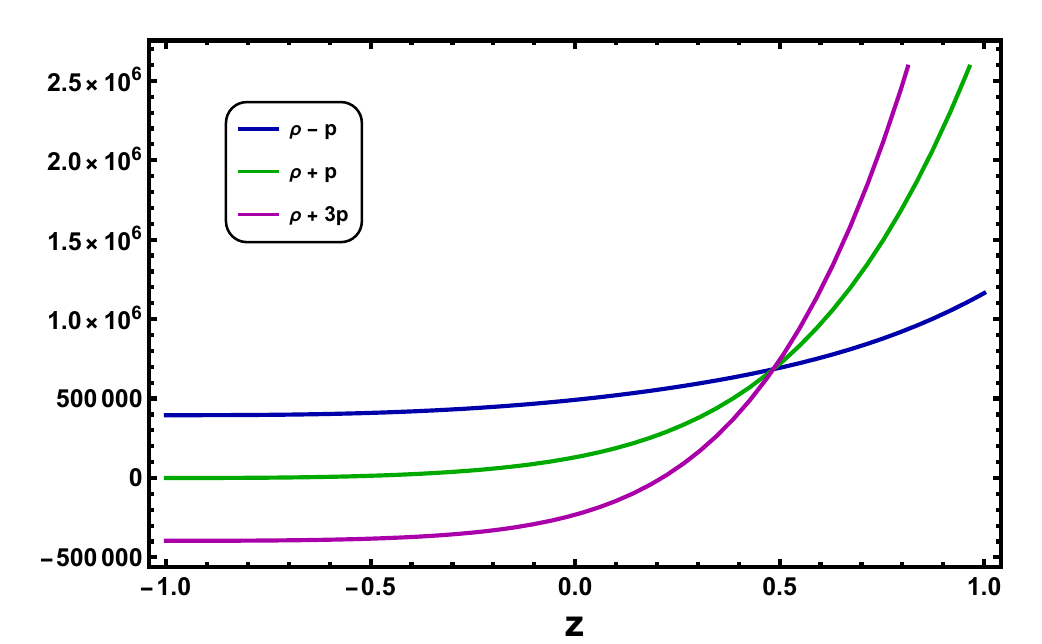}
        \caption{Energy conditions in redshift for $\text{OHD}+\text{Pantheon}^{+}+\text{BAO}$ data.}
        \label{ec3}
    \end{subfigure}%
    \caption{The behavior of energy conditions for the model reconstructed through non-coincident gauge.}
    \label{fig5.5}
\end{figure}
As illustrated in Fig. \ref{fig5.5}, across all datasets, the NEC decreases from early to late time, remains positive throughout, but approaches zero in late-time. In contrast, the DEC remains positive at all times without any violation. The SEC is satisfied in the early Universe but is violated in the late epoch. The parameter scheme for the plots is $\alpha_{1}=1.20$, $\alpha_{2}= 0.50$, $\alpha_{3}= -0.0001$, $\beta_{1}=0.90$ and $\beta_{2}=0.01$.\\
The dynamical behaviour and energy conditions are presented in the previous section offer valuable insights into the evolution of the models. To further understand the stability of the reconstructed model, we now turn to the analysis of scalar perturbations, which will provide a deeper understanding of the stability of the model.

\subsection{Analyzing stability with scalar perturbation} 
Our focus in this section will be on perturbations of homogeneous and isotropic FLRW metrics and their evolution, which ultimately determines whether cosmological solutions are stable in $f(Q)$ gravity. In order to study perturbations near the solutions $H(t)$ and $\rho(t)$, let us consider small deviations from the Hubble parameter and the evolution of the energy density defined as \cite{Cruz-Dombriz_2012_29_245014, Farrugia_2016_94_124054},
\begin{equation}\label{eq5.9}
    H(t)\rightarrow H(t)(1+\delta)~, \quad\quad \rho\rightarrow\rho(1+\delta_{m})~,
\end{equation}
where $\delta$ and $\delta_{m}$ represent the isotropic deviation of the Hubble parameter and the matter over density, respectively. In this case, $H(t)$ and $\rho(t)$ represent zero order quantities [in some references, these are sometimes designated as $H_{0}(t)$ and $\rho_{0}(t)$, but this notation is avoided here in order to distinguish from quantities evaluated at present], which satisfy Eqs. \eqref{eq5.1}, \eqref{eq5.2} and continuity equation provided in \eqref{continuityeqn}. The perturbation of the function $f$ and its derivatives are
\begin{equation*}
    \delta f = f_{Q}\delta Q~,\quad \delta f_{Q} = f_{QQ}\delta Q~, \quad \delta f_{QQ} = f_{QQQ}\delta Q~.
\end{equation*}
where $\delta x$ represents the first-order perturbation of the variable $x$. Here $\delta Q = (-12H^{2}+9H)\delta$. In this way, the perturbed equations of Eq. \eqref{eq5.1} and continuity equation become,
\begin{eqnarray}
    c_{1}\dot{\delta}+c_{0}\delta &=& \rho\delta_{m}~,\label{eq5.10}\\[5pt]
    \dot{\delta_{m}}+3H(1+\omega)\delta &=& 0~.\label{eq5.11}
\end{eqnarray}
where
    \begin{align*}
    c_{0} =& ~6f_{Q}H^{2}-72f_{QQ}H^{4}+108f_{QQ}H^{3}-\frac{81f_{QQ}H^{2}}{2}+36f_{QQ}H\dot{H}-\frac{27f_{QQ}\dot{H}}{2}-\frac{243}{2}f_{QQQ}H\dot{H}\\
    &-216f_{QQQ}H^{3}\dot{H}+324f_{QQQ}H^{2}\dot{H}~,\\[10pt]
    c_{1} =& ~18f_{QQ}H^{2}-\frac{27f_{QQ}H}{2}~.
\end{align*}
By solving Eqs. \eqref{eq5.10} and \eqref{eq5.11}, one can determine the stability of a particular FLRW cosmological solution in the context of $f(Q)$ gravity with connection-III. Since Eq. \eqref{eq5.10} has a linear character, the solution for $\delta(t)$ can generally be split into two branches: the first one corresponds to the solution of the homogeneous equation in \eqref{eq5.10}, which reflects perturbations induced by a particular gravitational Lagrangian. The second branch would correspond to the particular solution of that equation, which is merely affected by the growth of matter perturbations $\delta_{m}$. Considering the Eqs. \eqref{eq5.4}, \eqref{eq5.10}, \eqref{eq5.11} and the Hubble parameter, one can find the perturbation equations for the reconstructed model. 
\begin{figure}[ht]
    \centering
    \begin{subfigure}{0.5\textwidth}
        \centering
        \includegraphics[width=88mm]{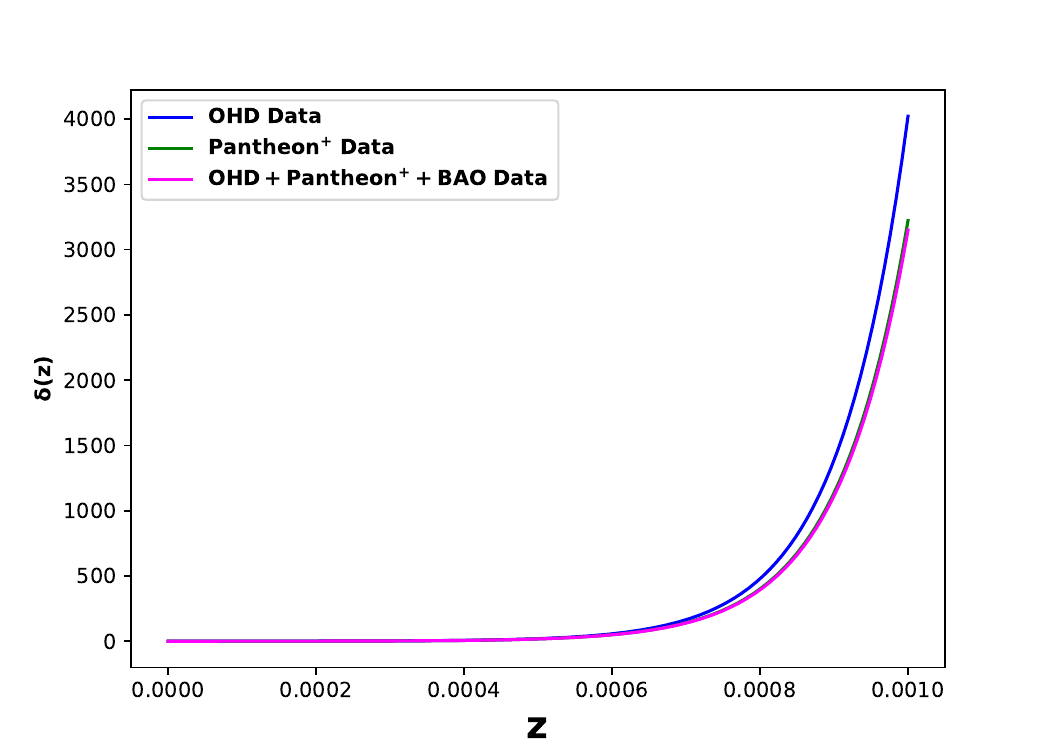}
        \caption{Evolution of $\delta$ in redshift.}
        \label{delta}
    \end{subfigure}%
    \hfill
    \begin{subfigure}{0.5\textwidth}
        \centering
        \includegraphics[width=88mm]{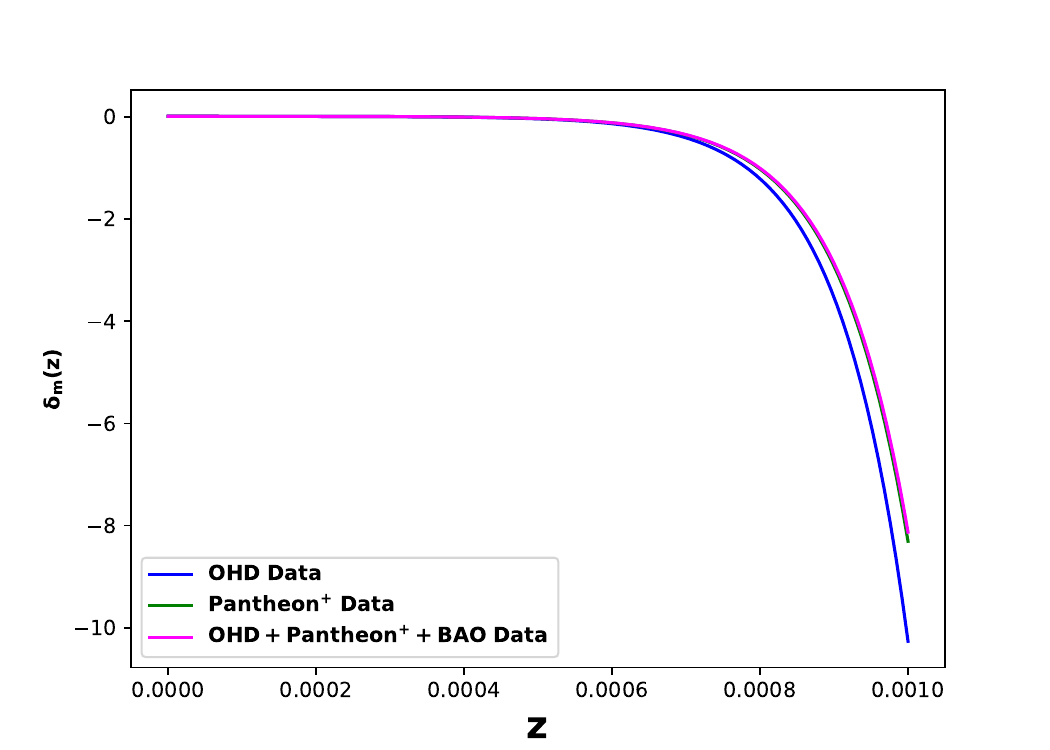}
        \caption{Evolution of $\delta_{m}$ in redshift.}
        \label{deltam}
    \end{subfigure}%
    \caption{Evolution of $\delta$ and $\delta_{m}$ versus redshift. Herein, we set the initial conditions $\delta(0) = 0.1$ and $\delta_{m}(0)$ = 0.01.}
    \label{fig5.6}
\end{figure}
The analysis of cosmological perturbations $\delta$ and $\delta_{m}$ as illustrated in Fig. \ref{fig5.6} provides valuable insights into the behavior of these perturbations across different redshifts. Perturbation $\delta$ is compared against three datasets: OHD dataset, Pantheon$^{+}$ dataset and the combined $\text{OHD}+\text{Pantheon}^{+}+\text{BAO}$ dataset. Similarly, perturbation $\delta_{m}$ is evaluated using the same datasets. Both perturbations exhibit small variations, demonstrating consistency across different observational constraints. Importantly, it can be seen that $\delta$ and $\delta_{m}$ converge to zero at the late epoch in the redshift, which confirms the stability of the model with respect to the Hubble parameter. This convergence highlights the robustness of the model in describing the expansion of the Universe over time.

\section{Conclusion}\label{sec5.5}
In this chapter, we applied the cosmological reconstruction method within the framework of $f(Q)$ gravity, using a non-trivial connection specifically, connection III. This introduces a non-metricity scalar that involves the Hubble parameter and an additional function. This unique formulation distinguishes the non-coincident gauge model, leading to novel cosmological dynamics and potential challenges related to stability. Through a comprehensive analysis utilizing the Hubble parameter $H(z)$ and various observational datasets, our models successfully capture the expansion history of the Universe. When comparing the two models, we found that both exhibit a decreasing energy density over time that does not vanish, but with different behaviors for the EoS parameter. The model within the coincident gauge results in an EoS parameter near $\omega_0 \sim -0.78$ at the present time, indicating a more consistent dynamical evolution. On the other hand, the non-coincident gauge model shows a slightly higher EoS parameter ($\omega_0 \sim -0.63$) and a more significant violation of the SEC at late time. 

Both models exhibit quintessence like behavior, but the non-coincident gauge model deviates more substantially from standard cosmology in the late Universe. This dual behavior highlights the flexibility of $f(Q)$ gravity in addressing the cosmological constant problem. Our detailed evaluation of the energy conditions for both models revealed that the NEC and the DEC are consistently satisfied, ensuring their physical plausibility. While the SEC is satisfied at early times, it is violated at later times, aligning with the observed accelerated expansion of the Universe. 

Furthermore, a stability analysis through scalar perturbations confirmed the robustness of the non-coincident gauge model, indicating that it is free from instabilities. These results position $f(Q)$ gravity as a compelling alternative to the $\Lambda$CDM model, providing a comprehensive explanation for the accelerating expansion of the Universe and offering new insights into fundamental cosmological issues. This work makes a significant contribution to our understanding of cosmological dynamics within modified gravity theories and sets the stage for future research into the implications of $f(Q)$ gravity in modern cosmology.
\chapter{Concluding Remarks and Future
perspectives} 

\label{Chapter6} 

\lhead{Chapter 6. \emph{Concluding Remarks and Future
perspectives}} 


This thesis delves into the cosmological implications of various $f(Q)$ gravity models, providing valuable insights into the accelerated expansion of the Universe and the role of DE. Through in-depth analyses and comparisons with observational data, we demonstrate that these models present viable alternatives to the standard $\Lambda$CDM model.

Chapter \ref{Chapter2} introduced a two $f(Q)$ gravity models, focusing on the log-square-root model and exponential model. The hybrid scale factor has been incorporated to analyze the evolutionary behavior of the models. The log-square-root model shows the quintessence behavior at present time and behaves like the $\Lambda$CDM at late times whereas exponential model shows the phantom behavior and the present value of the DE EoS parameter obtained to be $\omega_{DE}=-0.82$ and $\omega_{DE}=-1.01$ respectively. The dynamical system analysis has been performed for both the cosmological models obtained with the form of $f(Q)$ and the time varying deceleration parameter in the second phase. For both the models, the critical point (0, 0, 1) and (0, 0, 0) are unstable. Whereas, the curve ($x$, $1-x$, 0) has all eigenvalues for Jacobian matrix are negative real part and zero gives the stable node behavior. The behavior of the critical points are validated in the 2-D and 3-D phase portraits. Moreover, the evolutionary behavior for both the models shows the present value of density parameters as, $\Omega_{DE}\approx 0.7$ and $\Omega_{m}\approx 0.3$. Both the models shows early radiation era and late time DE era, transition through matter dominated era. The total EoS parameter obtained from hybrid scale factor and the total EoS parameter from the evolution plot shows same behavior of Universe for both the models.

In Chapter \ref{Chapter3}, by considering higher power of non-metricity in the expression for $f(Q)$ and performing some algebraic manipulations, we explicitly derived the expression for the Hubble function in terms of the redshift. Then, we constrained the model parameters and the EoS parameter using Bayesian statistical analysis with MCMC techniques, along with observational data from Hubble measurements and extended Pantheon+SH0ES dataset. The error bar plots reveal that both the curve for our models and the $\Lambda$CDM model pass through the range obtained from these datasets. Furthermore, the constrained parameters were validated using the BAO dataset and constraints on the geometrical parameters were derived to investigate the accelerating behavior of the Universe. This analysis shows that the cosmological model transitions from a deceleration to an acceleration phase. We derived a deceleration parameter $q_{0} \approx -0.61$ for the transiting Universe at the present epoch. The EoS parameter was found to be $-1.16\pm0.17$ and $-1.15\pm0.17$ respectively from the Hubble dataset and Pantheon$+$SH0ES dataset respectively, indicating the phantom behavior of the model. Utilizing the constrained value of the Hubble parameter, we have calculated the age of the Universe and the $Om(z)$ diagnostic analysis has been performed, which provides a null test to the $\Lambda$CDM model. If the model deviates from the $\Lambda$CDM model, there may be interactions between DE and DM components. Our results suggest that the behavior of the $Om(z)$ parameter may favor a phantom phase.

Chapter \ref{Chapter4} delve into the cosmological reconstruction in $f(Q)$ gravity with trivial connection. We have developed some novel form of deceleration parameter and studied its impact on the $f(Q)$ gravity in the context of accelerating Universe. The parametrization must to be carefully done, taking into account both theoretical and observational constraints, in order to ensure that it is reliable and predictive over all redshift ranges. The parametrized deceleration parameter presented in the paper offers a flexible framework for analyzing the dynamics of the Universe, enabling the exploration of various cosmological models and their implications. We constrained the cosmological parameters and validated the models utilizing the observational datasets such as OHD, SNe Ia and BAO. Bayesian statistical inference techniques and MCMC methods were employed for the analysis. Our results demonstrate that the parameters of our models match the observational data well. The power-law function, demonstrates compatibility with the latest high values of $H_{0}$, aligning with recent observational values and indicating a DE-dominated Universe. Moreover the reconstructed $f(Q)$ model has GR as a particular limit, it has the same number of free parameters as $\Lambda$CDM, but in a cosmological context it creates a scenario that does not have $\Lambda$CDM as a limit. Whereas, the logarithmic function, aligns with a slightly reduced matter density and accurately models the accelerated expansion. The reconstructed $f(Q)$ model for logarithmic function is presented as a novel approach for describing cosmic evolution, particularly focusing on avoiding early-time instabilities and phantom behaviors. In this model, the effective EoS parameter remains strictly non-phantom. Also, this model passes the Big Bang Nucleosynthesis constraints trivially, meaning it does not interfere with the delicate balance required for early Universe Nucleosynthesis, making it a stable and potentially viable model for both early and late Universe dynamics. Finally the comparative analysis against $\Lambda$CDM using AIC and BIC metrics shows that these models provide competitive fits, with model II offering a slight edge in simplicity. 

In Chapter \ref{Chapter5}, we presented a detailed implementation of the cosmological reconstruction method in $f(Q)$ gravity, utilizing non-trivial connections, specifically connection III within a symmetric teleparallel background. Through this approach, we derived an intriguing cosmological model, which is a noteworthy step in categorizing the association of cosmological models in modified gravity. Our model has been meticulously analyzed using the Hubble parameter $H(z)$ and various observational datasets, providing a robust representation of the expansion history of the Universe. Then the dynamical behaviour of the model within the coincident gauge, along with the reconstructed model has been studied. Both the models exhibits quintessence behavior at the current epoch, characterized by a dynamic DE component that evolves over time. This behavior transitions smoothly to the $\Lambda$CDM model at late time, which is consistent with the observed accelerated expansion of the Universe. The dual behavior of the reconstructed model underscores the versatility and potential of $f(Q)$ gravity in addressing the cosmological constant problem. We have also thoroughly evaluated the energy conditions for both the models. The NEC remains positive throughout the cosmic evolution, indicating that the energy density is non-negative and the speed of sound is real. The DEC is consistently satisfied, ensuring that the energy density exceeds the pressure and the energy propagation remains within causal bounds. The SEC, which is satisfied in the early Universe, is violated in the late epoch. This violation is consistent with the observed accelerated expansion and suggests a deviation from standard matter-dominated model. The stability analysis for the reconstructed model through scalar perturbations has confirmed the robustness of the model. 

In summary, this thesis has demonstrated that the models, based on the $f(Q)$ gravity, provide promising alternatives to the $\Lambda$CDM model. The research explored various forms of $f(Q)$ gravity, including models based on the log-square-root, exponential and higher powers of non-metricity, offering robust frameworks for understanding the accelerated expansion of the Universe. These models have shown strong agreement with current cosmological observations, including OHD, SNe Ia and BAO. Additionally, the thesis presented novel cosmological models reconstructed within the $f(Q)$ gravity framework, utilizing trivial connections and non-trivial connections. These reconstructed models offer an effective description of cosmic evolution, ensuring compatibility with both early and late Universe dynamics, while also providing a viable alternative to the $\Lambda$CDM model. The reconstructed models further demonstrated smooth transitions from deceleration to acceleration and matched observational data well. The findings also highlighted the versatility of $f(Q)$ gravity, with models exhibiting both quintessence and phantom behavior as well as smooth transitions between deceleration and acceleration phases of the Universe. Furthermore, the stability analysis and energy condition evaluations confirmed the physical viability of the models, with particular attention given to their ability to describe the nature of DE and the evolution of the Universe. The comparison with the $\Lambda$CDM model further supports the competitive and viable nature of $f(Q)$ gravity models in explaining the accelerating expansion of the Universe and the nature of DE.

In the future, the $f(Q)$ gravity models can be extended to explore their effects in strong gravitational regimes, particularly on the propagation of gravitational waves. This would provide a valuable platform for investigating how these models influence the propagation of gravitational waves, particularly in extreme cosmic environments such as black holes, neutron stars and the early Universe. Analyzing the full CMB, Large Scale Structure spectra, DESI data, weak lensing data and other relevant datasets can offer valuable insights into our findings. Such analyses are essential for assessing whether these models provide a better fit compared to the standard $\Lambda$CDM model or not. Further extending these models to higher-dimensional spacetime or including extra fields may provide new insights into cosmic inflation and DM interactions.




\addtocontents{toc}{\vspace{1em}} 
\backmatter



\label{References}
\lhead{\emph{References}}

\providecommand{\href}[2]{#2}\begingroup\raggedright\endgroup



\chapter{Appendices}
\lhead{\emph{Appendices}}
\section*{Cosmological datasets}\label{Appendices}
\subsection*{A) Hubble data with $32$ data points :}
\begin{center}
\begin{table}[H]
\centering 
\renewcommand{\arraystretch}{1.4}
\begin{tabular}{|p{0.10\linewidth}p{0.10\linewidth}p{0.10\linewidth}p{0.08\linewidth}||p{0.10\linewidth}p{0.10\linewidth}p{0.10\linewidth}p{0.08\linewidth}|} 
\hline 
 $z$-value & $H(z)$ & $\sigma_{H}$ & Ref. &  $z$-value & $H(z)$ & $\sigma_{H}$ & Ref.\\ [0.5ex] 
\hline 
 0.07 & 69.0 & 19.6 & \cite{Zhang_2014_14_1221}      &  0.4783 & 80.9 & 9.0 &  \cite{Moresco_2016_2016_014} \\ 
 0.09 & 69.0 & 12.0 & \cite{Simon_2005_71_123001}    &  0.48 & 97 & 62 &  \cite{Stern_2010_2010_008} \\ 
 0.12 & 68.6 & 26.2 & \cite{Zhang_2014_14_1221}      &  0.593 & 104 & 13 & \cite{Moresco_2012_2012_006} \\
 0.17 & 83 & 8 &  \cite{Simon_2005_71_123001}          &  0.68 & 92 & 8 & \cite{Moresco_2012_2012_006} \\
 0.179 & 75.0 & 4.0 & \cite{Moresco_2012_2012_006}    &  0.75 & 98.8 & 33.6 &  \cite{Borghi_2022_928_L4} \\
 0.199 & 75.0 & 5.0 &  \cite{Moresco_2012_2012_006}   &  0.781 & 105 & 12 &  \cite{Moresco_2012_2012_006}\\
 0.200 & 72.9 & 29.6 &  \cite{Zhang_2014_14_1221}    &  0.875 & 125 & 17 &  \cite{Moresco_2012_2012_006} \\
 0.27 & 77 & 14 &  \cite{Simon_2005_71_123001}         &  0.88 & 90 & 40 &  \cite{Stern_2010_2010_008}  \\
 0.28 & 88.8 & 36.6 & \cite{Zhang_2014_14_1221}      &  0.9 & 117 & 23 &  \cite{Simon_2005_71_123001} \\
 0.352 & 83 & 14 &  \cite{Moresco_2012_2012_006}      &  1.037 & 154 & 20 &  \cite{Moresco_2012_2012_006} \\ 
 0.38 & 83.0 & 13.5 &  \cite{Moresco_2016_2016_014}   &  1.3 & 168 & 17 &  \cite{Simon_2005_71_123001} \\
 0.4 & 95 & 17 &  \cite{Simon_2005_71_123001}          &  1.363 & 160 & 33.6 & \cite{Moresco_2015_450_L16} \\
 0.4004 & 77 & 10.2 &  \cite{Moresco_2016_2016_014}   &  1.43 & 177 & 18 &  \cite{Simon_2005_71_123001} \\
 0.425 & 87.1 & 11.2 &  \cite{Moresco_2016_2016_014}  &  1.53 & 140 & 14 &  \cite{Simon_2005_71_123001} \\
 0.445 & 92.8 & 12.9 &  \cite{Moresco_2016_2016_014}  &  1.75 & 202 & 40 &  \cite{Simon_2005_71_123001} \\
 0.47 & 89 & 49.6 & \cite{Ratsimbazafy_2017_467_3239} &  1.965 & 186.5 & 50.4 & \cite{Moresco_2015_450_L16} \\ 
\hline 
\end{tabular}
\caption{$H(z)$ measurements expressed in [km s$^{-1}$ Mpc$^{-1}$] units, were obtained using the CC technique, along with the associated errors.}
\label{table6.1} 
\end{table}
\end{center}

\subsection*{B) Pantheon$^{+}$ data with $1701$ samples :}
\begin{center}
\begin{table}[H]
\centering 
\renewcommand{\arraystretch}{1.5}
\begin{tabular}{|p{0.30\linewidth}p{0.15\linewidth}p{0.24\linewidth}p{0.12\linewidth}|}
\hline
 Source & $N_{\mathrm{SN}}~$/~$N_{\mathrm{Tot}}$ & $z$ range &  Ref.\\ [0.5ex]
\hline
 LOSS1 & $105/165$ & $0.0020-0.0948$ & \cite{Ganeshalingam_2010_190_418} \\
 LOSS2 & $48/78$ & $0.0008-0.082$ & \cite{Stahl_2019_490_3882} \\
 SOUSA & $57/121$ & $0.0008-0.0616$ & \cite{Peter_2014_354_89} \\
 CSP & $89/134$ & $0.0038-0.0836$ & \cite{Krisciunas_2017_154_211} \\
CfA1 & $13/22$ & $0.0031-0.123$ & \cite{Riess_1999_117_707} \\
CfA2 & $24/44$ & $0.0067-0.0542$ & \cite{Jha_2006_131_527} \\
 CfA3S + CfA3K & $92/185$ & $0.0032-0.084$ & \cite{Hicken_2009_700_331} \\
 CfA4 & $50/94$ & $0.0067-0.0745$ & \cite{Hicken_2012_200_12} \\
 LOWZ & $46/95$ & $0.0014-0.123$ & \cite{Jha_2007_659_122, Milne_2010_721_1627, Tsvetkov_2010_30_2, Zhang_2010_122_1, Contreras_2010_139_519, Krisciunas_2017_1_36, Stritzinger_2011_142_156, Wee_2018_863_90, Kawabata_2020_893_143} \\
 CNIa0.02 & $15/17$ & $0.0041-0.0303$ & \cite{Chen_2022_259_53} \\
 Foundation & $179/242$ & $0.0045-0.1106$ & \cite{Foley_2018_475_193} \\
 SDSS & $321/499$ & $0.0130-0.5540$ & \cite{Sako_2018_130_064002} \\
 PS1MD & $269/370$ & $0.0252-0.670$ & \cite{Scolnic_2018_859_101} \\
 SNLS & $160/239$ & $0.1245-1.06$ & \cite{Betoule_2014_568_A22} \\
 DES & $203/251$ & $0.0176-0.850$ & \cite{Brout_2022_938_110, Smith_2020_160_267} \\
 HDFN & $0/1$ & $1.755$ & \cite{Gilliland_1999_521_30, Riess_2001_560_49} \\
 SCP & $6/8$ & $1.014-1.415$ & \cite{Suzuki_2012_746_85} \\
 CANDELS + CLASH & $8/13$ &$ 1.03-2.26$ & \cite{Riess_2018_853_126} \\
 GOODS + PANS & $16/29$ & $0.460-1.390$ & \cite{Riess_2004_607_665, Riess_2007_659_98} \\
\hline
\end{tabular}
\caption{The Pantheon$^+$ compilation includes various samples, detailing the number of SNe Ia utilized in the cosmological sample ($N_{\mathrm{SN}}$) versus the total count from the complete dataset ($N_{\mathrm{Tot}}$) along with the specifics on the redshift range covered.}
\label{table6.2} 
\end{table}
\end{center}

\subsection*{C) BAO measurement samples :}
\begin{center}
\begin{table}[H]
\renewcommand\arraystretch{1.5}
\centering
\begin{tabular}{|p{0.12\linewidth}p{0.25\linewidth}p{0.20\linewidth}p{0.20\linewidth}p{0.10\linewidth}|}
\hline
 ~~$z_{BAO}$~~ & ~~$\frac{r_{s}(z_{d})}{D_{V}(z_{\text{BAO}})}$~~ & ~~$\frac{d_{A}(z_{*})}{D_{V}(z_{\text{BAO}})}$~~ & ~~$\frac{d_{A}(z_{*})}{D_{V}(z_{\text{BAO}})}\cdot\frac{r_{s}(z_{d})}{r_{s}(z_{*})}$ & ~~Ref.~~\\ [0.5ex] \hline
~~$0.106$~~ & $0.336 \pm 0.015$ & $30.95 \pm 1.46$ & ~~~~$32.35 \pm 1.45$~~ & ~~\cite{Beutler_2011_416_3017}~~ \\
~~$0.200$~~ & $0.1905 \pm 0.0061$ & $17.55 \pm 0.60$ & ~~~~$18.34 \pm 0.59$~~ & ~~\cite{Percival_2010_401_2148}~~ \\
~~$0.350$~~ & $0.1097 \pm 0.0036$ & $10.11 \pm 0.37$ & ~~~~$10.56 \pm 0.35$~~ & ~~\cite{Percival_2010_401_2148}~~ \\
~~$0.440$~~ & $0.0916 \pm 0.0071$ & $8.44 \pm 0.67$ & ~~~~$8.82 \pm 0.68$~~ & ~~\cite{Blake_2011_418_1707}~~ \\
~~$0.600$~~ & $0.0726 \pm 0.0034$ & $6.69 \pm 0.33$ & ~~~~$6.99 \pm 0.33$~~ & ~~\cite{Blake_2011_418_1707}~~ \\
~~$0.730$~~ & $0.0592 \pm 0.0032$ & $5.45 \pm 0.31$ & ~~~~$5.70 \pm 0.31$~~ & ~~\cite{Blake_2011_418_1707}~~ \\
\hline
\end{tabular}
\caption{The most recent BAO distance dataset includes measurements at six different redshifts.}
    \label{table6.3}
\end{table}
\end{center}

\cleardoublepage
\pagestyle{fancy}
\label{Publications}
\lhead{\emph{List of Publications and Presentations}}
\chapter{List of Publications and Presentations}
\section*{Thesis Publications}
\begin{enumerate}
    \item \textbf{S. A. Narawade}, Shashank P. Singh and B. Mishra, ``Accelerating cosmological models in $f(Q)$ gravity and the phase space analysis", \textit{Phys. of the Dark Universe} \textbf{42} (2023) 101282.

    \item \textbf{S. A. Narawade} and B. Mishra, ``Phantom cosmological model with observational constraints in $f(Q)$ gravity", \textit{Ann. der Phys.} \textbf{535} (2023) 2200626.

        \item \textbf{S. A. Narawade} and B. Mishra, ``Insights into $f(Q)$ gravity: modeling through deceleration parameter", \textit{J. High Energy Astrop.} \textbf{45} (2025) 409.

    \item \textbf{S. A. Narawade}, Santosh V Lohakare and B. Mishra, ``Stable $f(Q)$ gravity model through non-trivial connection", \textit{Annals Phys.} \textbf{474} (2025) 169913.

     \item \textbf{S. A. Narawade}, S. H. Shekh, B. Mishra, Wompherdeiki Khyllep and Jibitesh Dutta, ``Modelling the Accelerating Universe with $f(Q)$ Gravity: Observational Consistency", \textit{Eur. Phys. J. C} \textbf{84} (2024) 773. 
\end{enumerate}

\section*{Other Publications}
\begin{enumerate}
\item \textbf{S. A. Narawade}, Laxmipriya Pati, B. Mishra and S. K. Tripathy, ``Dynamical system analysis for accelerating models in non-metricity $f(Q)$ gravity", \textit{Phys. of the Dark Universe} \textbf{36} (2022) 101020.
\item \textbf{S. A. Narawade}, M. Koussour and B. Mishra, ``Constrained $f(Q,T)$ gravity accelerating cosmological model and its dynamical system analysis", \textit{Nuc. Phys. B} \textbf{992} (2023) 116233.
\item \textbf{S. A. Narawade}, M. Koussour and B. Mishra, ``Observational Constraints on Hybrid Scale Factor in $f(Q,T)$ Gravity with Anisotropic Space-Time", \textit{Ann. der Phys.} \textbf{535} (2023) 2300161.
\item \textbf{S. A. Narawade}, S. K. Tripathy, Raghunath Patra and B. Mishra, ``Baryon asymmetry constraints on Extended Symmetric Teleparallel Gravity", \textit{Gravitation and Cosmology} \textbf{30} (2024) 135.
\item Laxmipriya Pati, \textbf{S. A. Narawade}, S. K. Tripathy and B. Mishra, ``Evolutionary behaviour of cosmological parameters with dynamical system analysis in $f(Q,T)$ gravity", \textit{Eur. Phys. J. C} \textbf{83} (2023) 445.
\item Rahul Bhagat, \textbf{S. A. Narawade} and B. Mishra, ``Weyl type $f(Q,T)$ gravity observational constrained cosmological model", \textit{Phys. of the Dark Universe} \textbf{41} (2023) 101250.
\item Rahul Bhagat, \textbf{S. A. Narawade}, B. Mishra and S. K. Tripathy, ``Constrained cosmological model in $f(Q,T)$ gravity with non-linear non-metricity", \textit{Phys. of the Dark Universe} \textbf{42} (2023) 101358.
\item Muhammad Azzam Alwan, Tomohiro Inagaki, B. Mishra and \textbf{S. A. Narawade}, ``Neutron Star in Covariant $f(Q)$ gravity", \textit{J. Cosmol. Astrophys.} \textbf{2024} (2024) 011.
\item Y. Kalpana Devi, \textbf{S. A. Narawade} and B. Mishra, ``Constraining Parameters for the Accelerating Universe in $f(R,\mathcal{L}_{m})$ Gravity", \textit{Phys. of the Dark Universe} \textbf{46} (2024) 101640.
\item Muhammad Azzam Alwan, Tomohiro Inagaki, \textbf{S. A. Narawade} and B. Mishra, ``Exploring the universal $\bar{\mathcal{I}}-\mathcal{C}$ relations for relativistic stars in $f(Q)$ gravity", \textit{arXiv:2501.11296.}
\end{enumerate}

\section*{Conferences and Presentations}
\begin{enumerate}
     \item	Participated in the $27^{th}$ International Conference of International Academy of Physical Sciences on Advances in Relativity and Cosmology (PARC-2021) organized by BITS Pilani, Hyderabad Campus, India (October 26 – 28, 2021).

     \item Presented a paper entitled \emph{“Phantom cosmological model with observational constraints in $f(Q)$ gravity"} in the \textit{Physical Interpretations of Relativity Theory (PIRT-2023)} organized by Bauman Moscow State Technical University, Moscow, Russia (July 3 - 6, 2023).

     \item Presented a paper entitled \emph{“Accelerating cosmological models in $f(Q)$ gravity and the phase space analysis"} in the \textit{The Indian Mathematical Society (\textsc{IMS-2023})} organized by BITS-Pilani, Hyderabad Campus, India (December 23 - 25, 2023).
    
    \item	Presented a paper entitled \emph{“Stable $f(Q)$ gravity model through non-trivial connection"} in \textit{II International Scientific Conference Space. Time. Civilization. (STC -- 2024)} organized by BITS Pilani, Hyderabad Campus, India (November 02 – 07, 2024).
\end{enumerate}

\section*{Workshop/ International visit (In person)}
\begin{enumerate}

    \item Participated in the workshop on \emph{General Relativity and Cosmology (GRC - 2022)} organized by GLA University, Mathura. (November 24 - 26, 2022).
    
    \item Participated in the \emph{Teacher’s Enrichment Workshop} organized by BITS-Pilani Hyderabad Campus (January 09 - 14, 2023).

    \item A research visit as a Special Research Student under the BITS-HU joint supervision program to the \emph{Hiroshima University} at Hiroshima, Japan (04 January – 29 February, 2024).
\end{enumerate}

\cleardoublepage

\pagestyle{fancy}
\lhead{\emph{Biography}}
\chapter{Biography}
\textbf{\Large Brief Biography of the Candidate :}\\
\textbf{Mr. Shubham Atmaram Narawade} completed his M.Sc. from National Institute of Technology, Manipur, Manipur, in 2019. He achieved the 31$^{st}$ rank in the joint CSIR-UGC NET all-India exams in June 2019. He also secured 581$^{st}$ all-India rank in the GATE exam in 2019. He published several research papers in reputed national and international journals. He has presented research papers at several national and international conferences. He visited Hiroshima University, Japan as a Special Research Student under BITS-HU joint supervision program during the period January 04, 2024 to February 29, 2024.\\

\textbf{\Large Brief Biography of the Supervisor :}\\
\textbf{Prof. Bivudutta Mishra} received his Ph.D. degree from Sambalpur University, Odisha, India, in 2003. His main research areas are Geometrically Modified Theories of Gravity, Theoretical Aspects of Dark Energy and Wormhole Geometry. He has published over 170 research papers in national and international journals, presented papers at conferences in India and abroad, supervised six Ph.D. students and is currently guiding six more. He has also organized academic and scientific events in the department. He has become a member of the scientific advisory committee of national and international academic events. He has successfully completed multiple sponsored projects funded by Government Funding agencies and is at present working on three projects funded by CSIR, SERB-DST (MATRICS) and SERB-DST (CRG-ANRF). He is also an awardee of DAAD-RISE, 2019, 2022. He has also reviewed several research papers in highly reputed journals, is a Ph.D. examiner and is a BoS member of several universities. He has been invited by many foreign universities to share his research in scientific events, some of which are Canada, Germany, the Republic of China, Russia, Australia, Switzerland, Japan, the UK, Poland, etc. As an academic administrator, he was Head of the Department of Mathematics from September 2012 to October 2016 and was Associate Dean of International Programmes and Collaborations from August 2018 to September 2024. He is also a visiting professor at Bauman Moscow State Technical University, Moscow, a visiting associate at Inter-University Centre for Astronomy and Astrophysics, Pune, a Fellow of the Royal Astronomical Society, UK and a Fellow of the Institute of Mathematics and Applications, UK. Foreign member of the Russian Gravitational Society, Moscow. He has been listed among the top $2\%$ of scientists according to the Stanford University author database of standardized citation indicators.
\end{document}